\documentclass{jaa}
\usepackage{graphicx}
\usepackage{natbib}
\def\pasp{Publ. Astron. Soc. Pacific} 
\def\aj{Astron. J.}
\def\araa{Ann. Rev. Astron. Astrophys.}
\def\apj{Astrophys. J.}
\def\aap{Astron. Astrophys.}
\def\aaps{Astron. Astrophys. Suppl.}
\def\mnras{Mon. Not. R. Astron. Soc.}
\def\nat{Nature}
\def\pasp{Publ. Astron. Soc. Pacific} 

\def\aaps{A\&AS}

\begin{document}\sloppy

\title{Intra-night optical variability monitoring of $\gamma$-ray emitting blazars\\}


\author{K.Subbu Ulaganatha Pandian\textsuperscript{1,*}, A.Natarajan\textsuperscript{1}, C.S.Stalin\textsuperscript{2}, 
Ashwani Pandey\textsuperscript{2}, S. Muneer\textsuperscript{2}  and B. Natarajan\textsuperscript{3}}
\affilOne{\textsuperscript{1}Research and Development Centre, Bharathiyar University, Coimbatore - 641046, India\\}
\affilTwo{\textsuperscript{2}Indian Institute of Astrophysics, Block II, Koramangala, Bangalore-560034, India\\}
\affilThree{\textsuperscript{3}Government Arts and Science College, Sivakasi – 626124 , India \\}


\twocolumn[{

\maketitle

\corres{subbuathoor@gmail.com}

\msinfo{10 December 2021}{13 February 2022}

\begin{abstract}
We present the results obtained from our campaign to characterize the 
intra-night-optical variability properties of 
blazars detected by the {\it Fermi} Large Area Telescope.
This involves R-band monitoring observations of a sample of 
18 blazars, that includes five flat spectrum radio quasars (FSRQs) and
thirteen BL Lac objects (BL Lacs) covering the redshift range z = 0.085$-$1.184. Our observations, carried out using
the 1.3 m J.C. Bhattacharya Telescope cover a total of 
40 nights ($\sim$200 hrs) between the period 2016 December 
and 2020 March. We characterized variability using  
the power enhanced $F-$test. We found duty cycle (DC) of variability
of about 11\% for FSRQs and 12\% for BL Lacs. Dividing
the sample into different sub-classes based on the position of the synchrotron
peak in their broad band spectral energy distribution (SED), 
we found DC of $\sim$16\%, $\sim$10\% and $\sim$7\% for low-synchrotron 
peaked (LSP), intermediate synchrotron peaked (ISP)  and high synchrotron 
peaked (HSP) blazars. Such high DC of variability in LSP blazars  
could be understood in the context of the 
R-band tracing the falling part (contributed 
by high energy electrons) of
the synchrotron component of the broad band SED. 
Also, the R-band
tracing the rising synchrotron part (produced by low energy electrons) 
in the case of ISP and HSP blazars, could cause lesser variability in them.   
Thus, the observed high DC of variability in LSP blazars relative to ISP and 
HSP blazars is in accordance with the leptonic model of emission from blazar jets. 
\end{abstract}


}]


\doinum{12.3456/s78910-011-012-3}
\artcitid{\#\#\#\#}
\volnum{000}
\year{0000}
\pgrange{1--}
\setcounter{page}{1}
\lp{1}

\section{Introduction}
Blazars are are a peculiar category of active galactic nuclei (AGN), with their
relativistic jets pointed close to the observer \citep{1995PASP..107..803U}. They 
are the dominant extragalactic population as seen by the {\it Fermi} Large Area Telescope (LAT) and radiate over the entire accessible
electromagnetic spectrum. They emit copiously in the radio band, have
a compact core jet morphology and are known to show rapid and 
high amplitude flux variations over a range of wavelengths 
\citep{1995ARA&A..33..163W}. They have high optical polarization 
\citep{1966ApJ...146..964K,1980ARA&A..18..321A} and also show
polarization variability \citep{2017ApJ...835..275R}. 
Blazars are sub-divided in flat spectrum radio
quasars (FSRQs) and BL Lac objects (BL Lacs) with FSRQs characterised
by broad optical emission lines with equivalent width greater than 5 \AA.   
It is suggested that the presence of broad emission lines in FSRQs is due
to them having a luminous broad line region (BLR), and a high
and efficient accretion process \citep{2012MNRAS.421.1764S}. On the other hand, the absence or presence
of weak emission lines in BL Lac objects is attributed to them
having low and inefficient accretion along with the dominance of the 
non-thermal relativistic jet emission \citep{2010MNRAS.402..497G}. 

The broad band spectral energy distribution (SED) of blazars shows a two peak
structure, with the low energy peak (at UV/optical/X-ray energies) attributed to 
synchrotron emission process and the high energy peak (at MeV/GeV energies)  
attributed to inverse Compton process. Based on the position of the peak of the 
synchrotron component of the broad band SED, blazars are further divided into
low synchrotron peaked (LSP) blazars with the rest frame synchrotron peak 
$\nu_{peak}$ $\le$ 10$^{14}$ Hz, intermediate synchrotron peaked (ISP) blazars 
with  $10^{14} < \nu_{peak} < 10^{15}$ Hz and high synchrotron peaked (HSP) 
blazars with $\nu_{peak} > 10^{15}$ Hz \citep{2010ApJ...716...30A}.  The observed flux variations in blazars 
is well explained by the shock-in-jet model \citep{1985ApJ...298..114M}. 
Though blazars have been studied
for intra-night optical variability (INOV) for more than two decades
\citep{1989Natur.337..627M,2004MNRAS.348..176S,2017ApJ...844...32P}, the 
exact causes for their flux variations are not well understood, in particular
the association of optical flux variations to variations in other bands of
the electromagnetic spectrum. Also, on intra-night time scales, varied 
correlation are found between optical flux and polarization variations 
\citep{2020MNRAS.498.5128R,2021MNRAS.504.1772R}.
The detection of $\gamma-$ray emission from a large population of blazars
has enabled characterising their INOV characteristics among the 
different populations of $\gamma$-ray emitting blazars. Such a comparative
study has been carried out by \cite{2017ApJ...844...32P}. In this work 
we present our results on the monitoring observations carried out 
on a sample of $\gamma$-ray emitting blazars. In Section 2 we present 
our sample, the details of the observations and reduction procedures,  
the results are discussed in Section 3, notes on 
individual sources are given in Section 4 followed by
the summary in Section 5. 

\begin{table*}

\caption{The blazars  monitored in this program. Information such as the 3FGL name, optical type,
SED class, right ascension ($\alpha$), declination ($\delta$) and redshift ($z$) are from 
\cite{2015ApJ...810...14A}, except the R-band magnitude which is from
the USNO-B1 catalog \citep{2003AJ....125..984M}\label{table1}. For J1427.0+2347,the redshift$^{\dag}$ is from
\cite{2016A&A...589A..92R}, while for J1555.7+1111 the redshift$^{\dag\dag}$ is from \cite{2018ApJ...854...11T}}. 
\begin{tabular}{ccccrlcc}
\topline

3FGL Name &  optical type & SED type & $\alpha_{2000}$ & $\delta_{2000}$ & $z$ & R (mag) & $\nu_{peak}$ \\\midline

J0050.6$-$0929   & BL Lac  & ISP  & 00:50:41.32 & $-$09:29:05.21 & 0.635  & 16.14  & 14.400 \\
J0109.1+1816     & BL Lac  & HSP  & 01:09:08.18 & 18:16:07.50    & 0.443  & 16.30  & 14.860 \\
J0112.1+2245     & BL Lac  & ISP  & 01:12:05.82 & 22:44:38.80    & 0.265  & 15.47  & 14.325 \\
J0217.2+0837     & BL Lac  & LSP  & 02:17:17.12 & 08:37:03.90    & 0.085  & 14.68  & 13.760 \\
J0303.4$-$2407   & BL Lac  & HSP  & 03:03:26.50 & $-$24:07:11.42 & 0.260  & 16.50  & 15.314 \\
J0738.1+1741     & BL Lac  & LSP  & 07:38:07.39 & 17:42:19.01    & 0.424  & 15.78  & 13.830 \\
J0739.4+0137     & FSRQ    & ISP  & 07:39:18.03 & 01:37:04.62    & 0.189  & 16.19  & 14.050 \\
J0825.9$-$2230   & BL Lac  & ISP  & 08:26:01.57 & $-$22:30:27.22 & 0.911  & 15.80  & 14.160 \\
J0846.7$-$0651   & BL Lac  & LSP  & 08:47:56.74 & $-$07:03:16.92 & ----   & 15.35  & 13.530 \\
J0854.8+2006     & BL Lac  & LSP  & 08:54:48.88 & 20:06:30.64    & 0.306  & 15.56  & 13.616 \\
J0912.9$-$2104   & BL Lac  & HSP  & 09:13:00.22 & $-$21:03:21.01 & 0.198  & 16.42  & 16.424 \\
J0927.9$-$2037   & FSRQ    & LSP  & 09:27:51.82 & $-$20:34:51.24 & 0.348  & 16.00  & 13.115 \\
J1015.0+4925     & BL Lac  & HSP  & 10:15:04.14 & 49:26:00.71    & 0.212  & 14.58  & 15.550 \\
J1129.9$-$1446   & FSRQ    & LSP  & 11:30:07.05 & $-$14:49:27.37 & 1.184  & 16.00  & 12.650 \\
J1224.9+2122     & FSRQ    & LSP  & 12:24:54.46 & 21:22:46.38    & 0.435  & 18.2   & 13.720 \\
J1229.1+0202     &  FSRQ   & LSP  & 12:29:06.70 & 02:03:08.60    & 0.158  & 14.11  & 13.460 \\
J1427.0+2347     & BL Lac  & HSP  & 14:27:00.39 & 23:48:00.04    & 0.604$^{\dag}$  & 14.50  & 15.344 \\
J1555.7+1111     & BL Lac  & HSP  & 15:55:43.04 & 11:11:24.36    & 0.500$^{\dag\dag}$  & 13.99  & 15.467 \\

\hline
\end{tabular}
\end{table*}

\begin{figure*}
\vbox{
\hbox{
\includegraphics[scale=0.53]{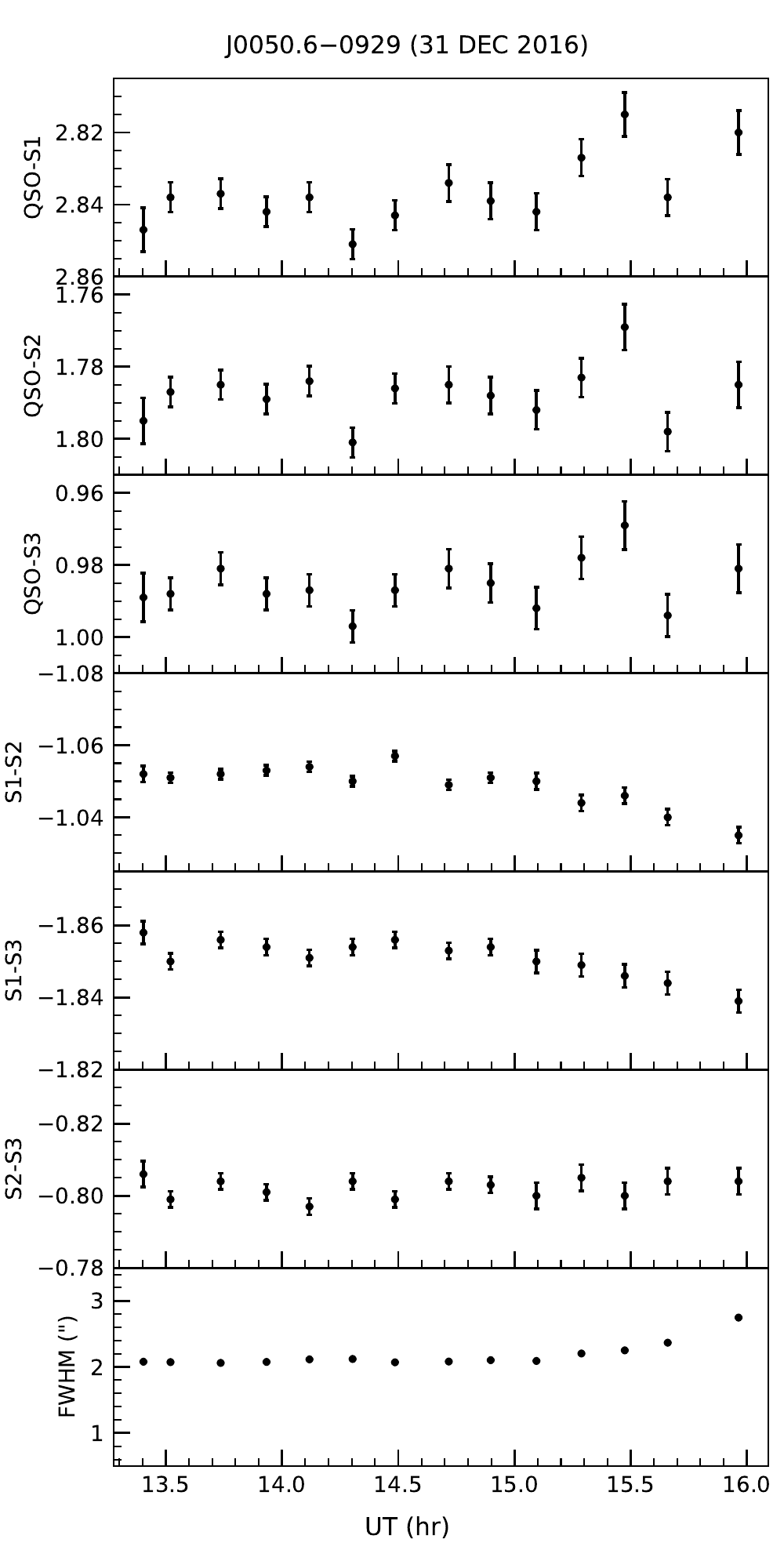}
\includegraphics[scale=0.53]{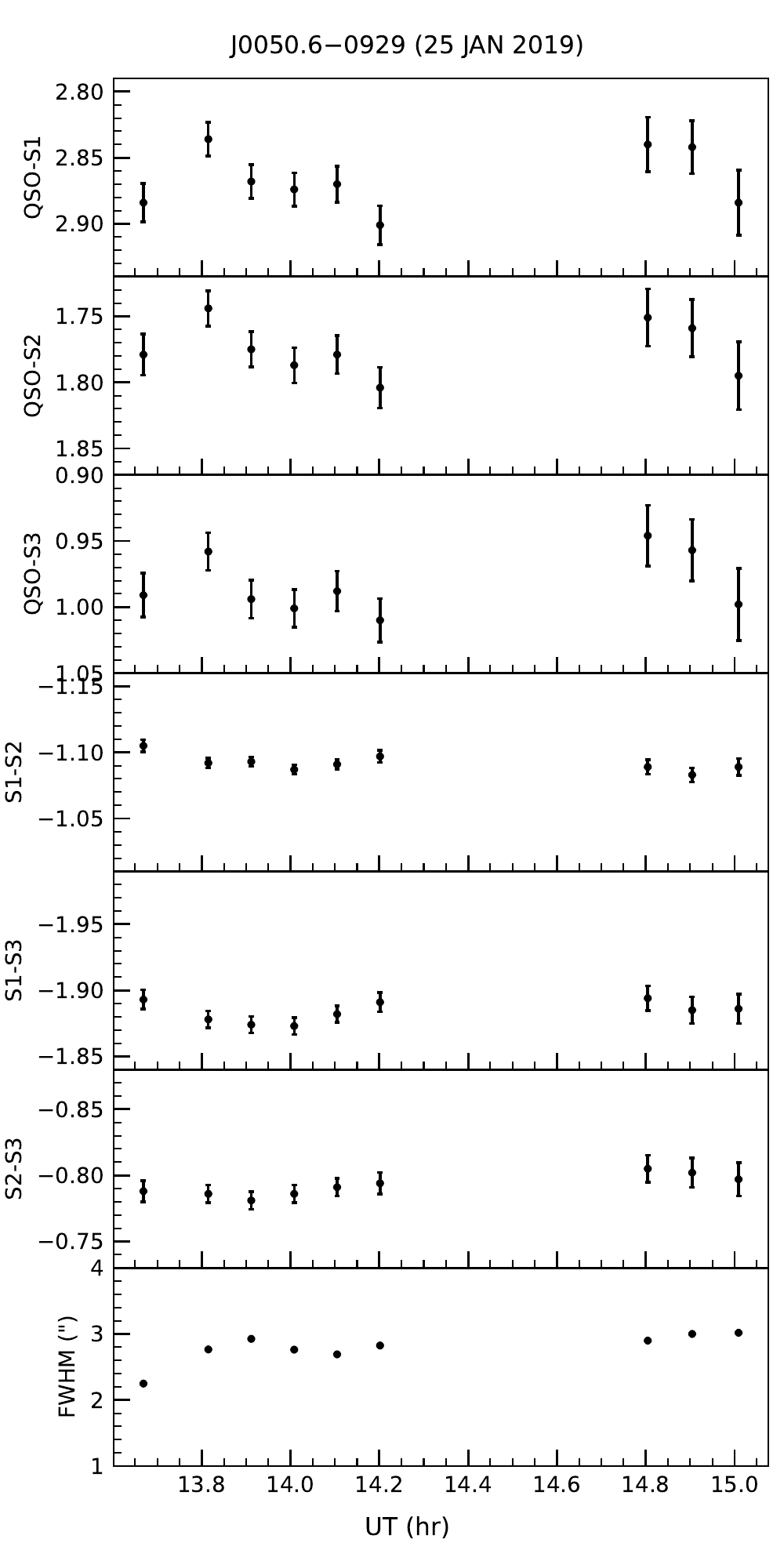}
\includegraphics[scale=0.53]{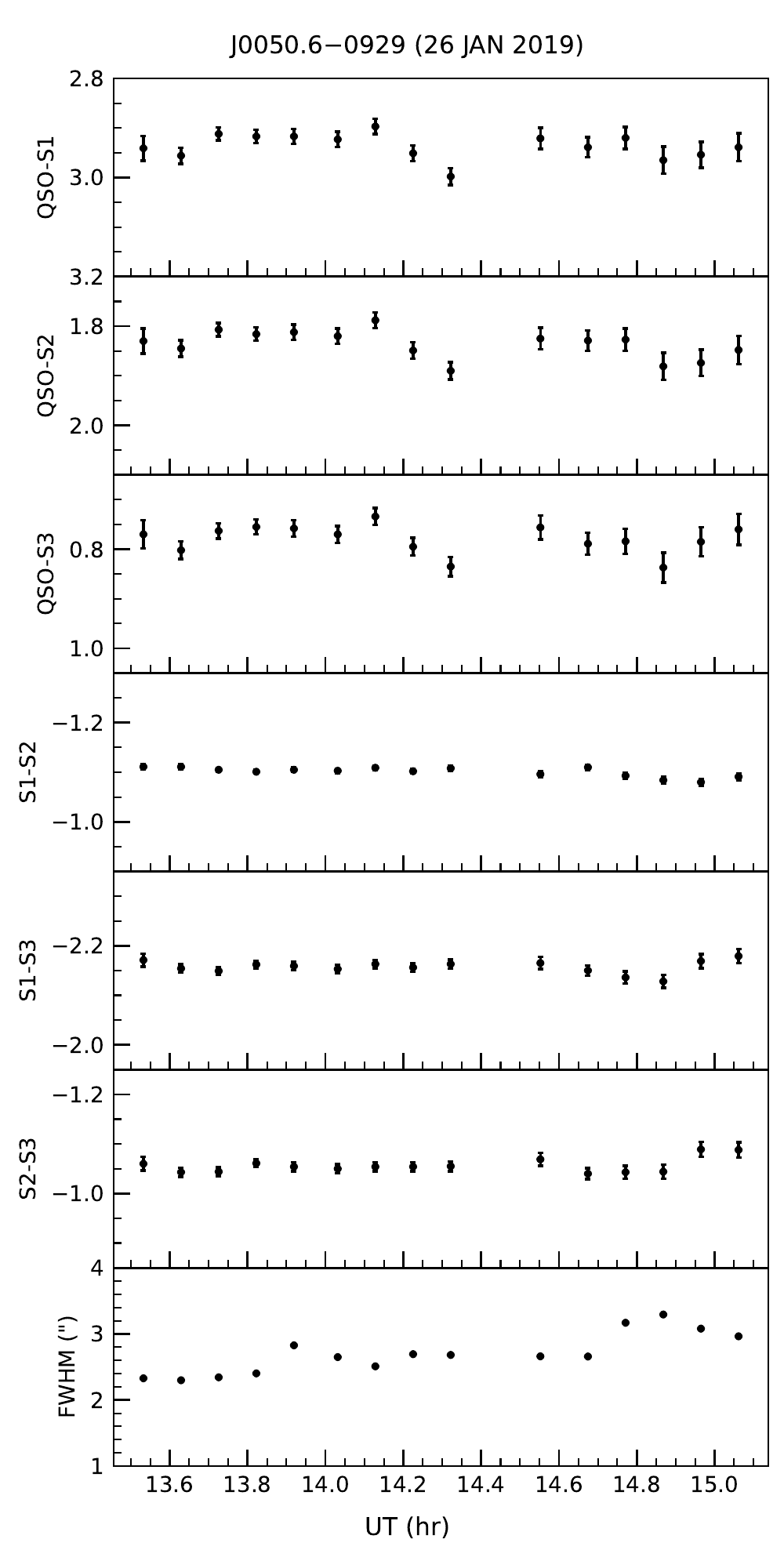}
}
\includegraphics[scale=0.53]{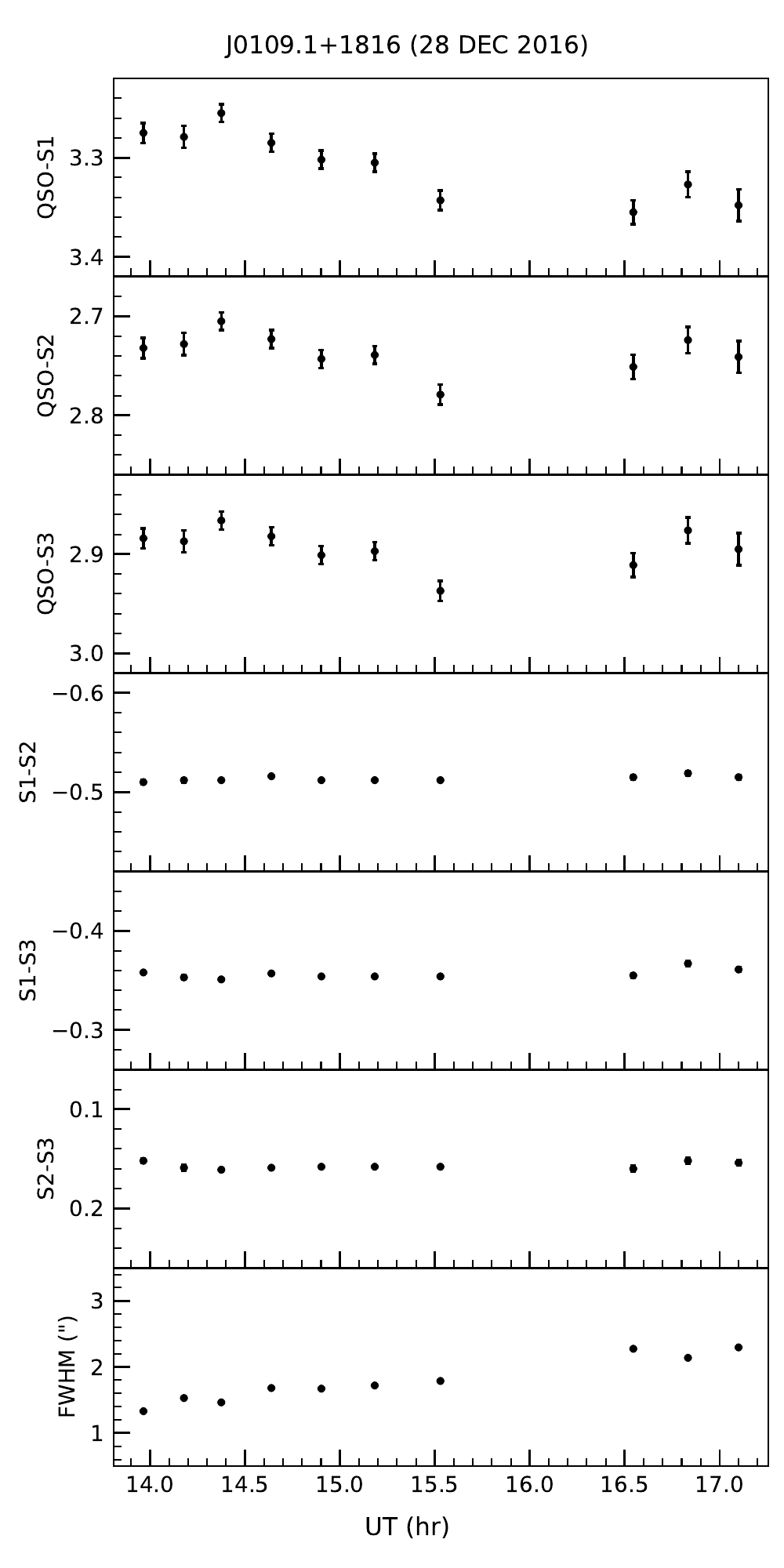}
\includegraphics[scale=0.53]{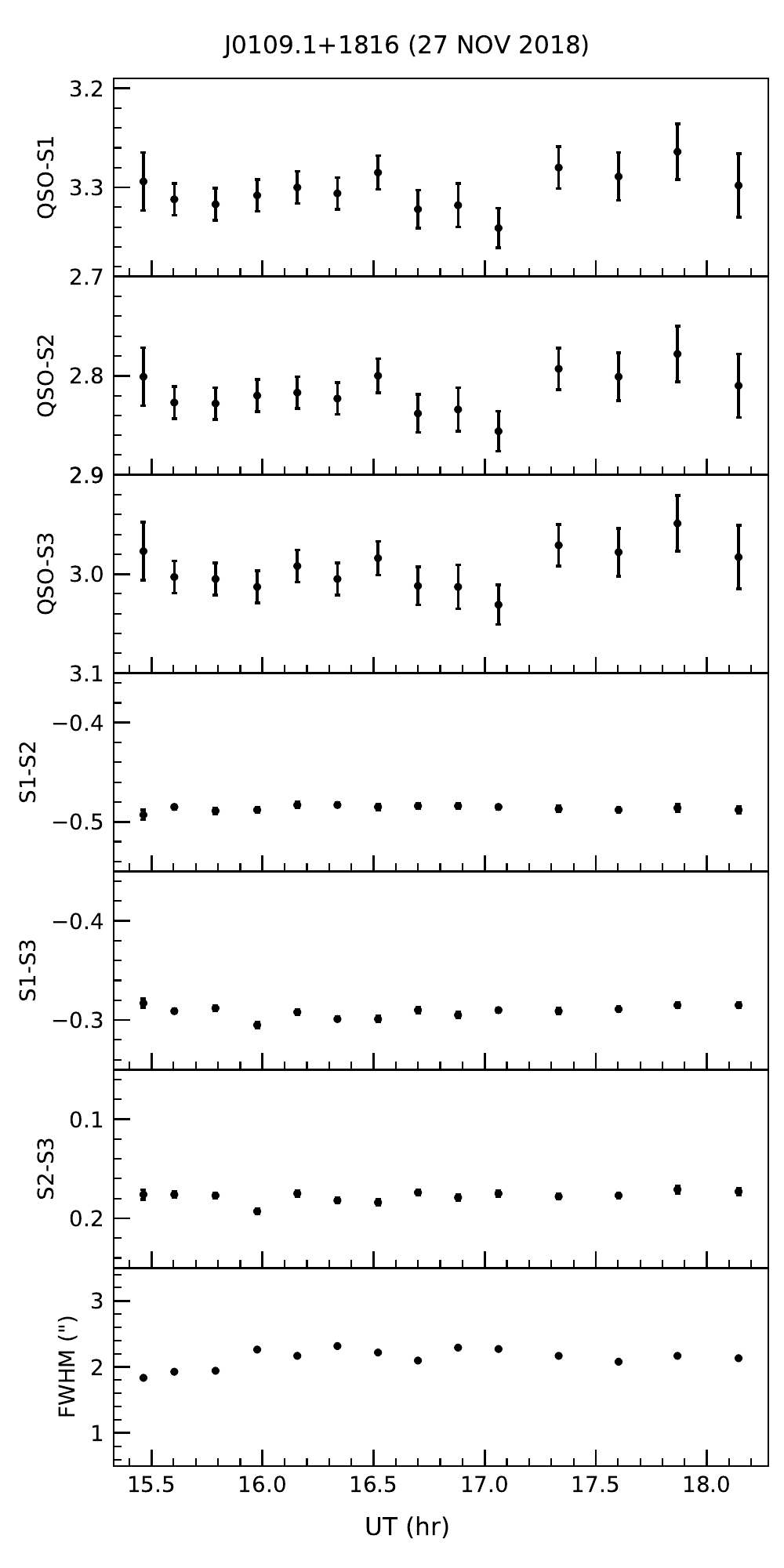}
\includegraphics[scale=0.53]{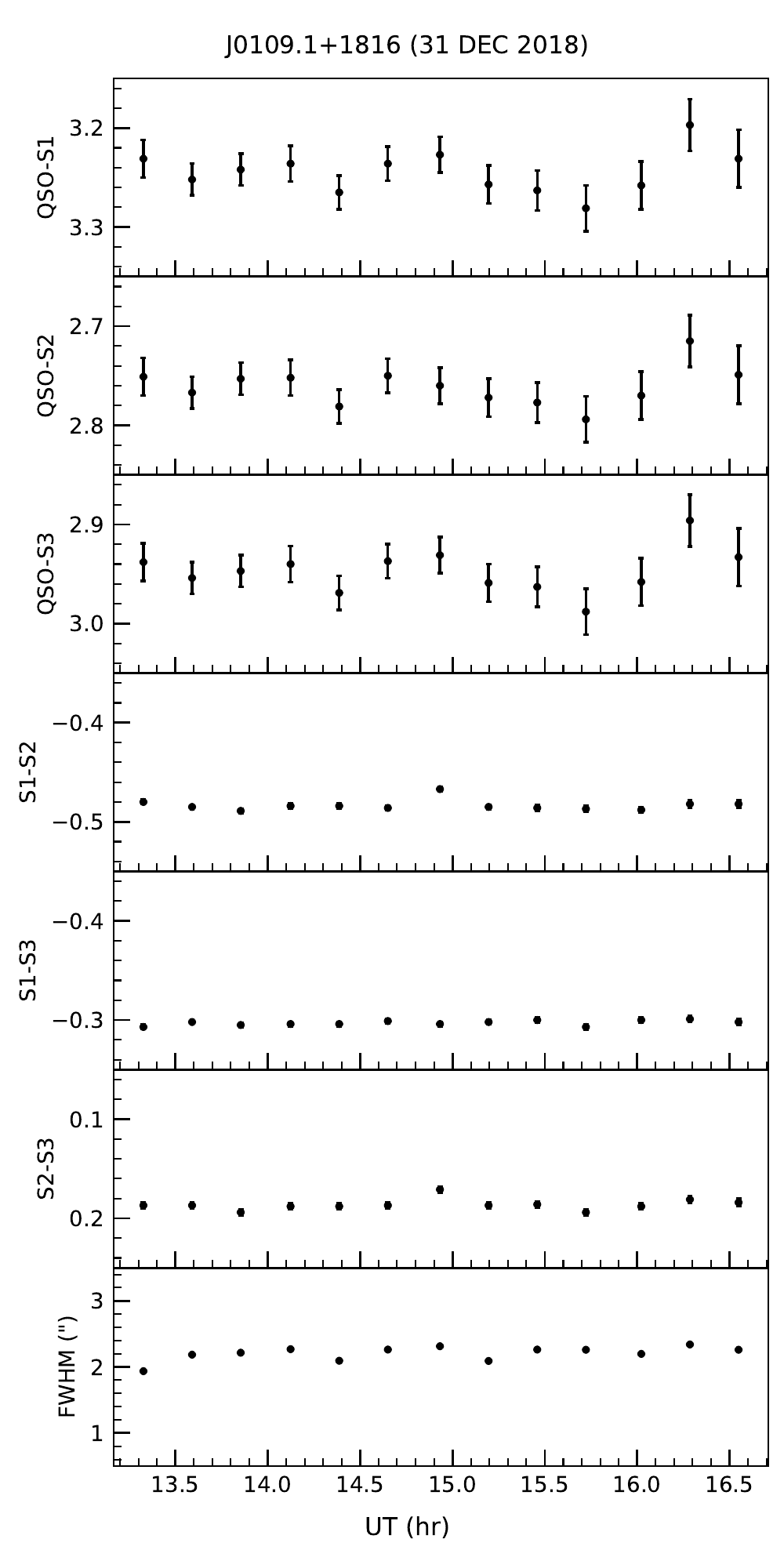}
}
\caption{DLCs of the ISP BL Lac J0050.6$-$0929 (top panels) and the HSP 
BL Lac J0109.1+1816 (bottom panels).Here, S1, S2 and S3 refer to the comparison 
stars and QSO refer to the blazar. The variation of the FWHM of the 
stellar light distribution during each night of observation is also given. }\label{fig-1}
\end{figure*}

\begin{figure*}
\vbox{
\hbox{
\includegraphics[scale=0.53]{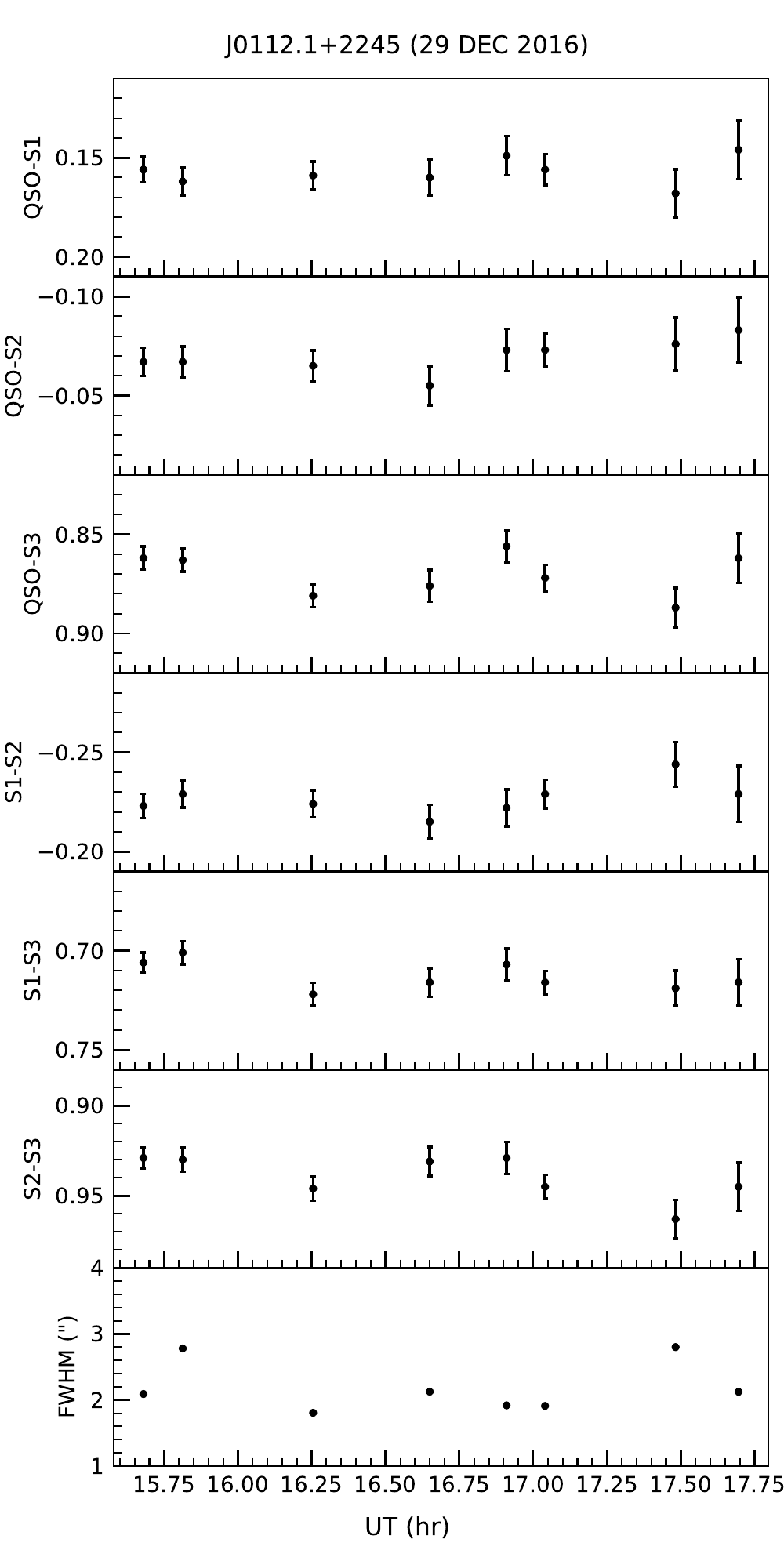}
\includegraphics[scale=0.53]{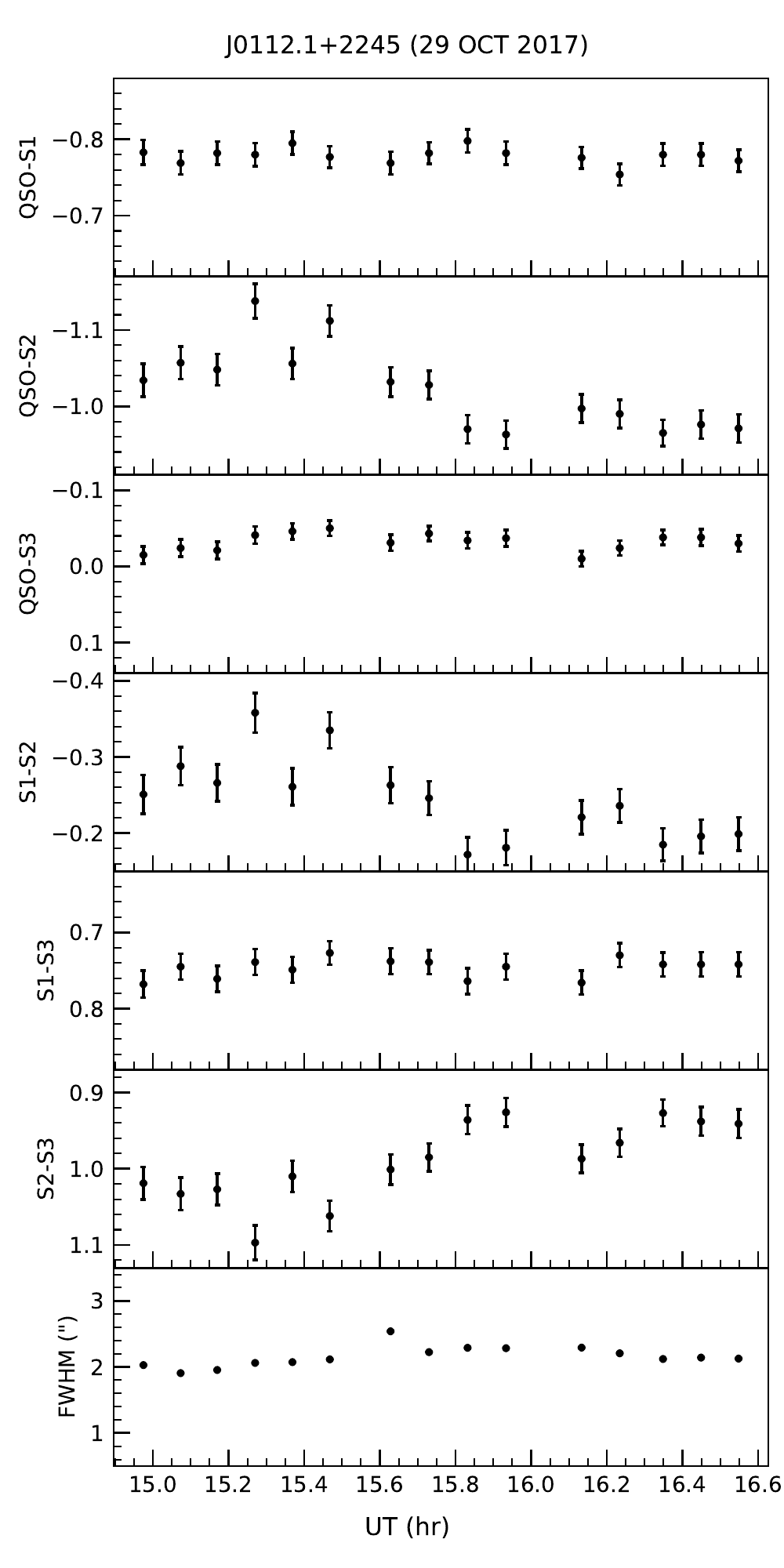}
\includegraphics[scale=0.53]{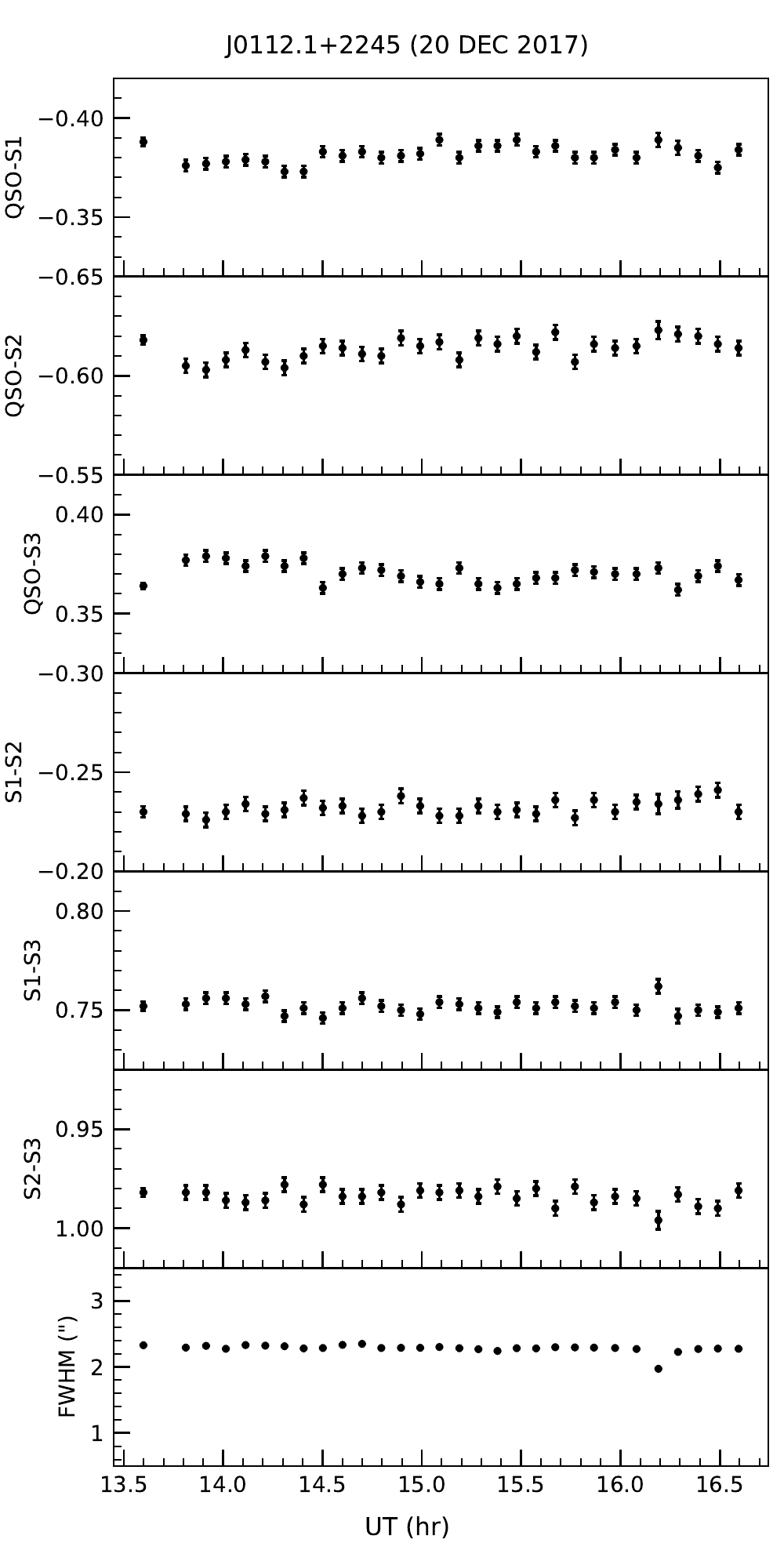}
}
\includegraphics[scale=0.53]{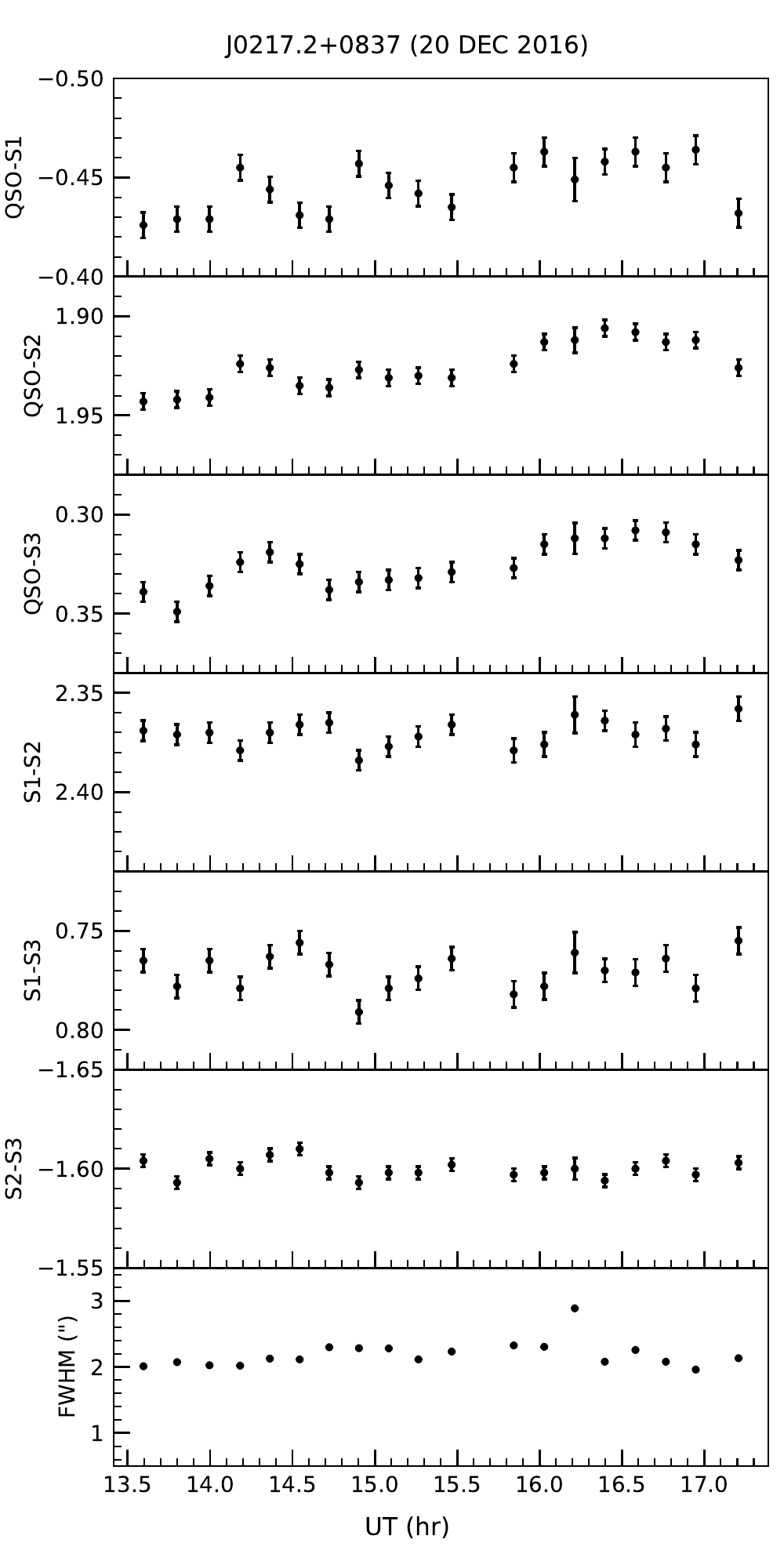}
\includegraphics[scale=0.53]{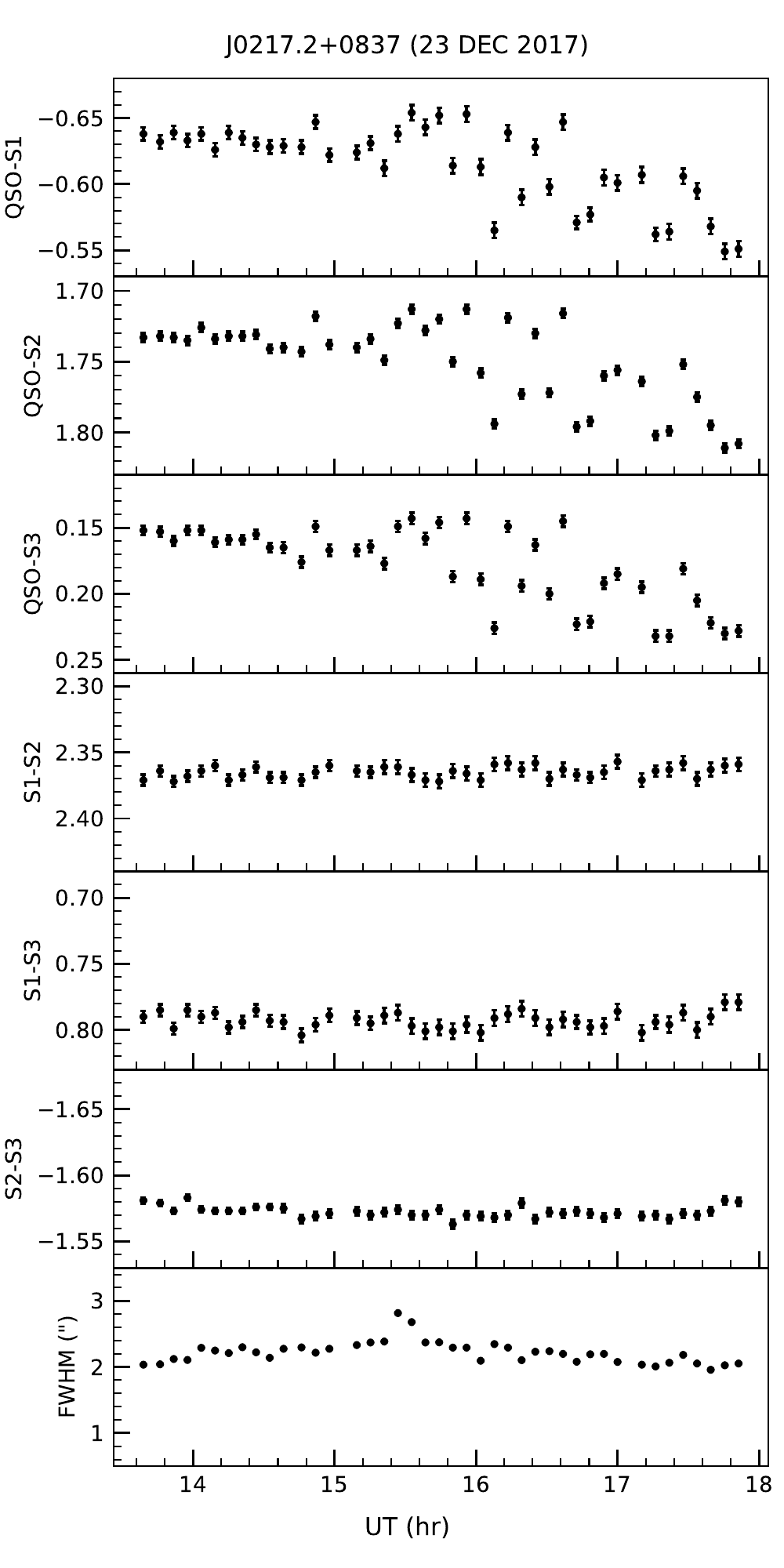}
\includegraphics[scale=0.53]{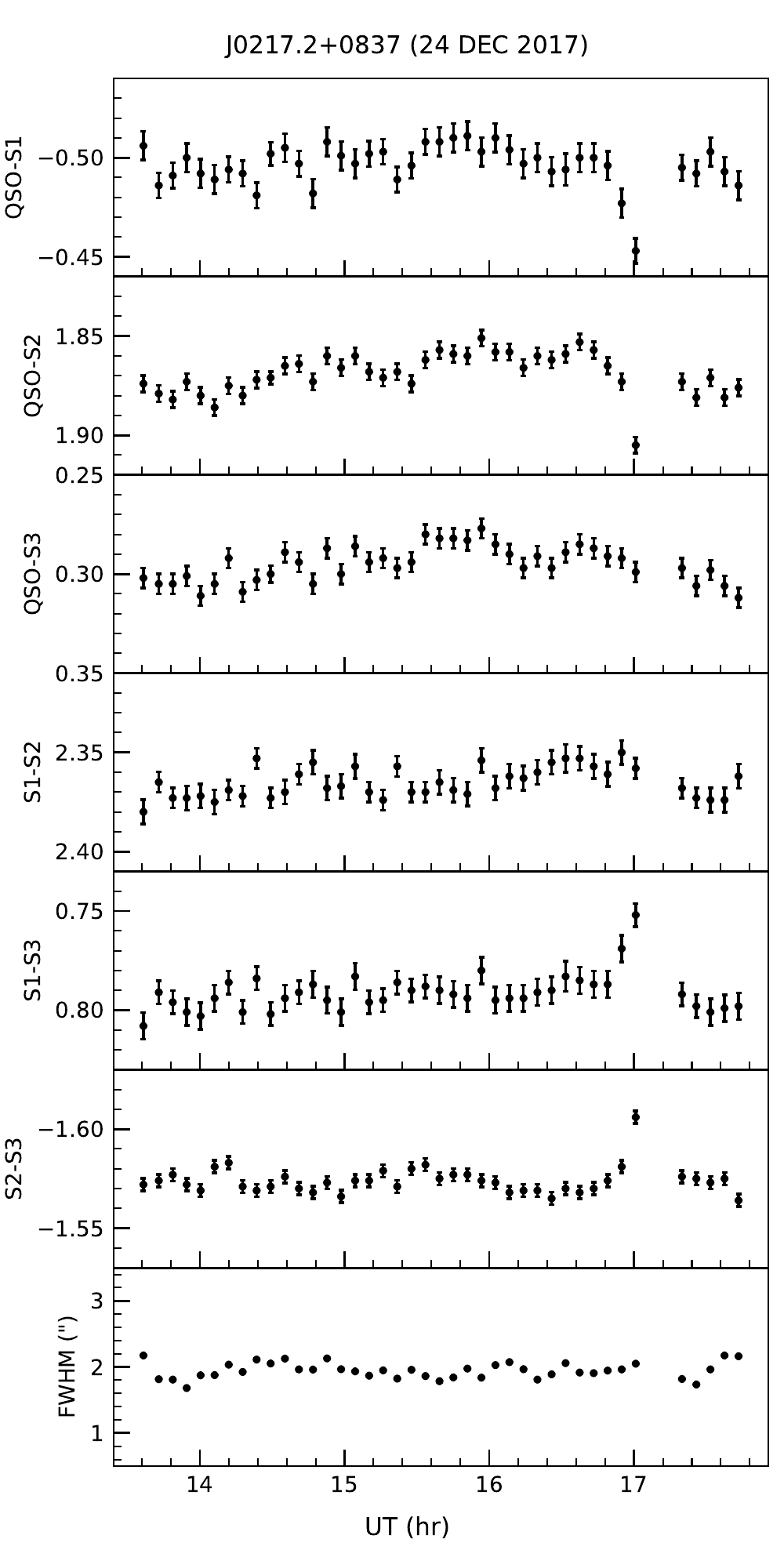}
}
\caption{DLCs of J0112.1+2245 (ISP BL Lac; top panels) and J0217.2+0837 (LSP BL Lac; bottom panels).
Labels have the same meaning as in Fig.1}\label{fig-2}
\end{figure*}

\begin{figure*}
\vbox{
\hbox{
\includegraphics[scale=0.53]{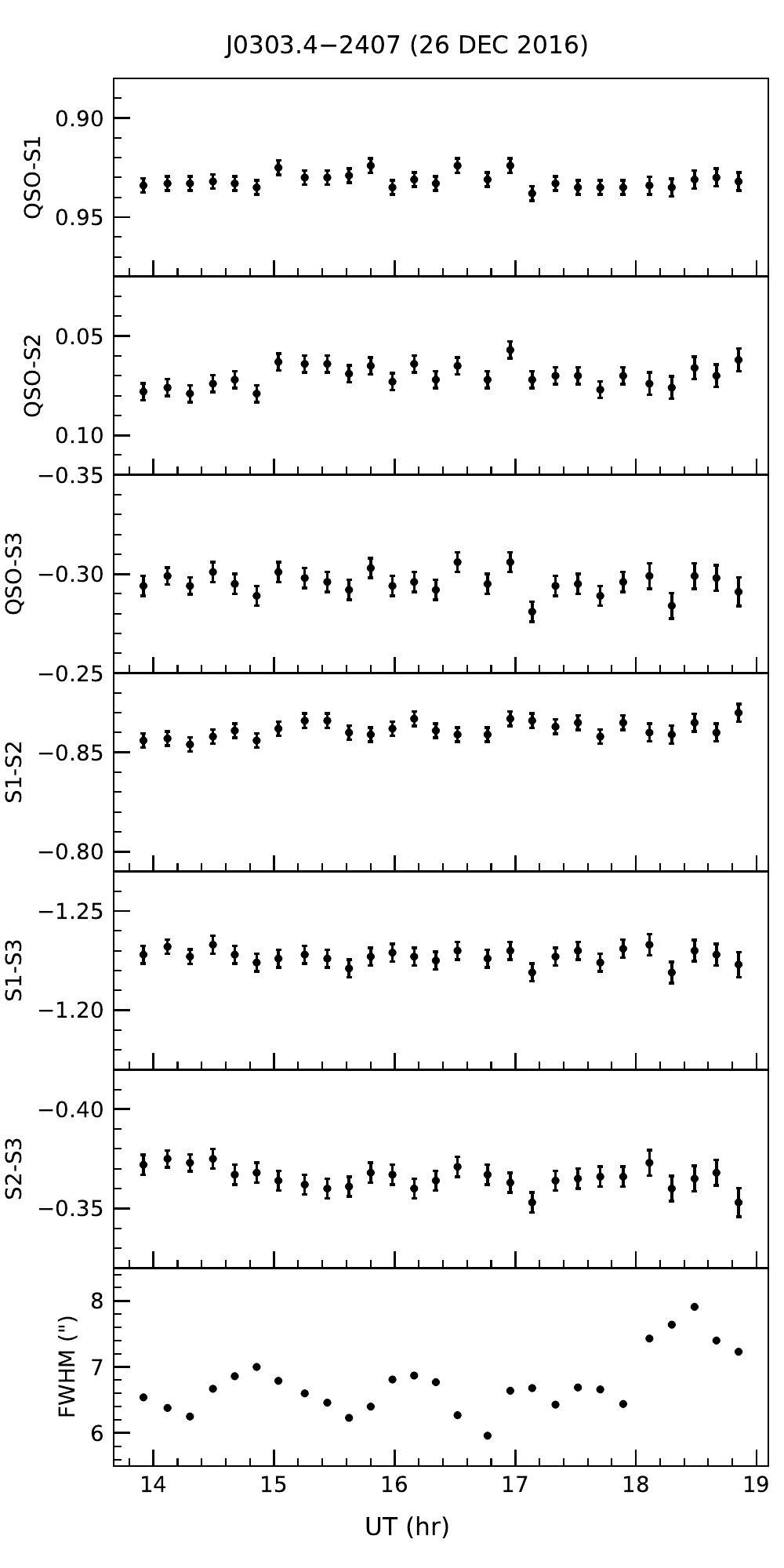}
\includegraphics[scale=0.53]{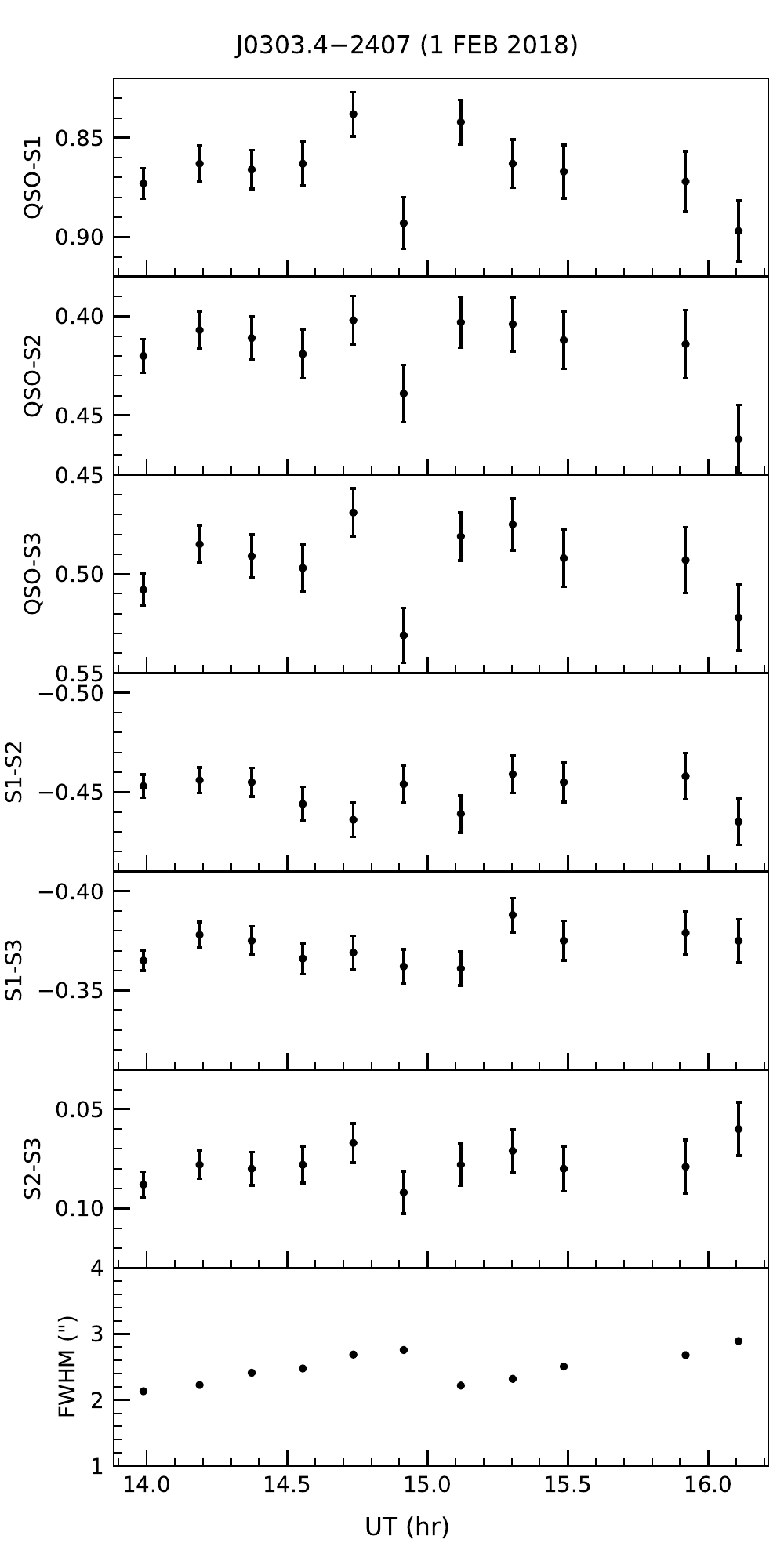}
\includegraphics[scale=0.53]{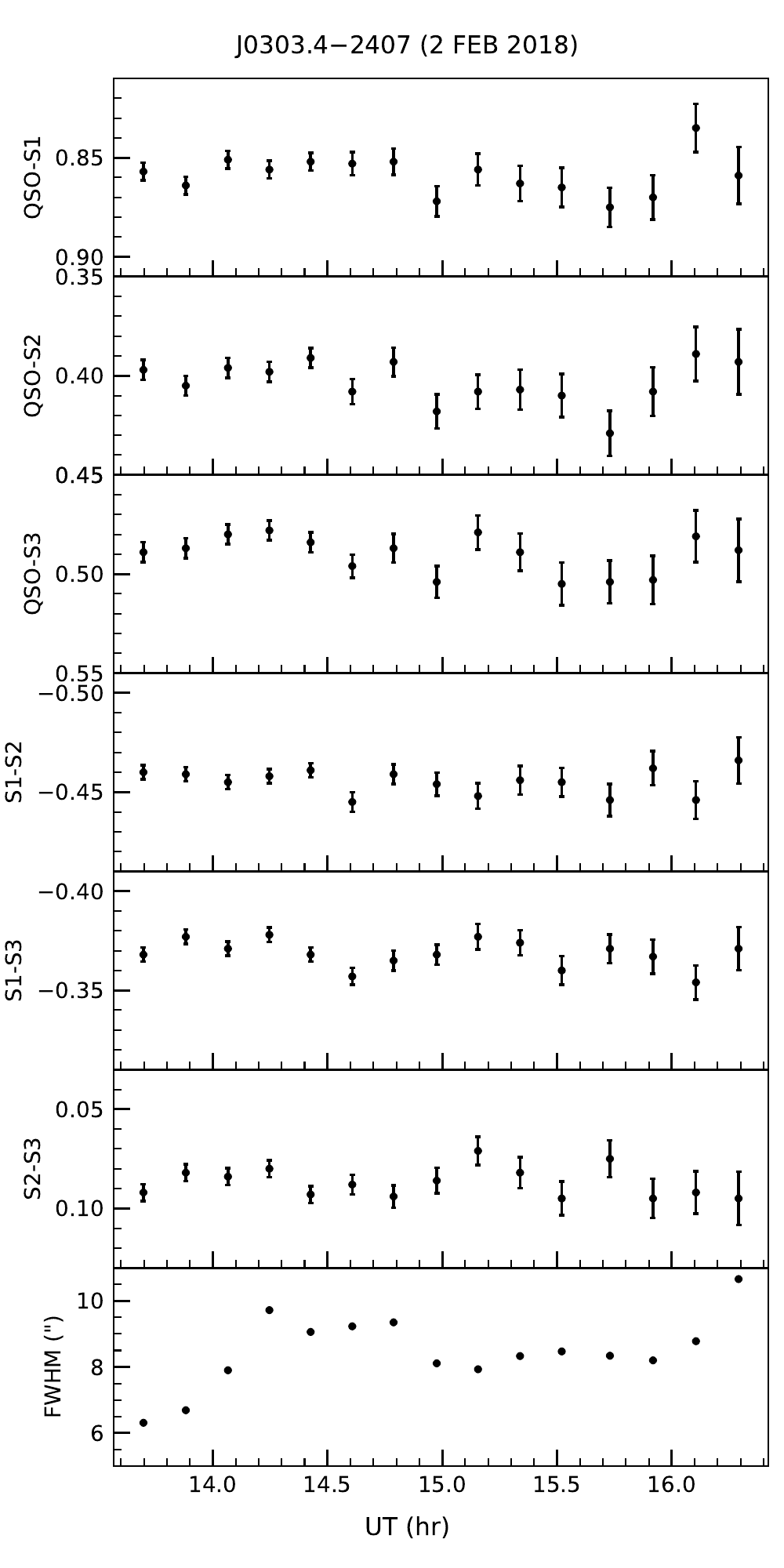}
}
\includegraphics[scale=0.53]{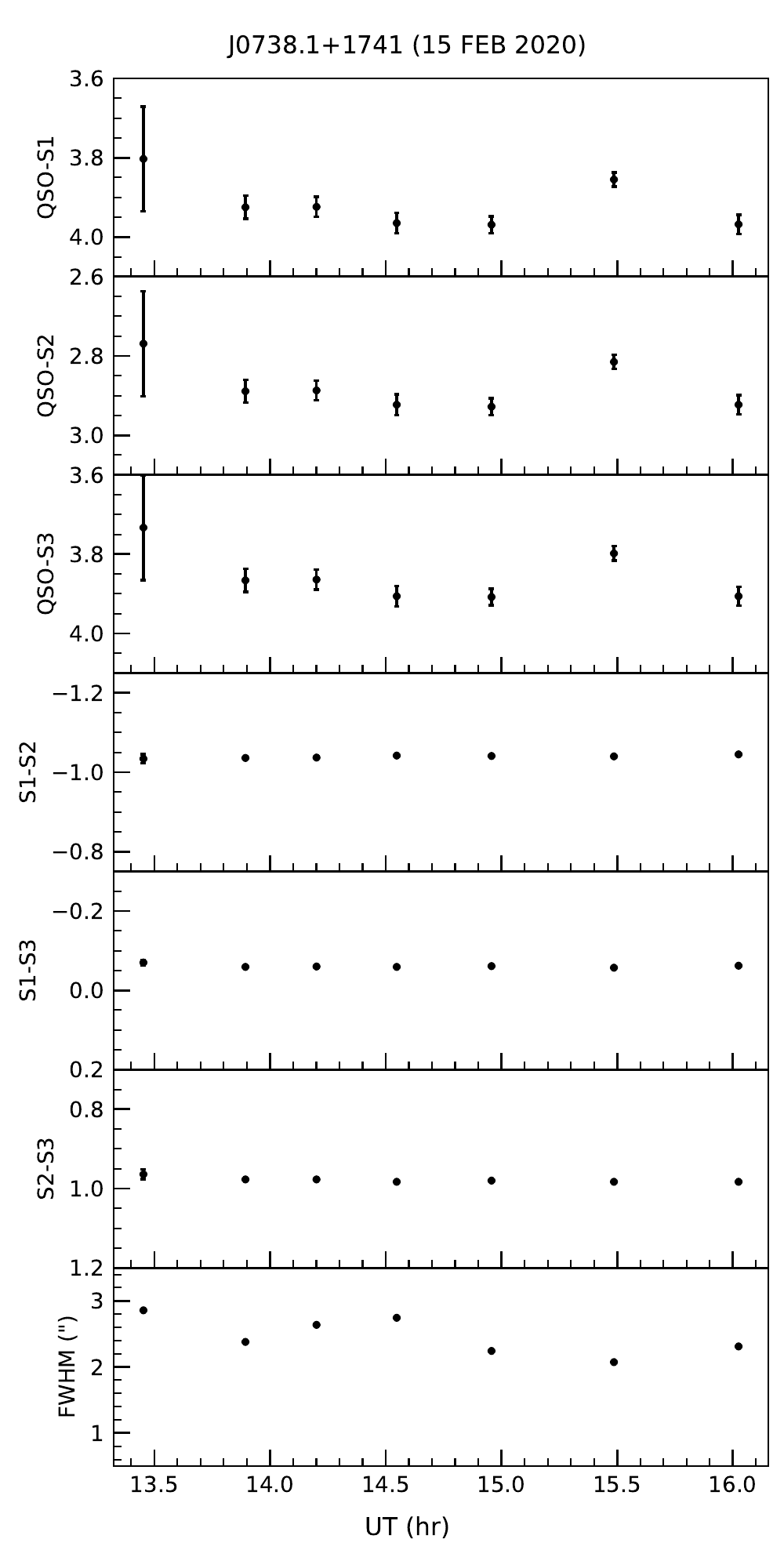}
\includegraphics[scale=0.53]{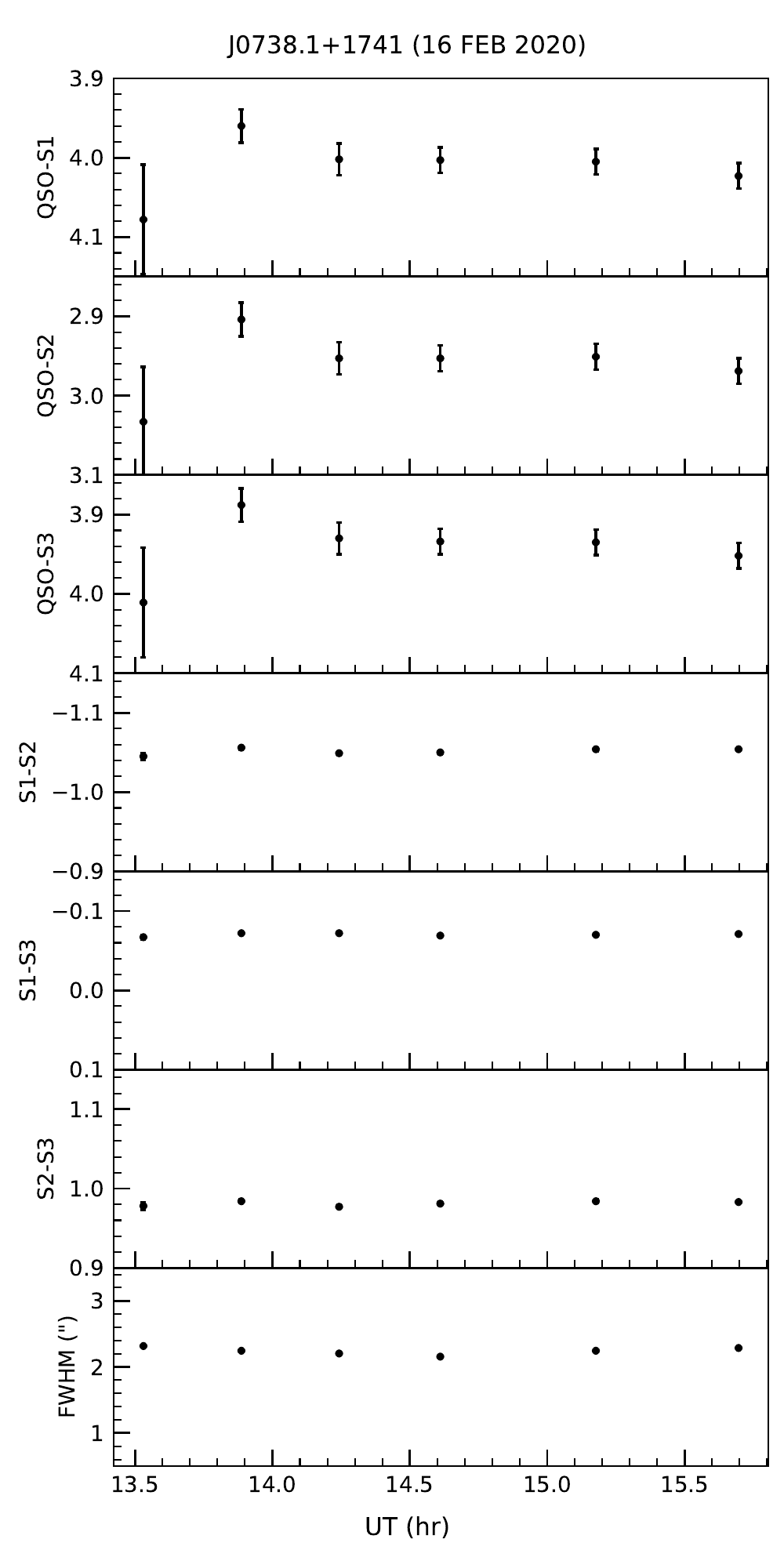}
\includegraphics[scale=0.53]{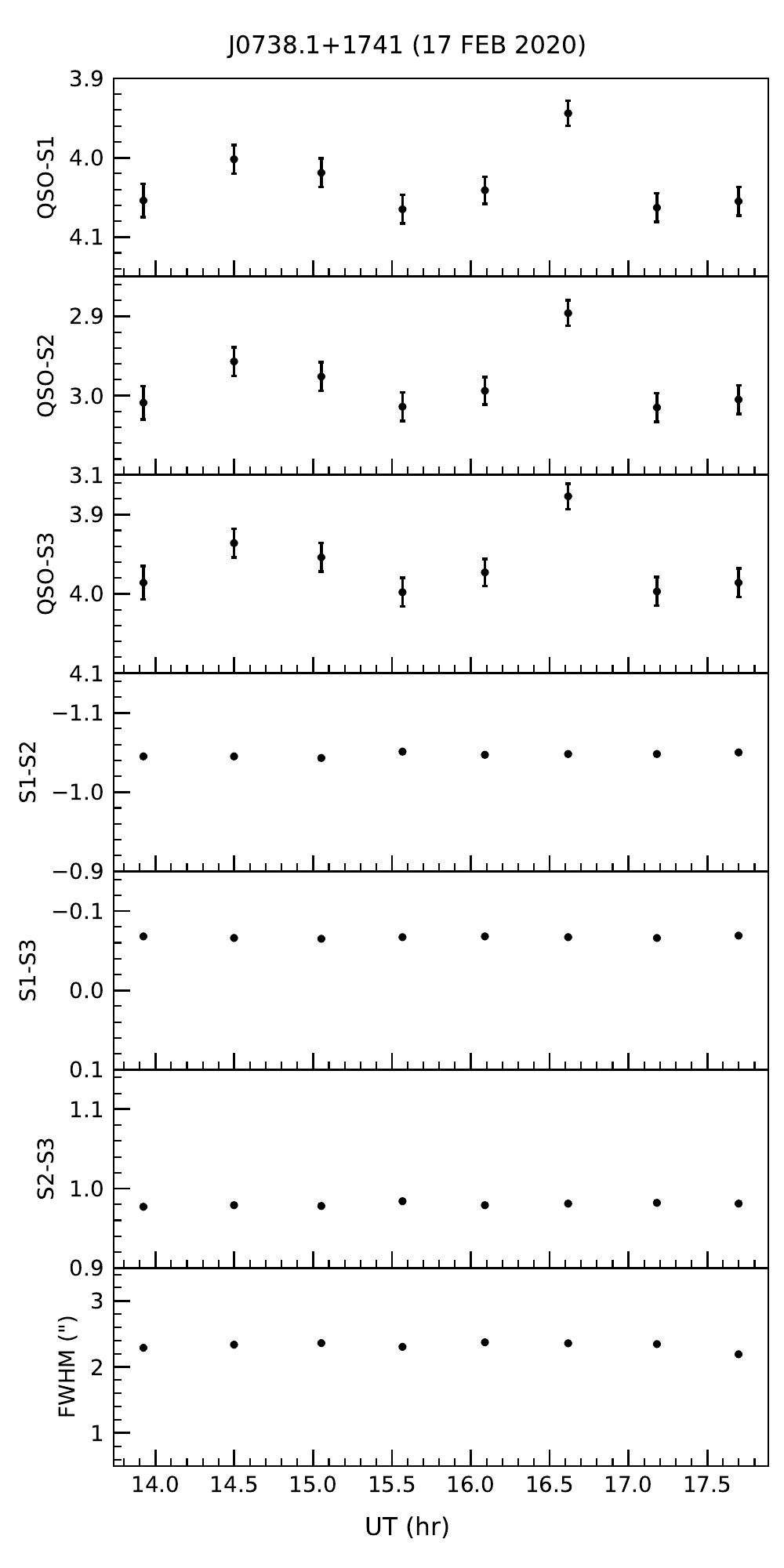}
}
\caption{The top panels show the DLCs of the HSP BL Lac J0303.4$-$2407 and the bottom panels show the
DLCs of J0738.1+1741, a LSP BL Lac object. The meanings of the labels are similar to that of Fig. 1}\label{fig-3}
\end{figure*}

\begin{figure*}
\vbox{
\hbox{
\includegraphics[scale=0.53]{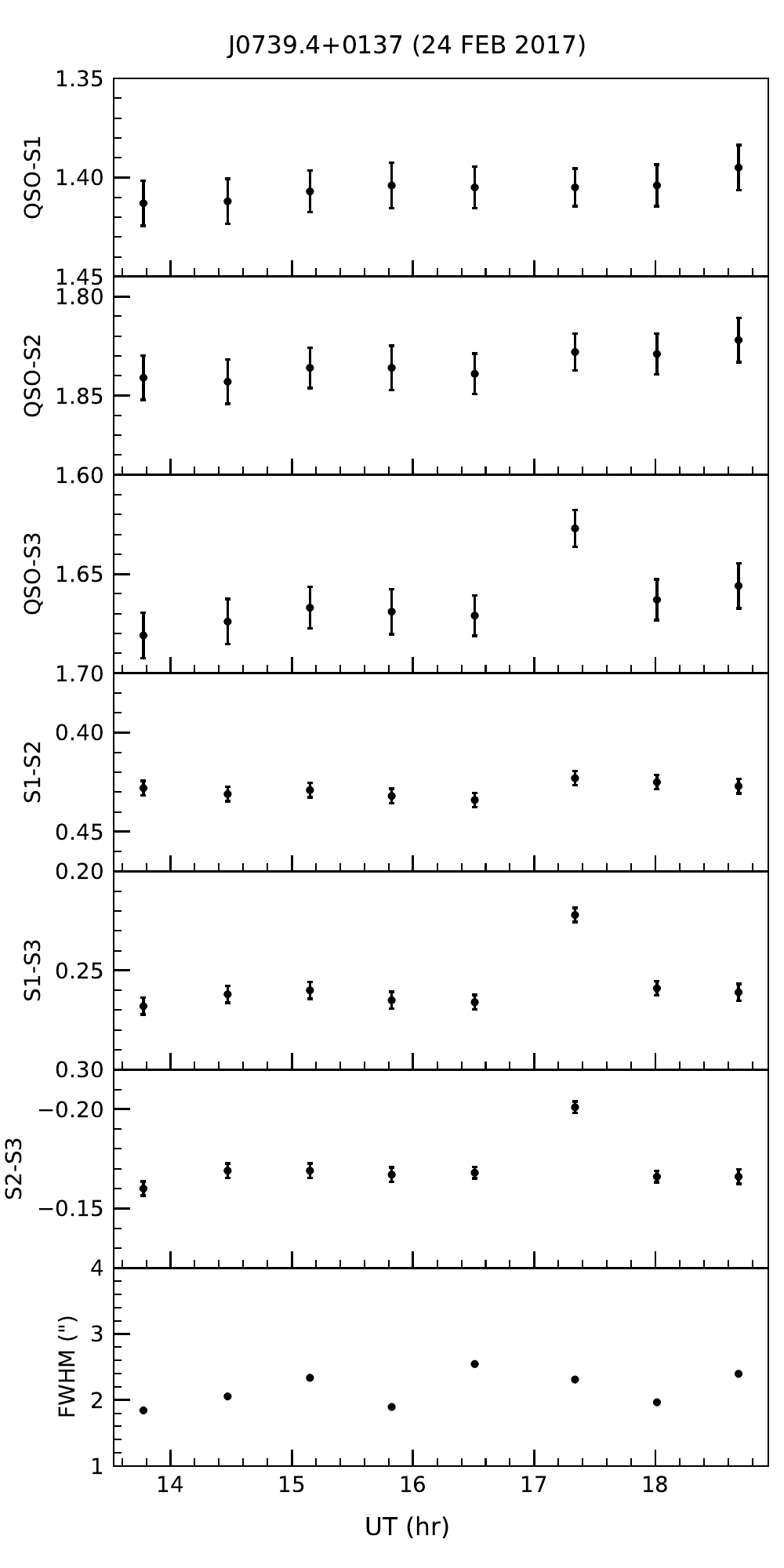}
\includegraphics[scale=0.53]{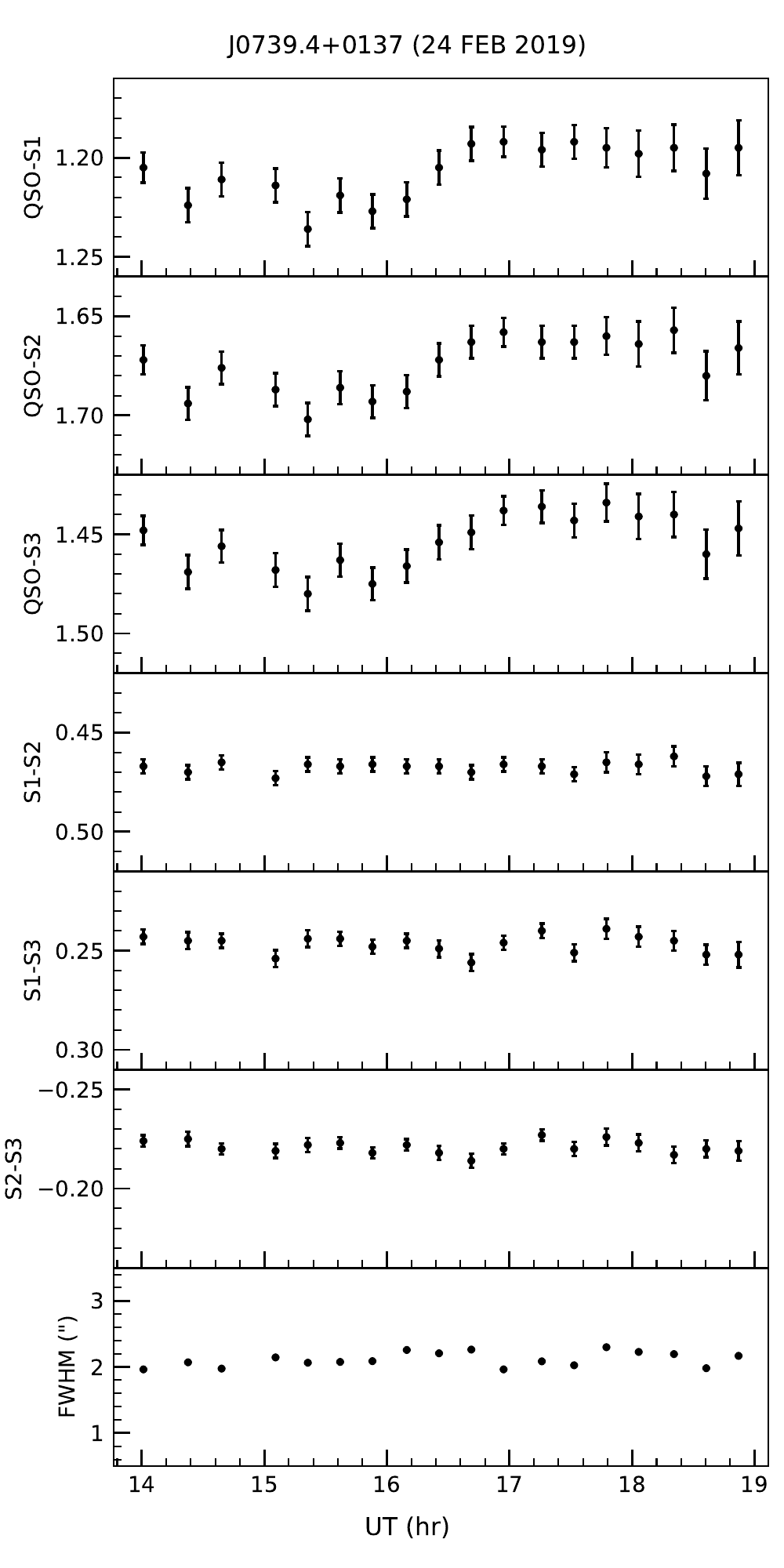}
\includegraphics[scale=0.53]{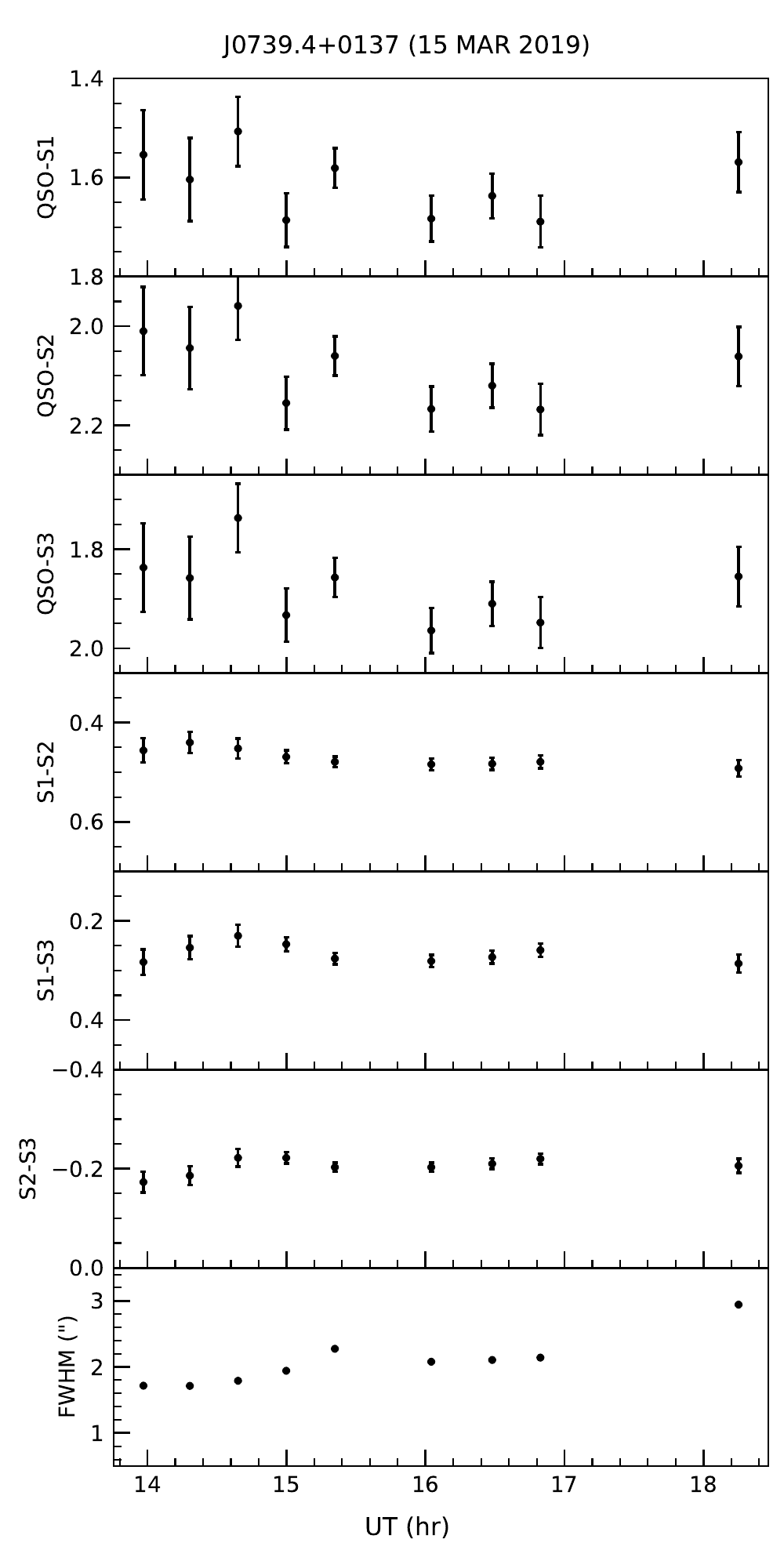}
}
\includegraphics[scale=0.53]{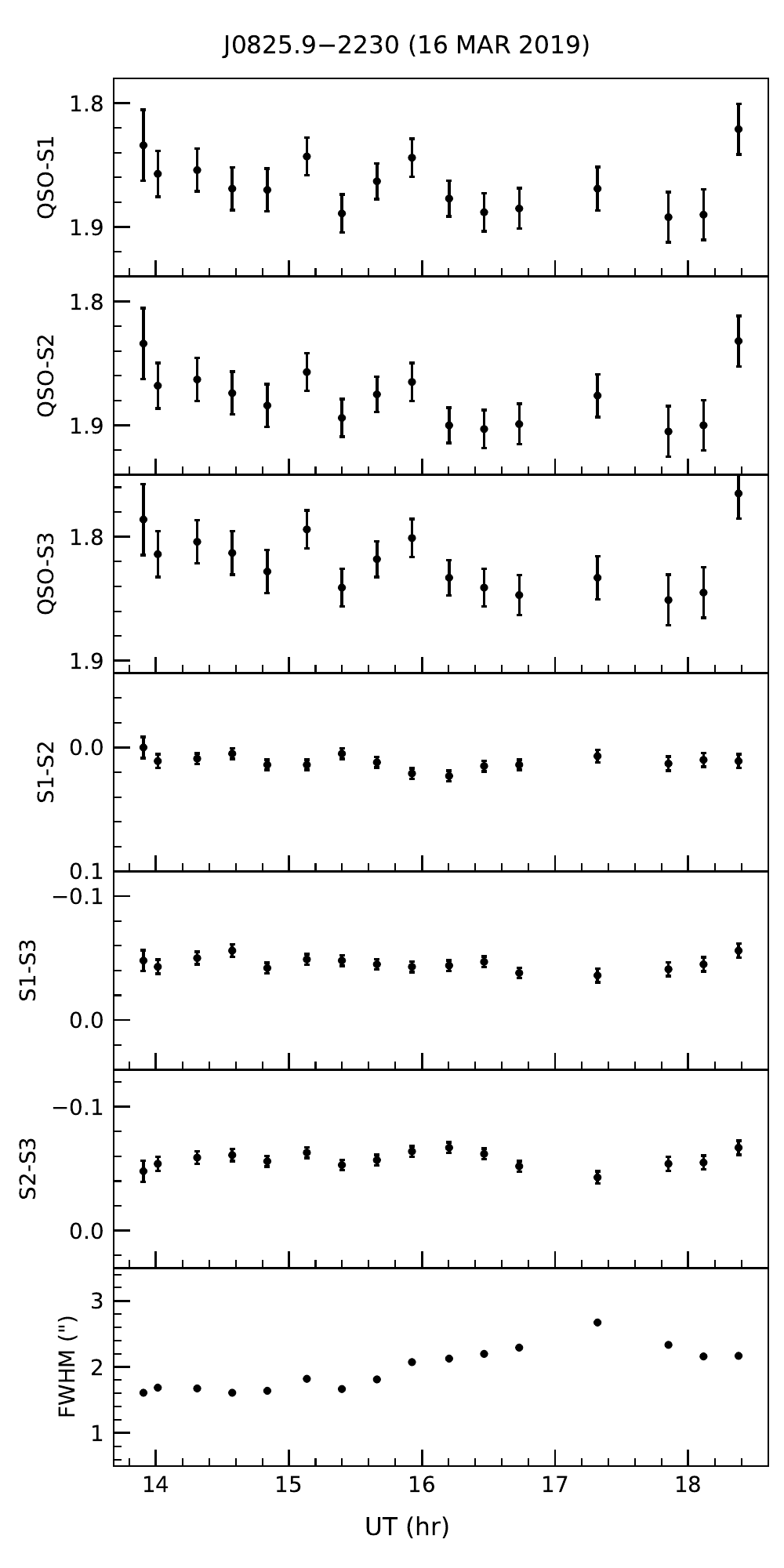}
\includegraphics[scale=0.53]{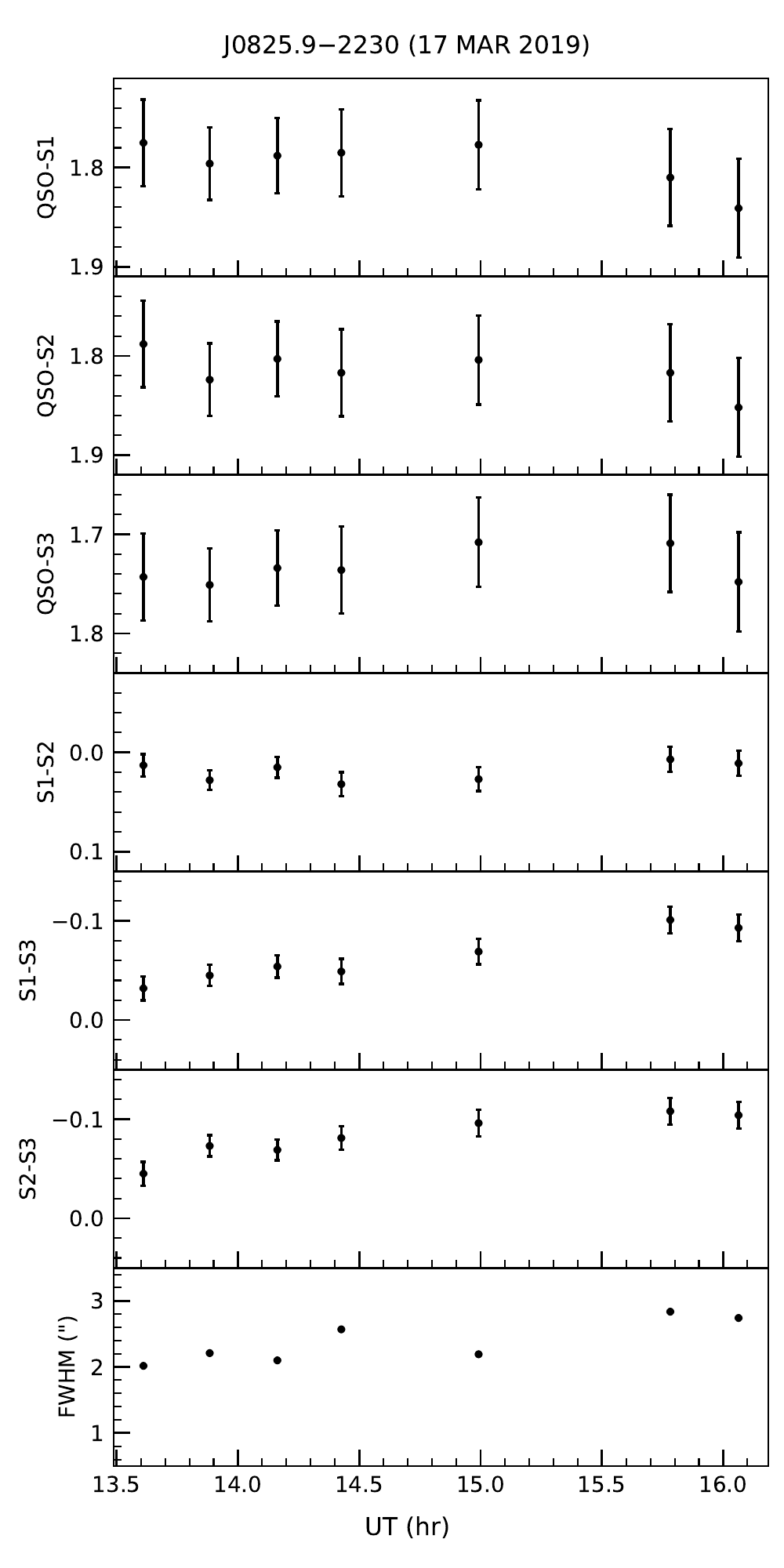}
\includegraphics[scale=0.53]{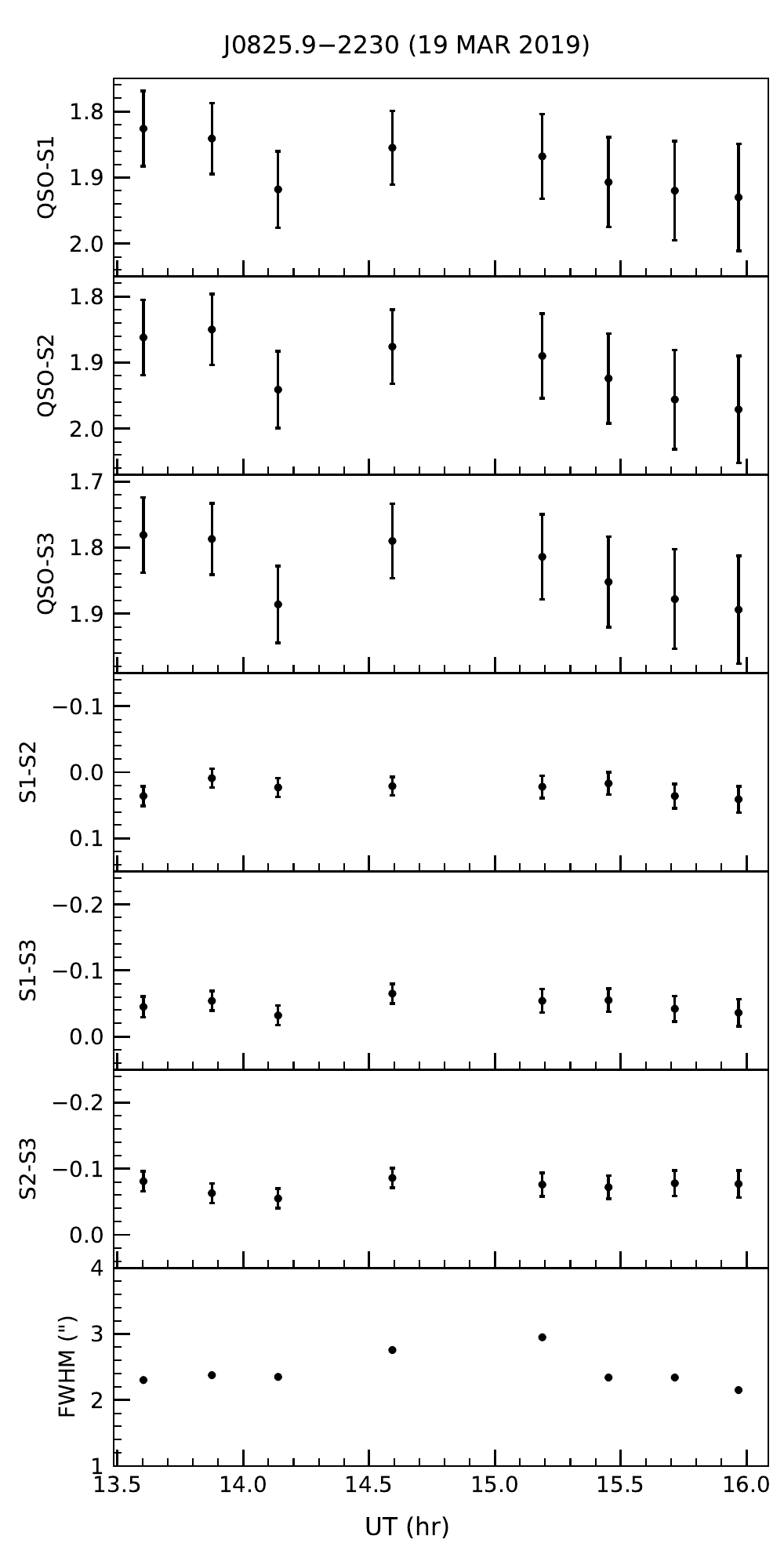}
}
\caption{DLCs of the FSRQ J0739.4+0137 (top panels) and the BL Lac J0825.9$-$2230. 
The labels have the same meaning as in Fig. 1}\label{fig-4}
\end{figure*}

\begin{figure*}
\vbox{
\hbox{
\includegraphics[scale=0.53]{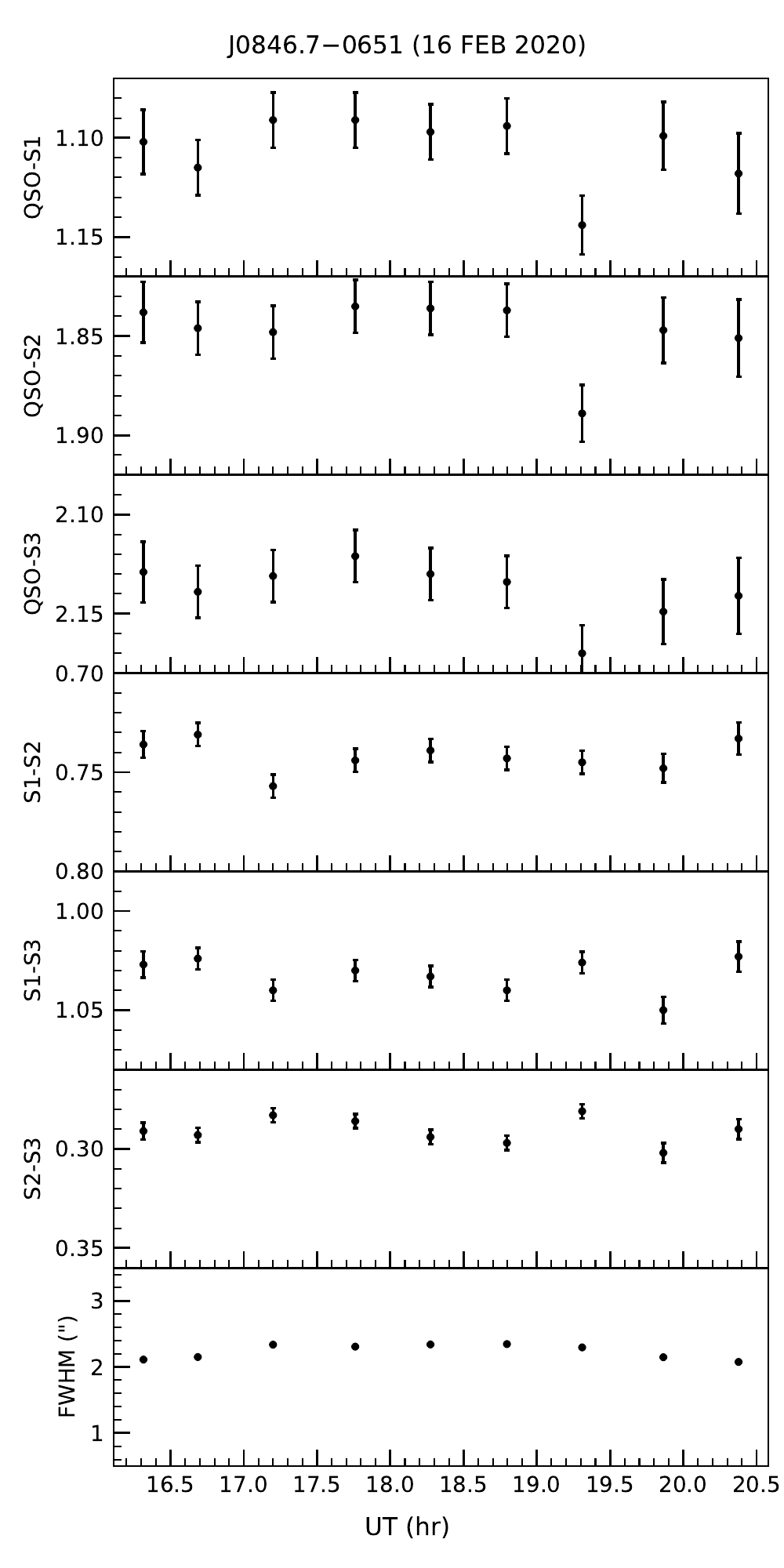}
\includegraphics[scale=0.53]{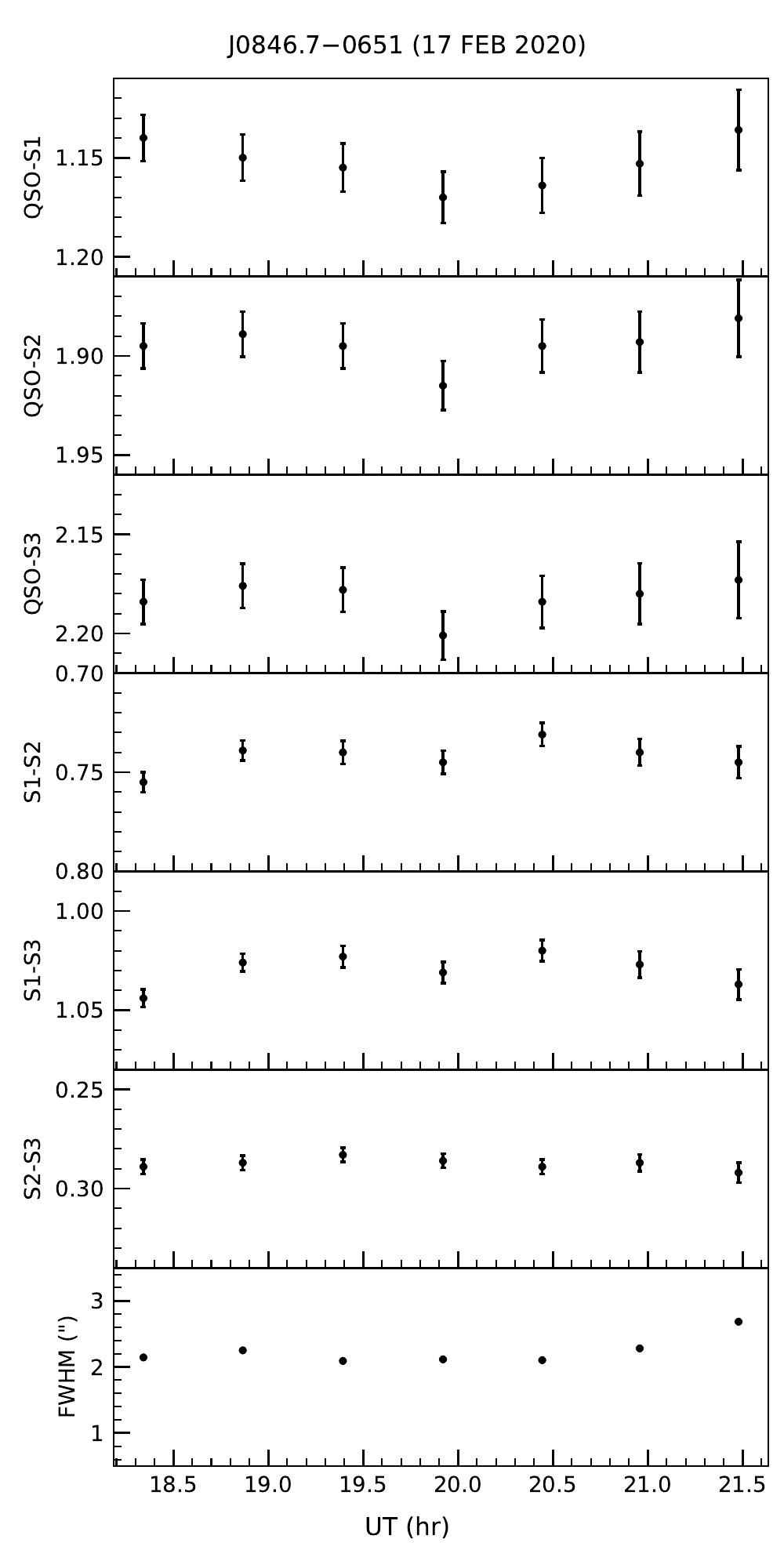}
\includegraphics[scale=0.53]{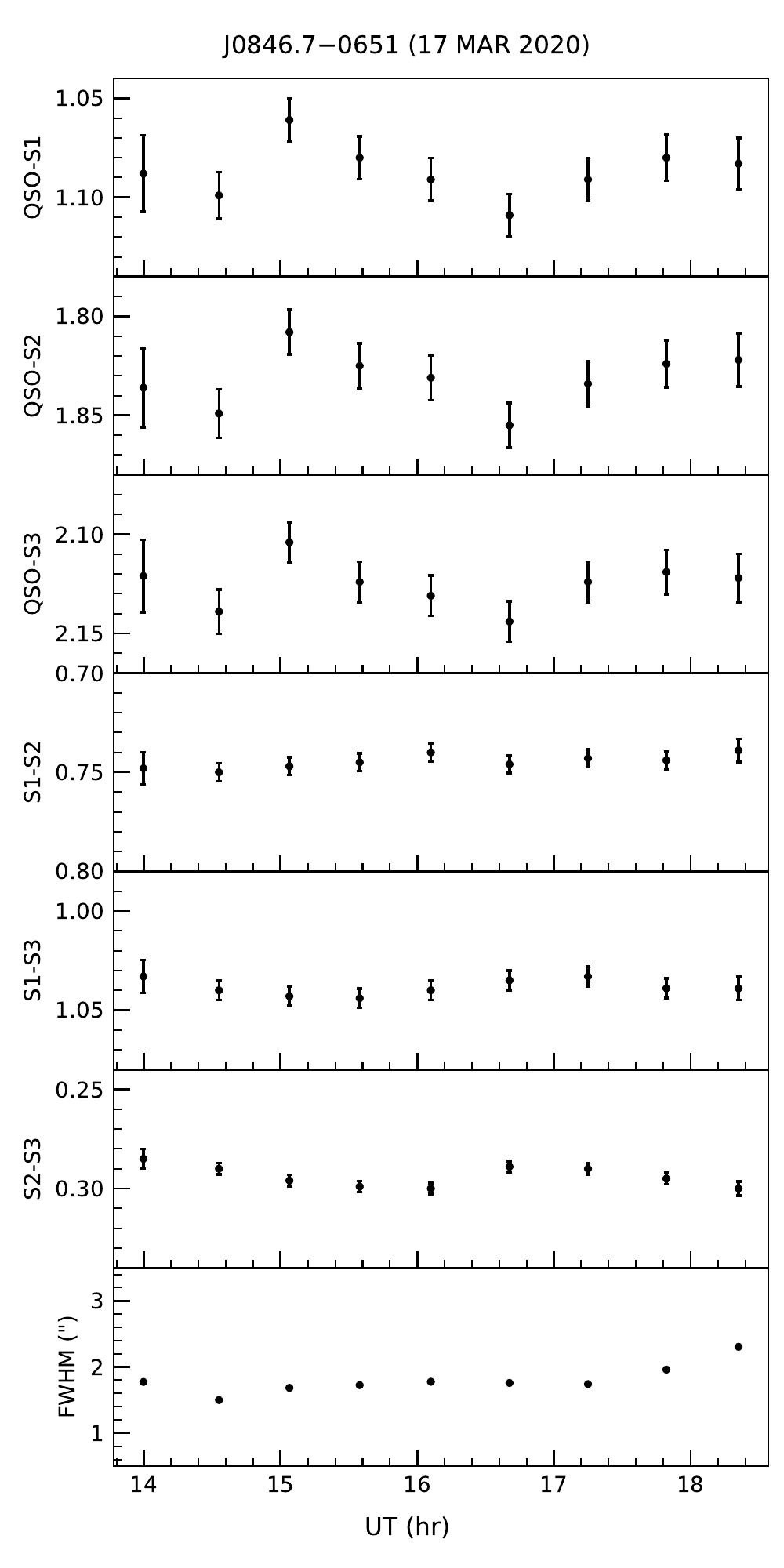}
}
\includegraphics[scale=0.53]{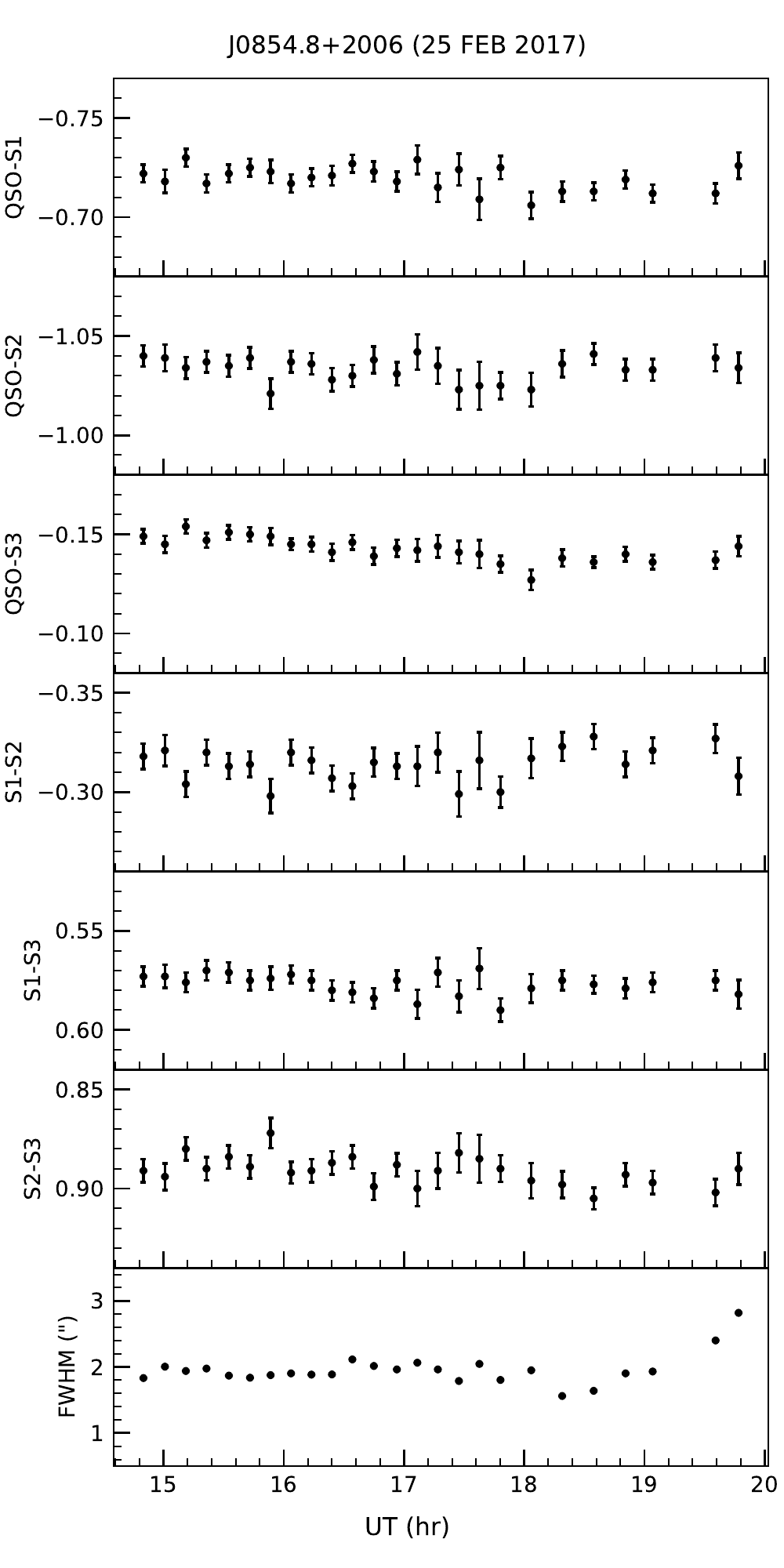}
\includegraphics[scale=0.53]{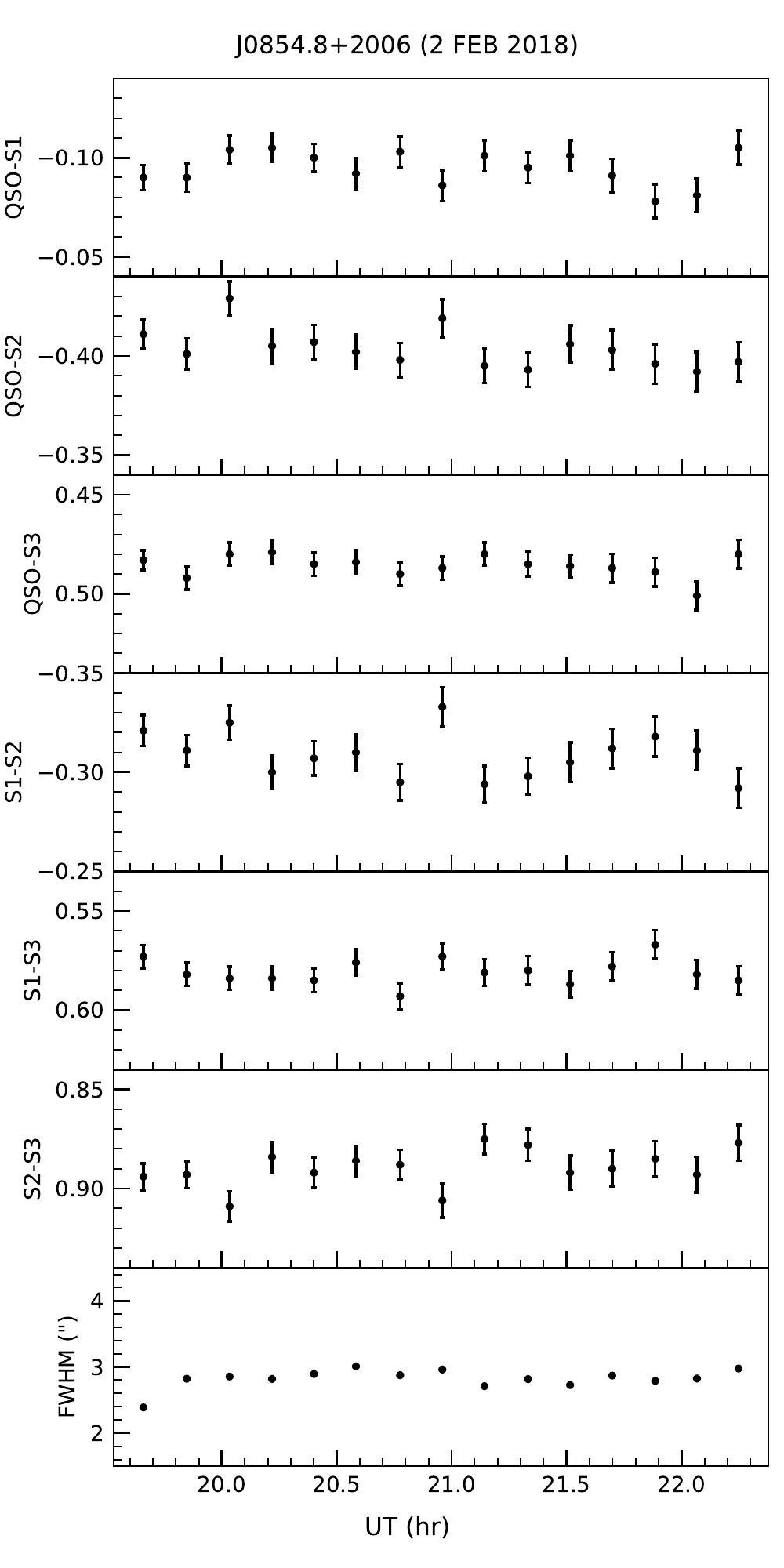}
\includegraphics[scale=0.53]{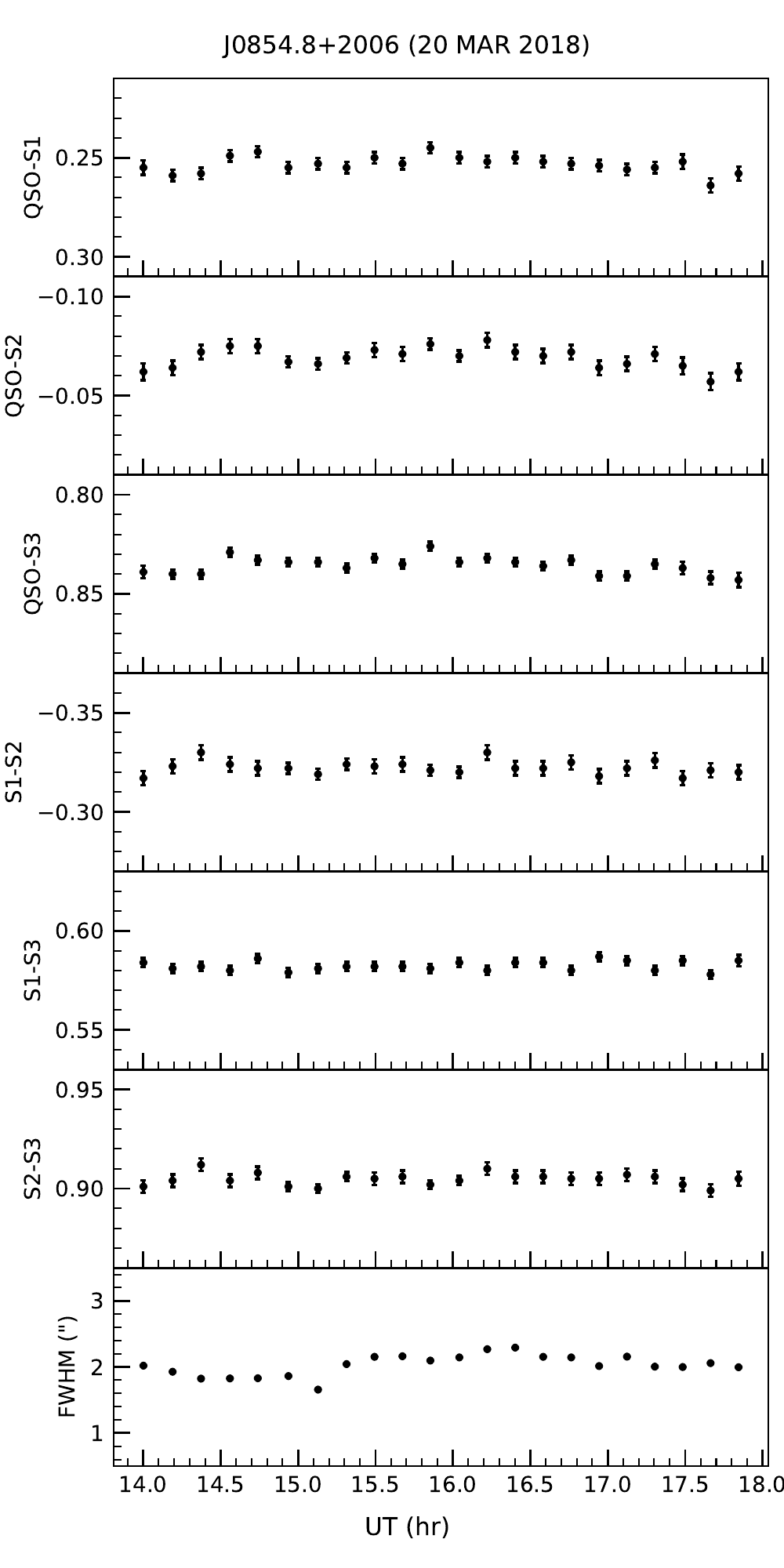}
}
\caption{DLCs of the BL Lac objects J0846.7$-$0651 (top panels) and J0854.8+2006 (bottom panels)}\label{fig-5}
\end{figure*}

\begin{figure*}
\vbox{
\hbox{
\includegraphics[scale=0.53]{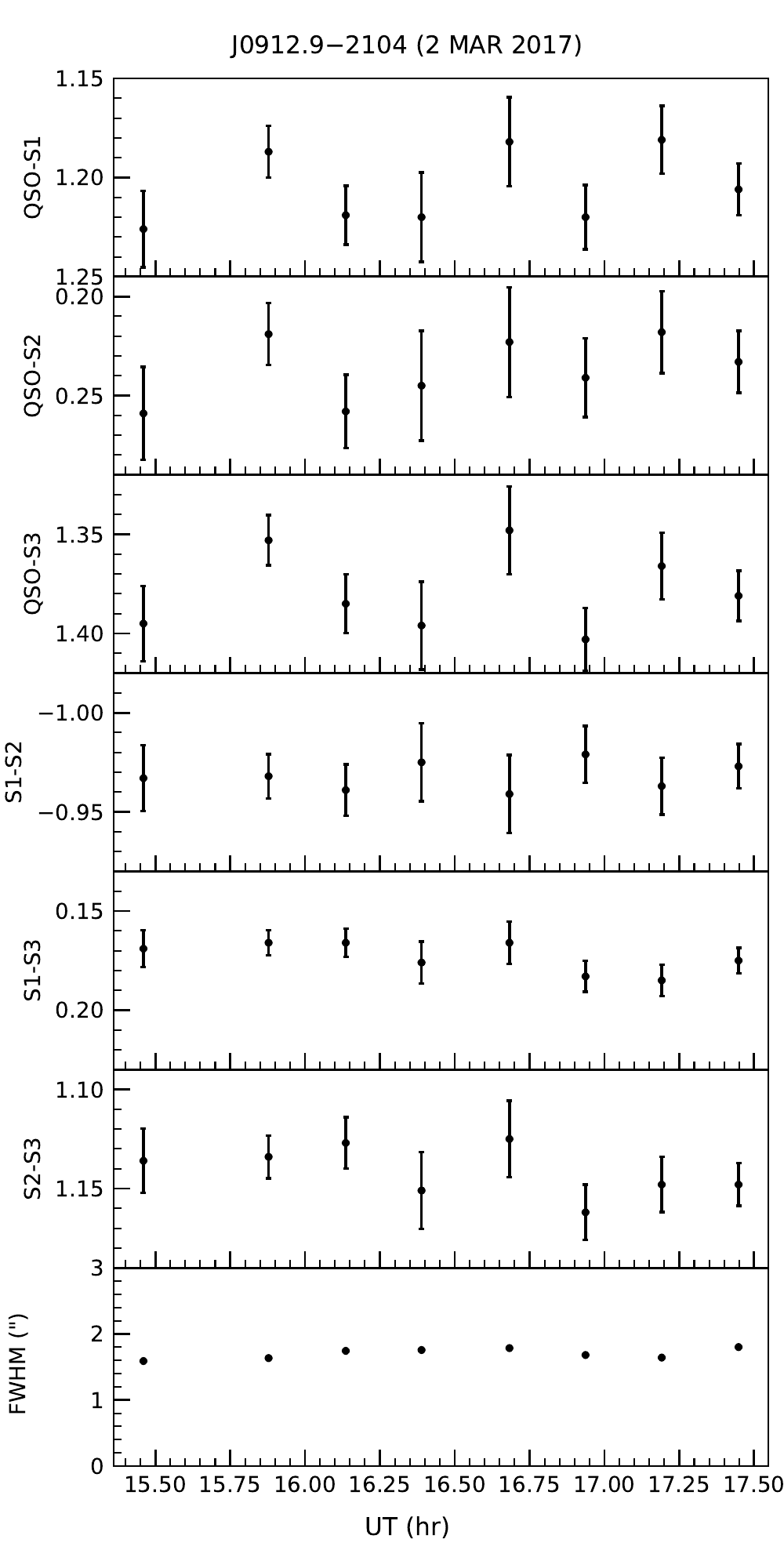}
\includegraphics[scale=0.53]{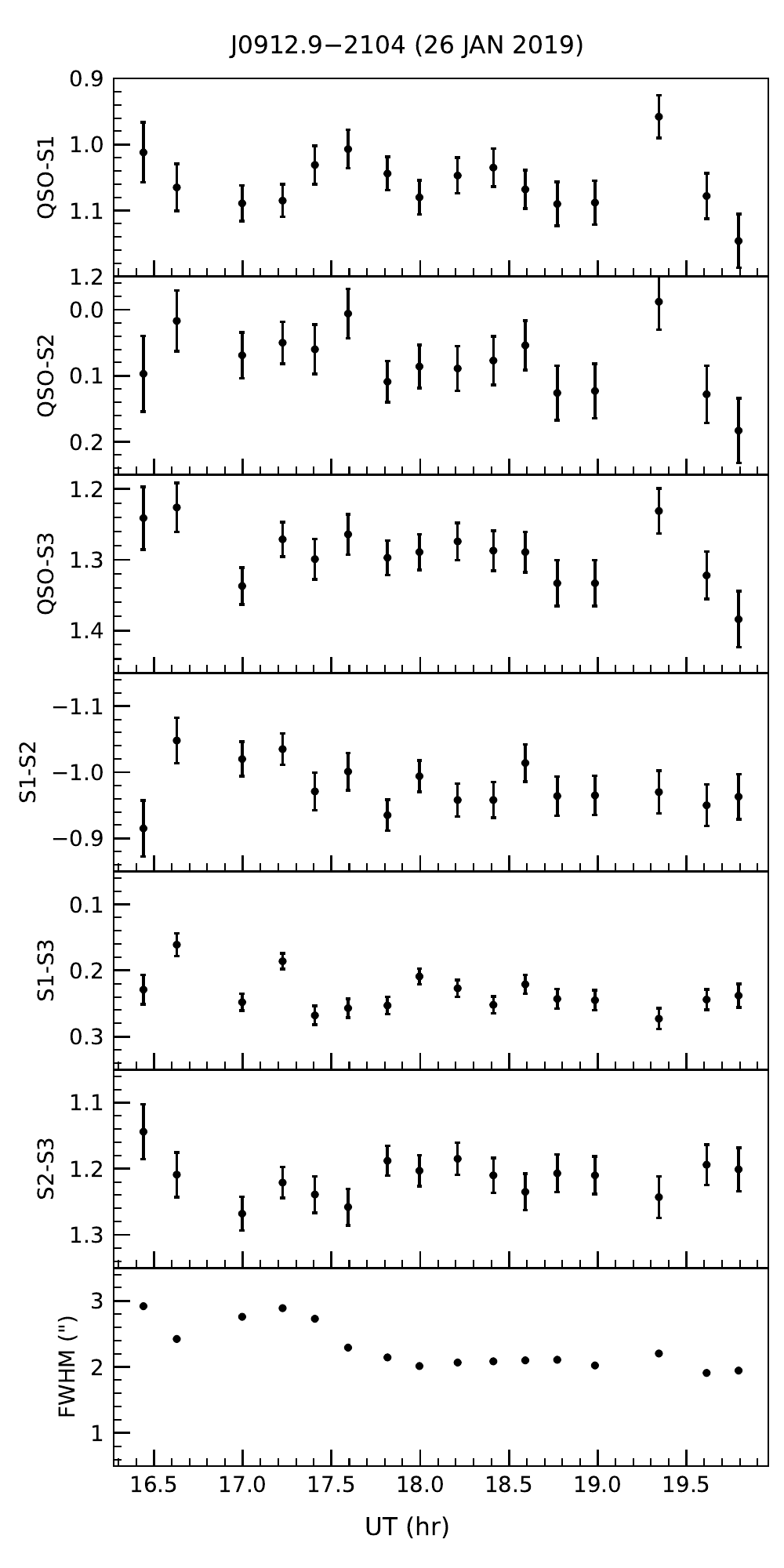}
\includegraphics[scale=0.53]{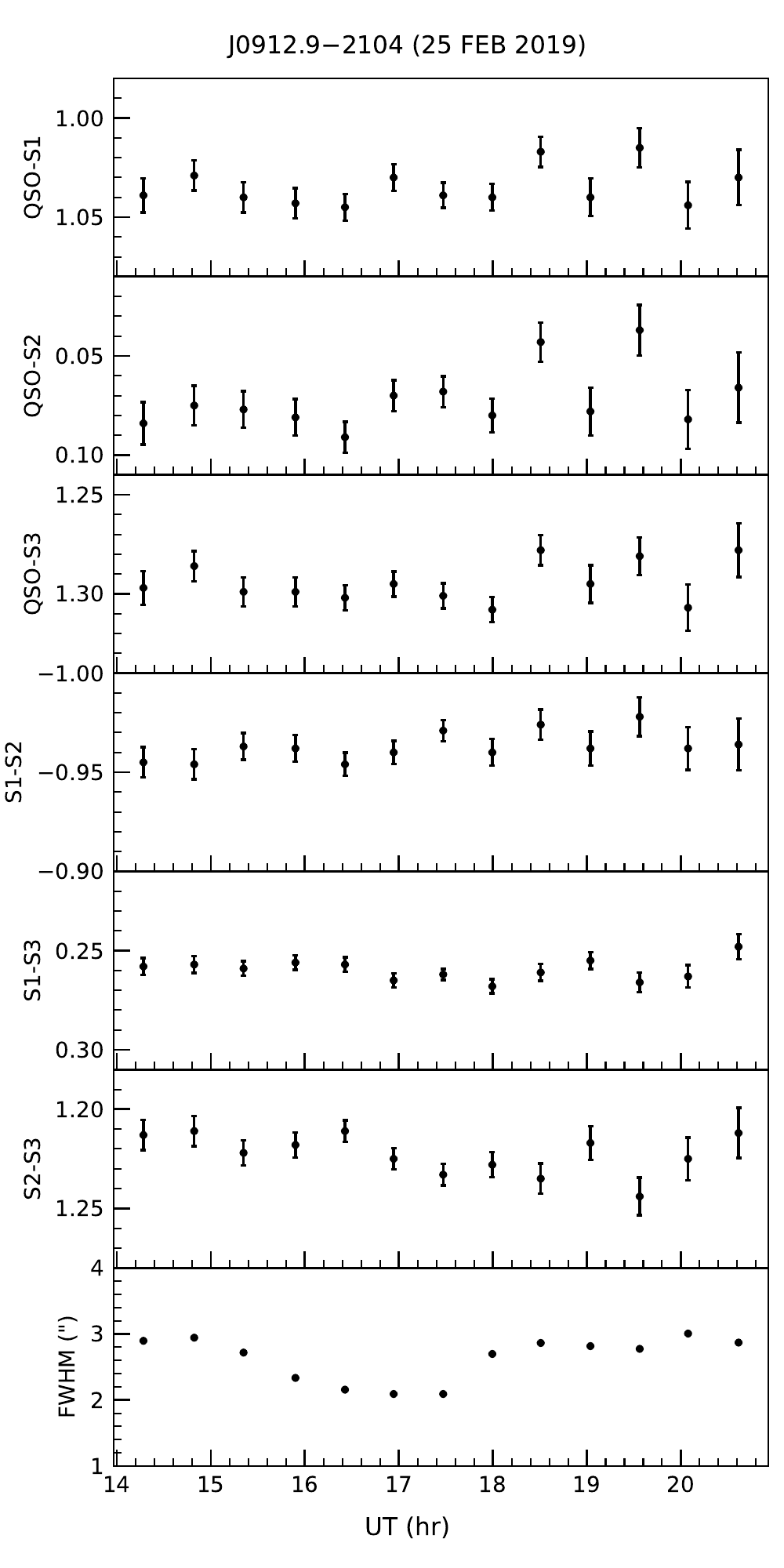}
}
\includegraphics[scale=0.53]{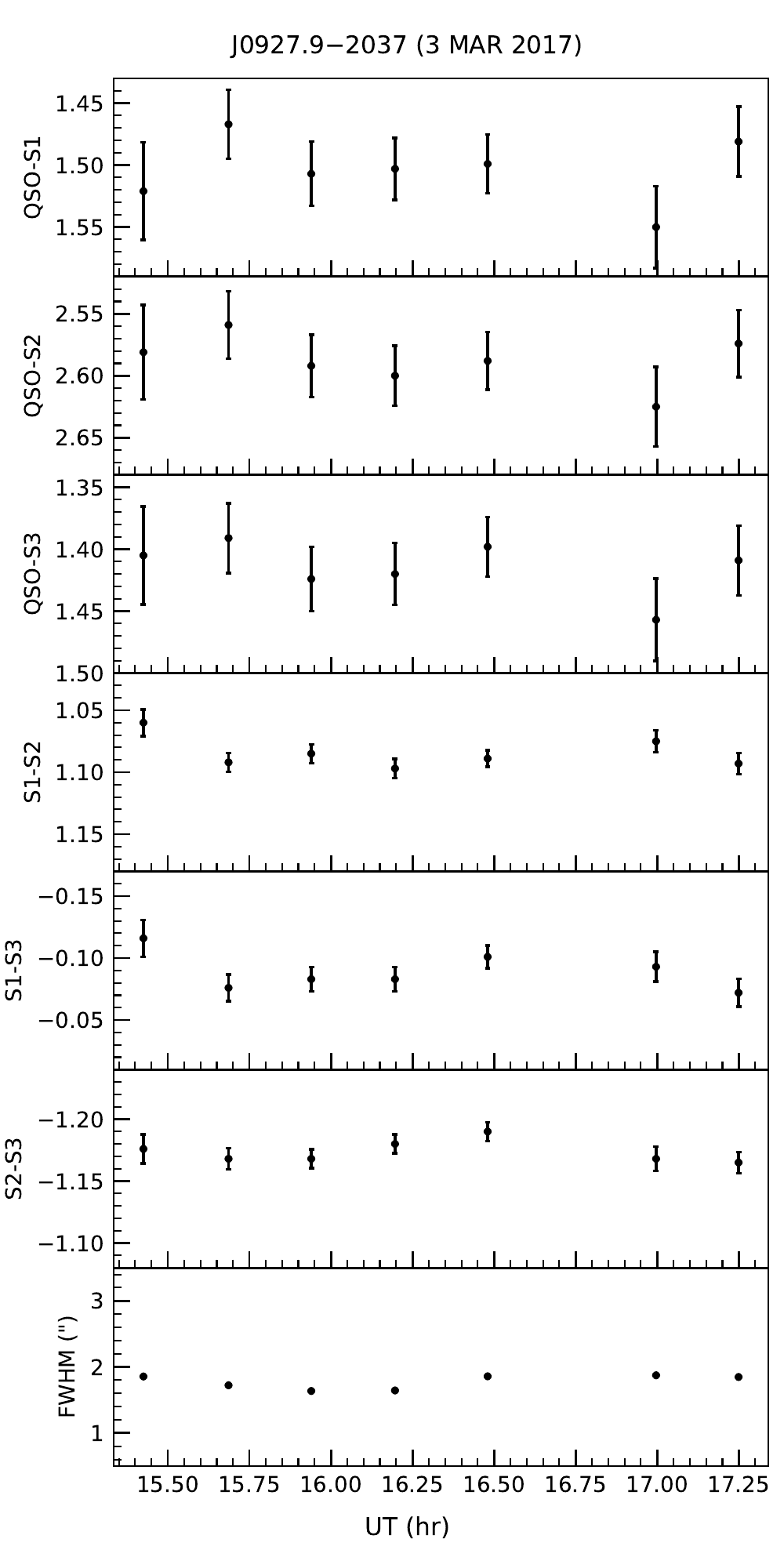}
\includegraphics[scale=0.53]{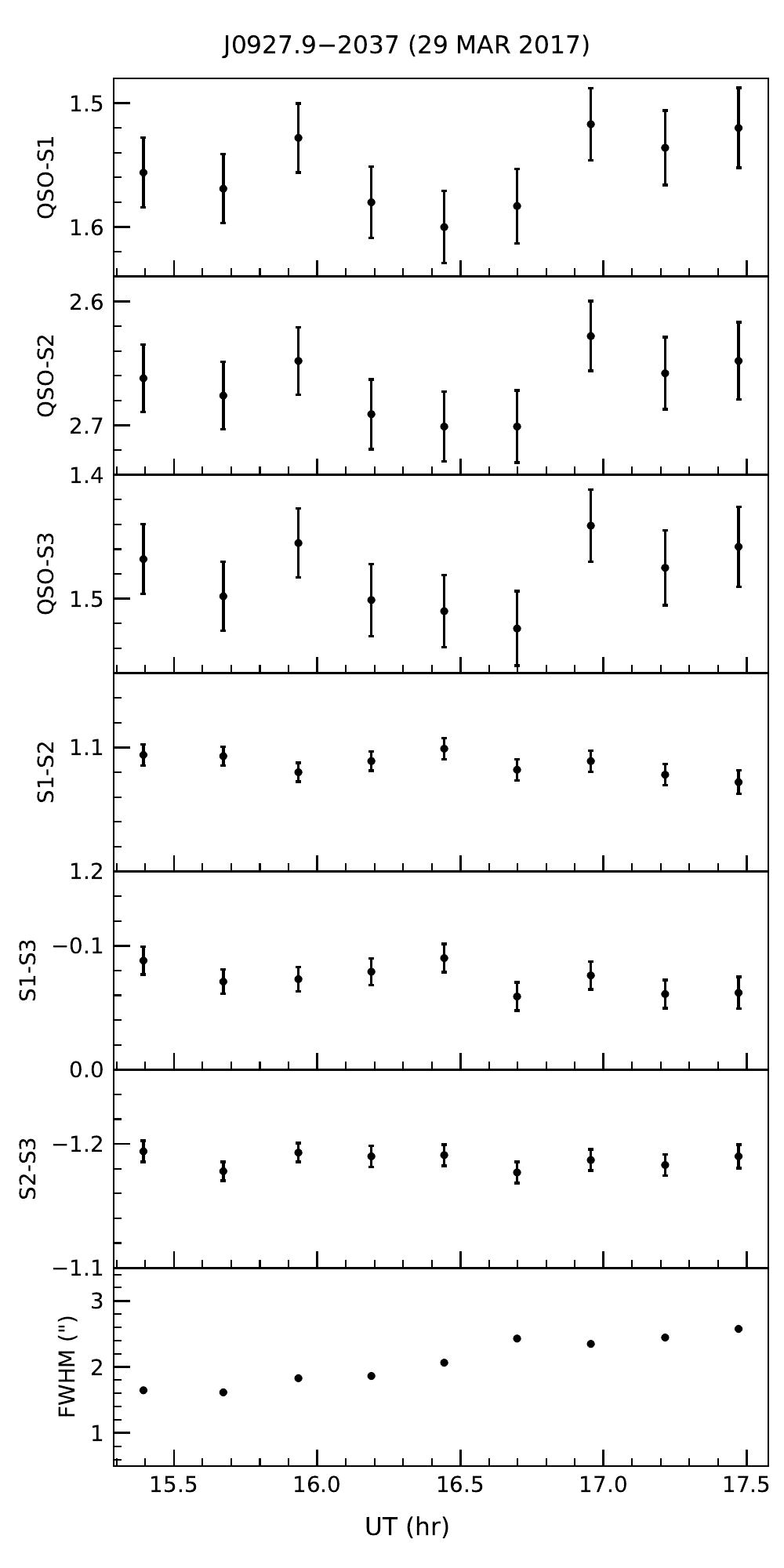}
\includegraphics[scale=0.53]{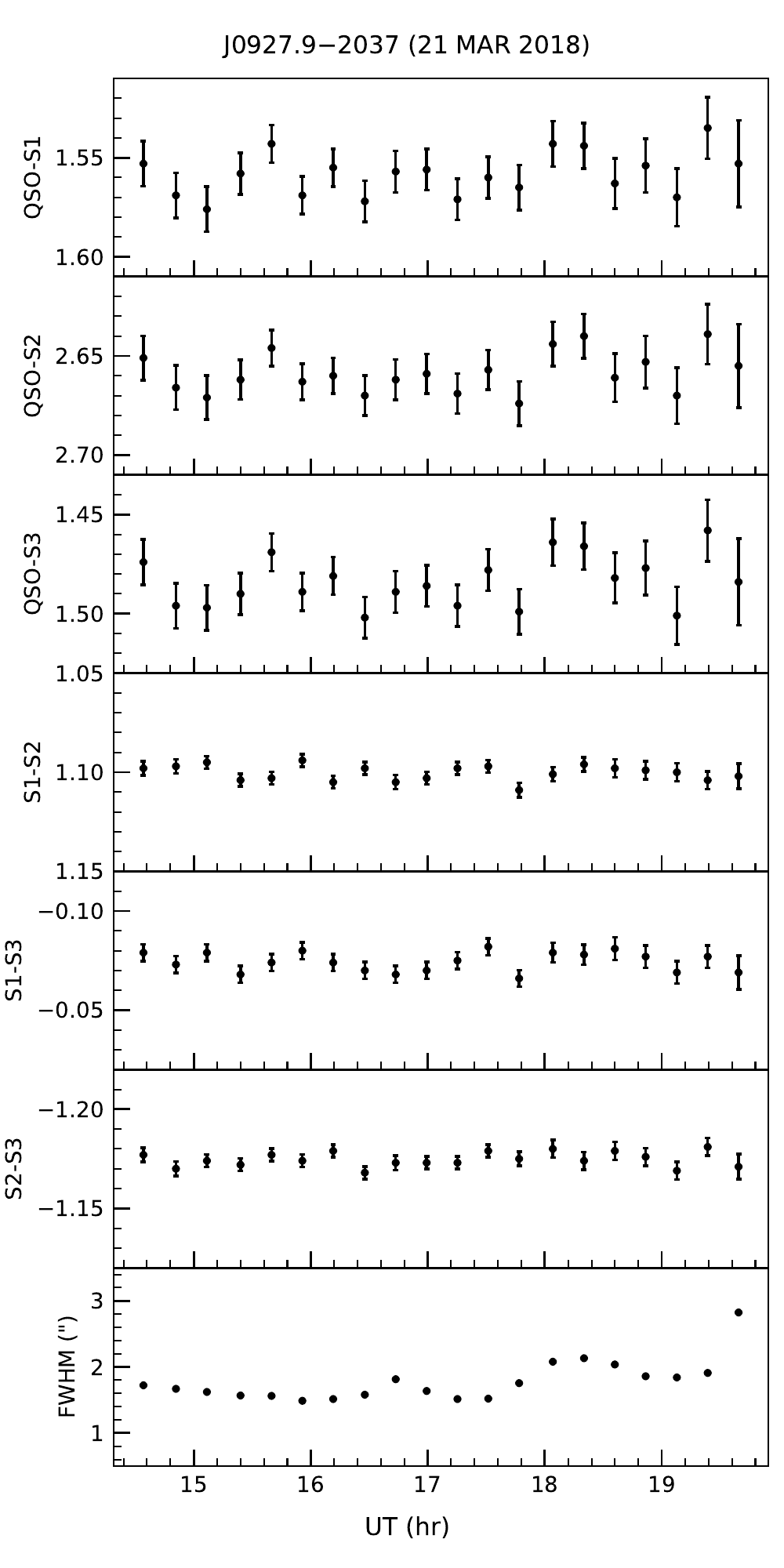}
}
\caption{Top panels: DLCs of the HSP BL Lac J0912.9$-$2104. Bottom panels: DLCs of the FSRQ J0927.9$-$2037. Labels
have the same meaning as that of Fig. 1}\label{fig-6}
\end{figure*}

\begin{figure*}
\vbox{
\hbox{
\includegraphics[scale=0.53]{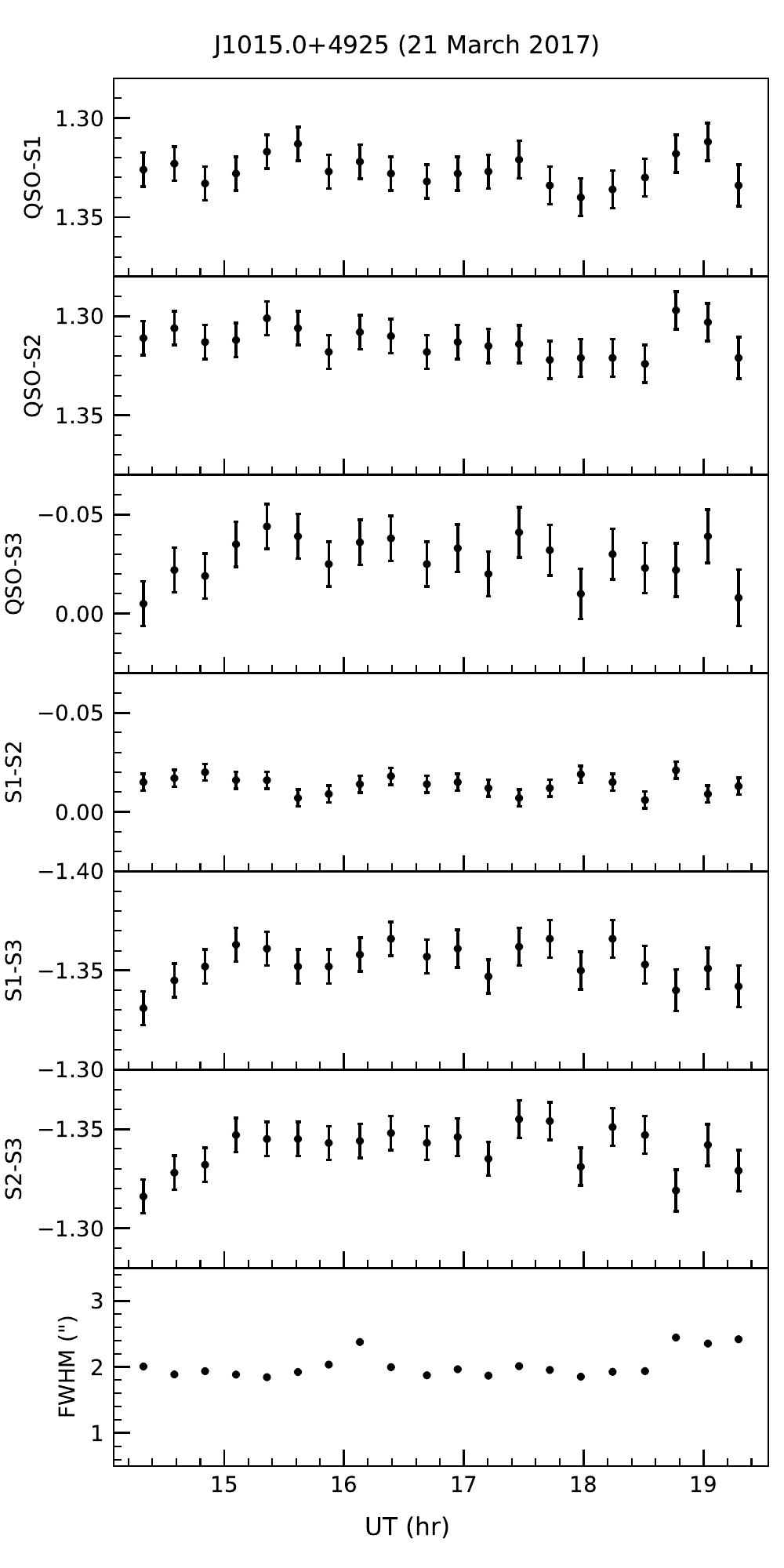}
\includegraphics[scale=0.53]{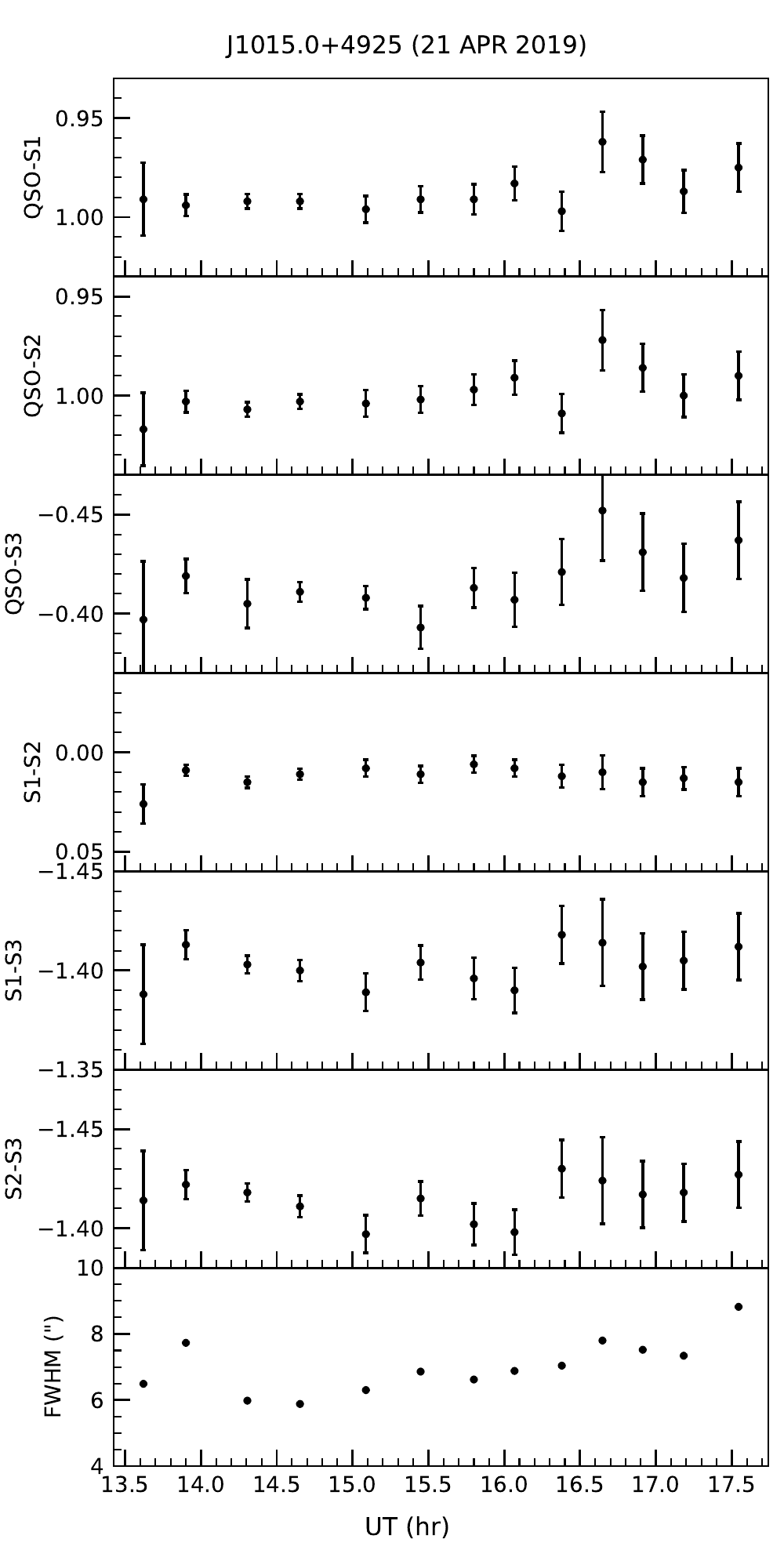}
\includegraphics[scale=0.53]{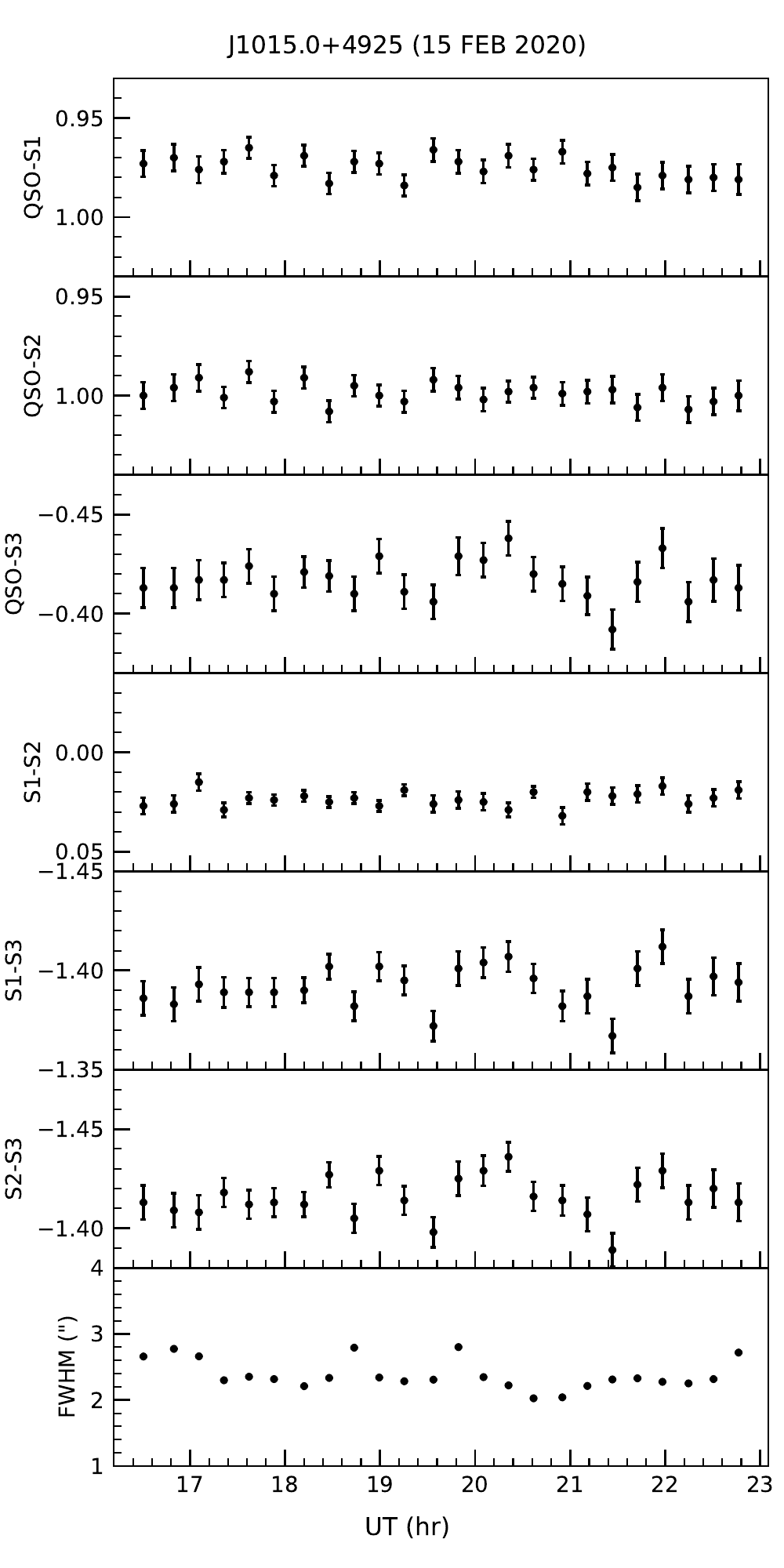}
}
\includegraphics[scale=0.53]{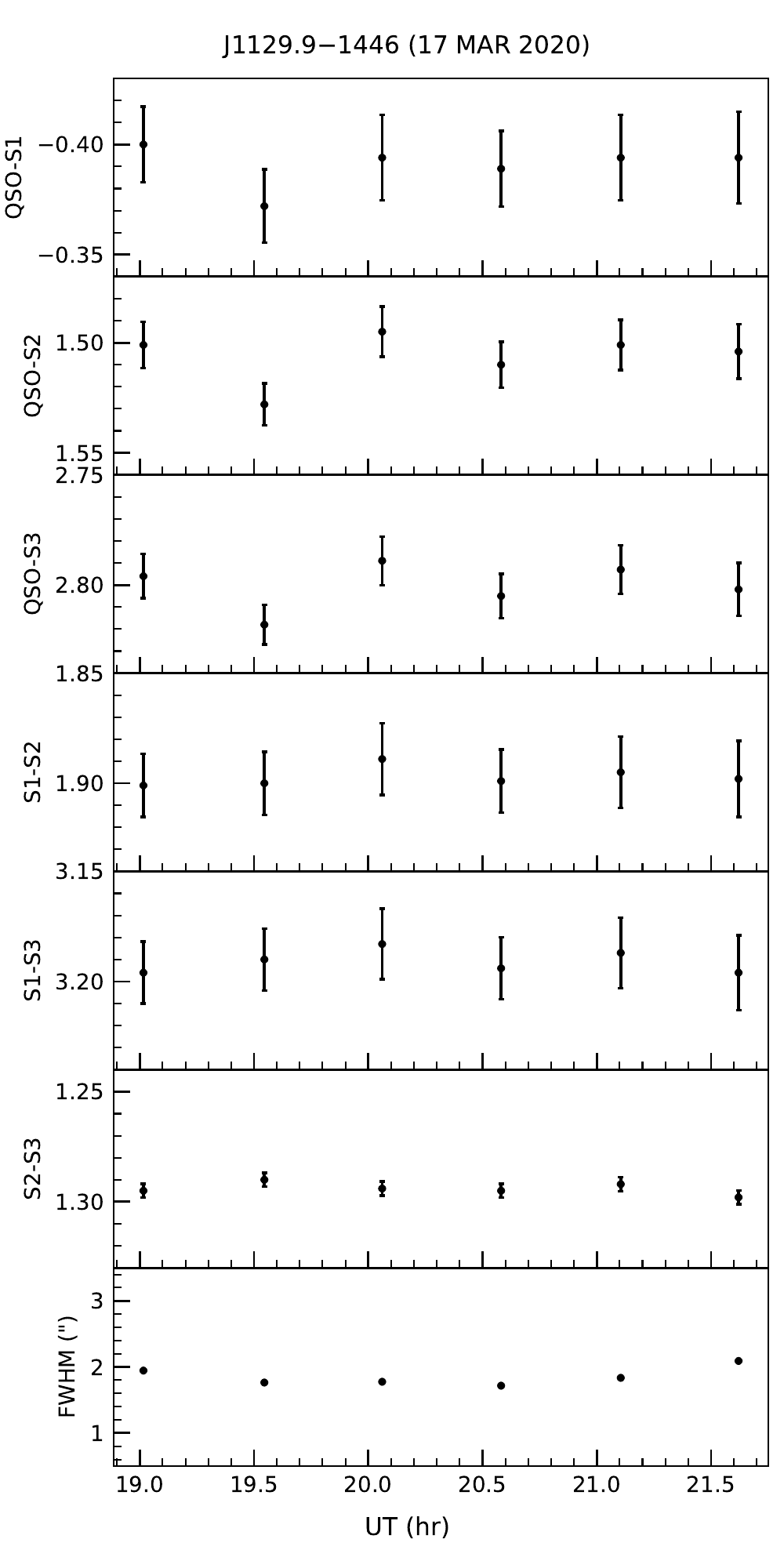}
\includegraphics[scale=0.53]{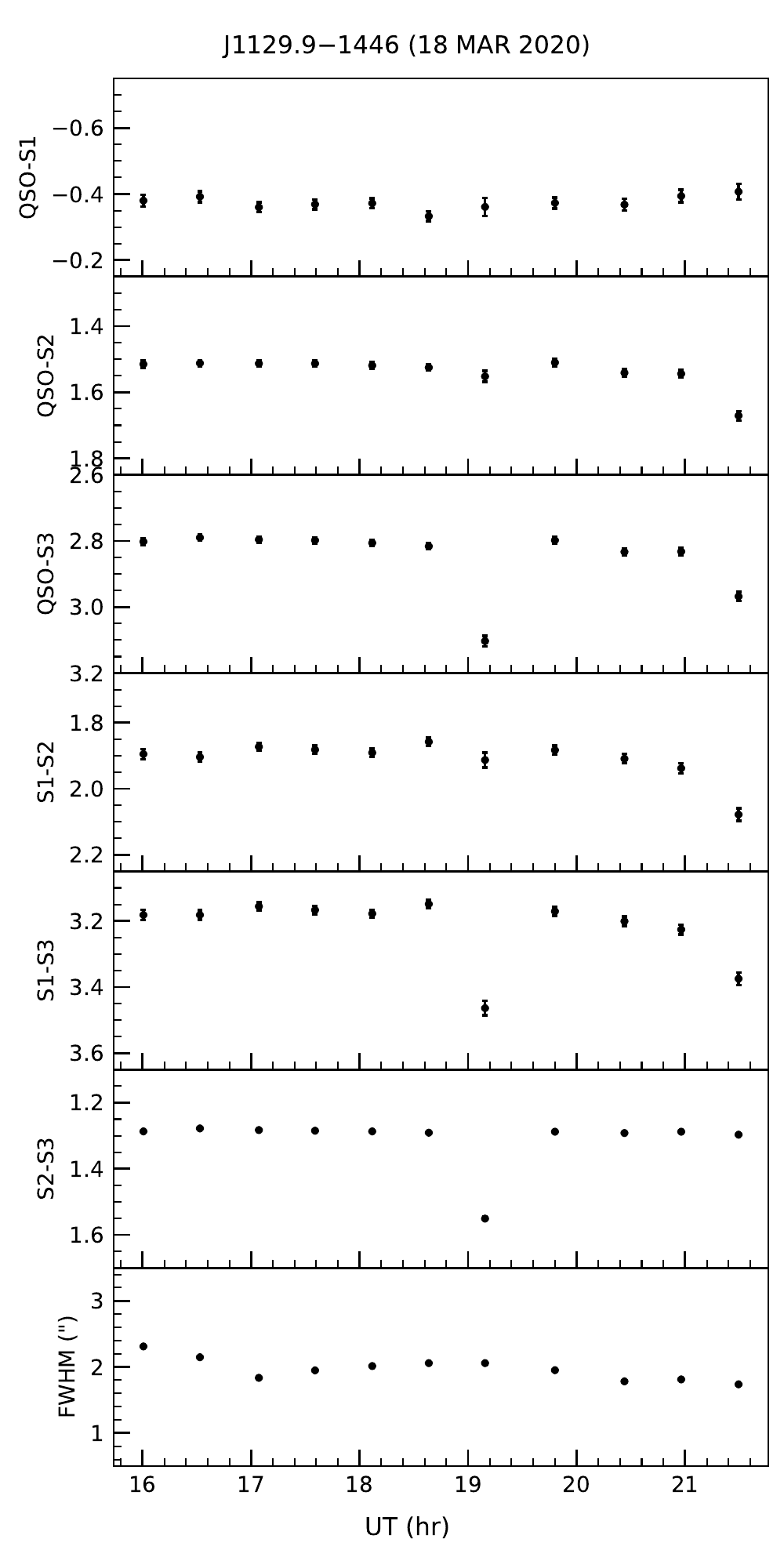}
}
\caption{DLCs of the HSP BL Lac J1015.0+4925 (top panels) and the FSRQ J1129.9$-$1446 (bottom panels).}\label{fig-7}
\end{figure*}

\begin{figure*}
\vbox{
\hbox{
\includegraphics[scale=0.53]{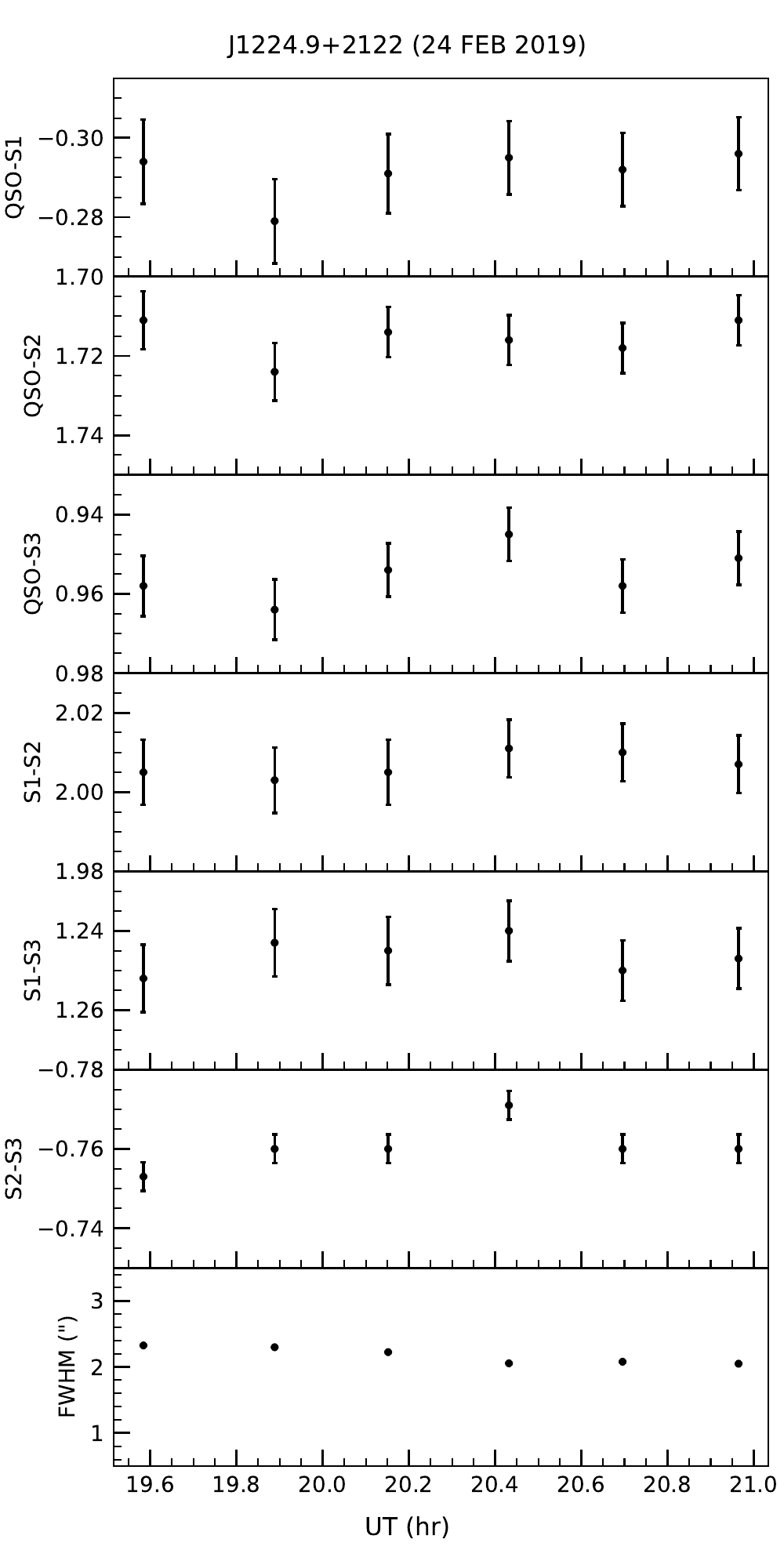}
\includegraphics[scale=0.53]{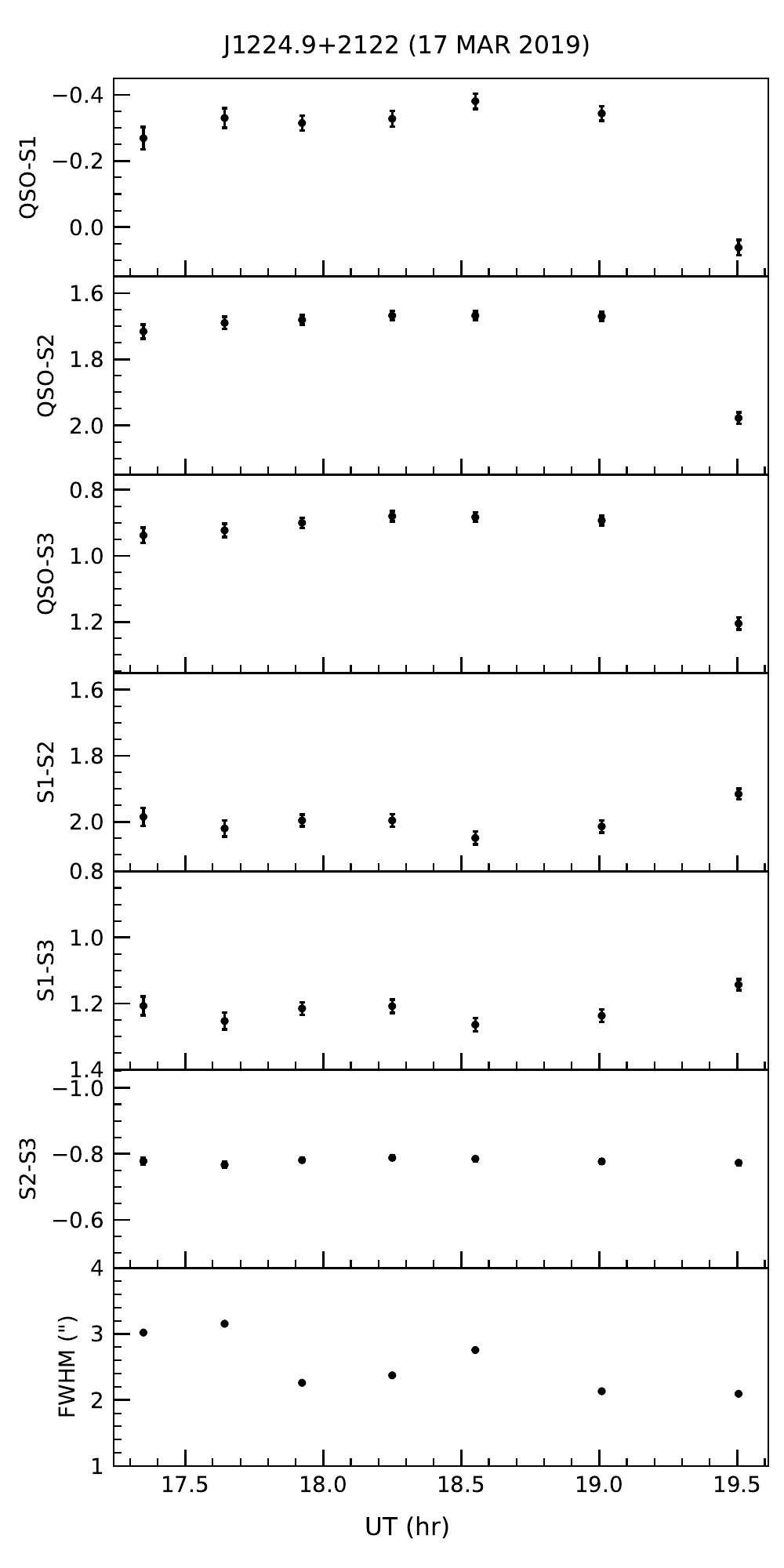}
\includegraphics[scale=0.53]{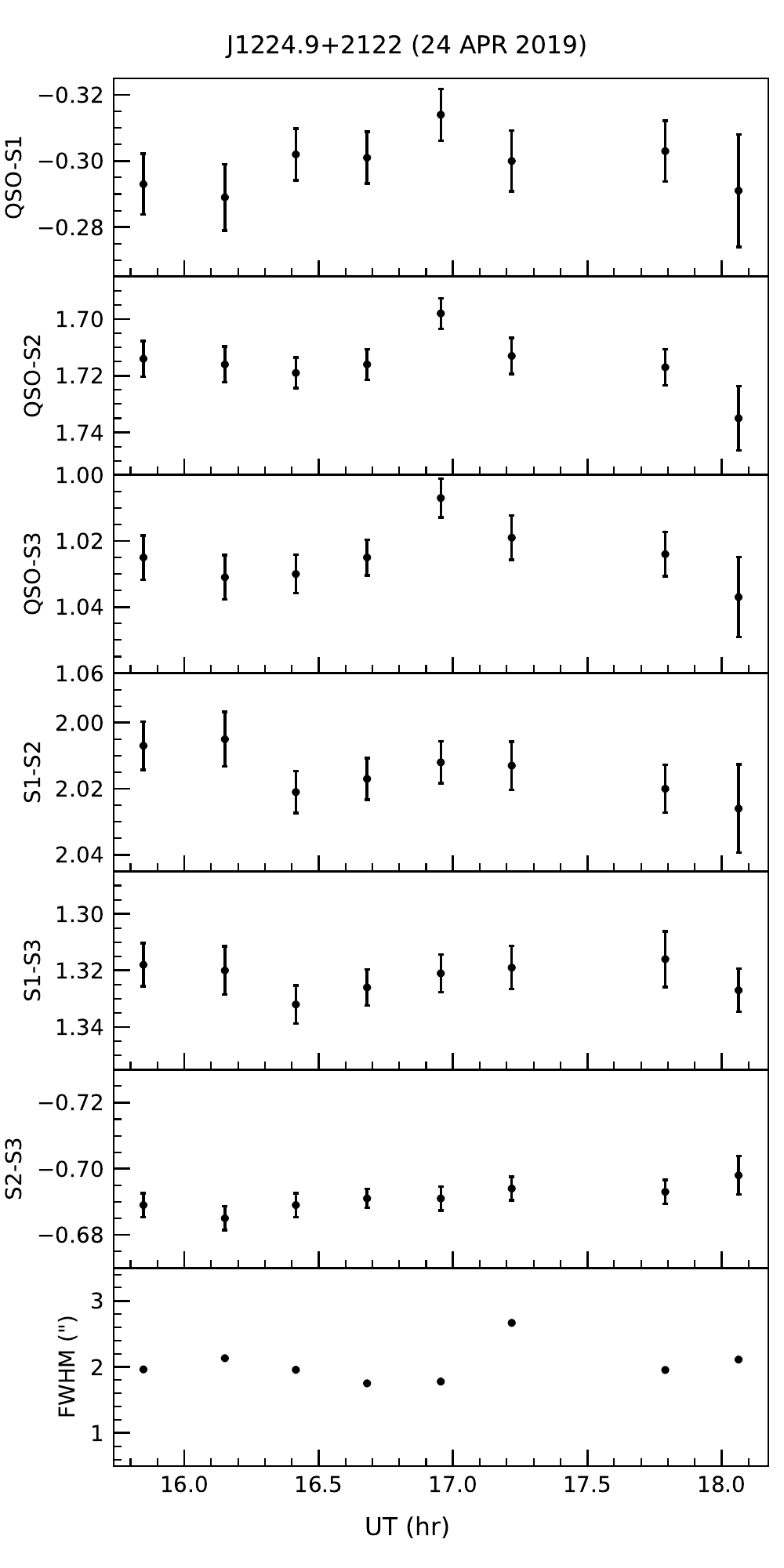}
}
\includegraphics[scale=0.53]{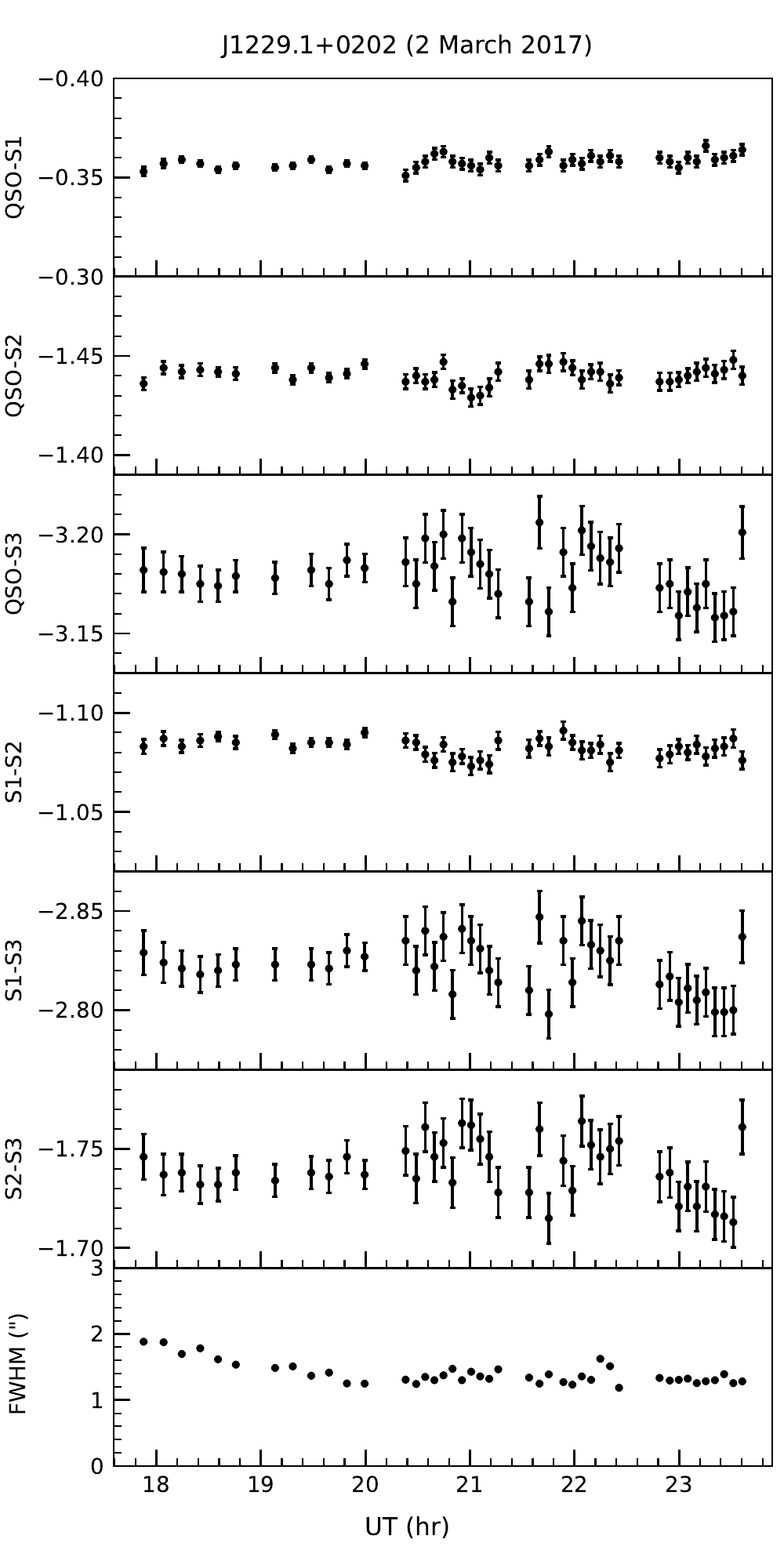}
\includegraphics[scale=0.53]{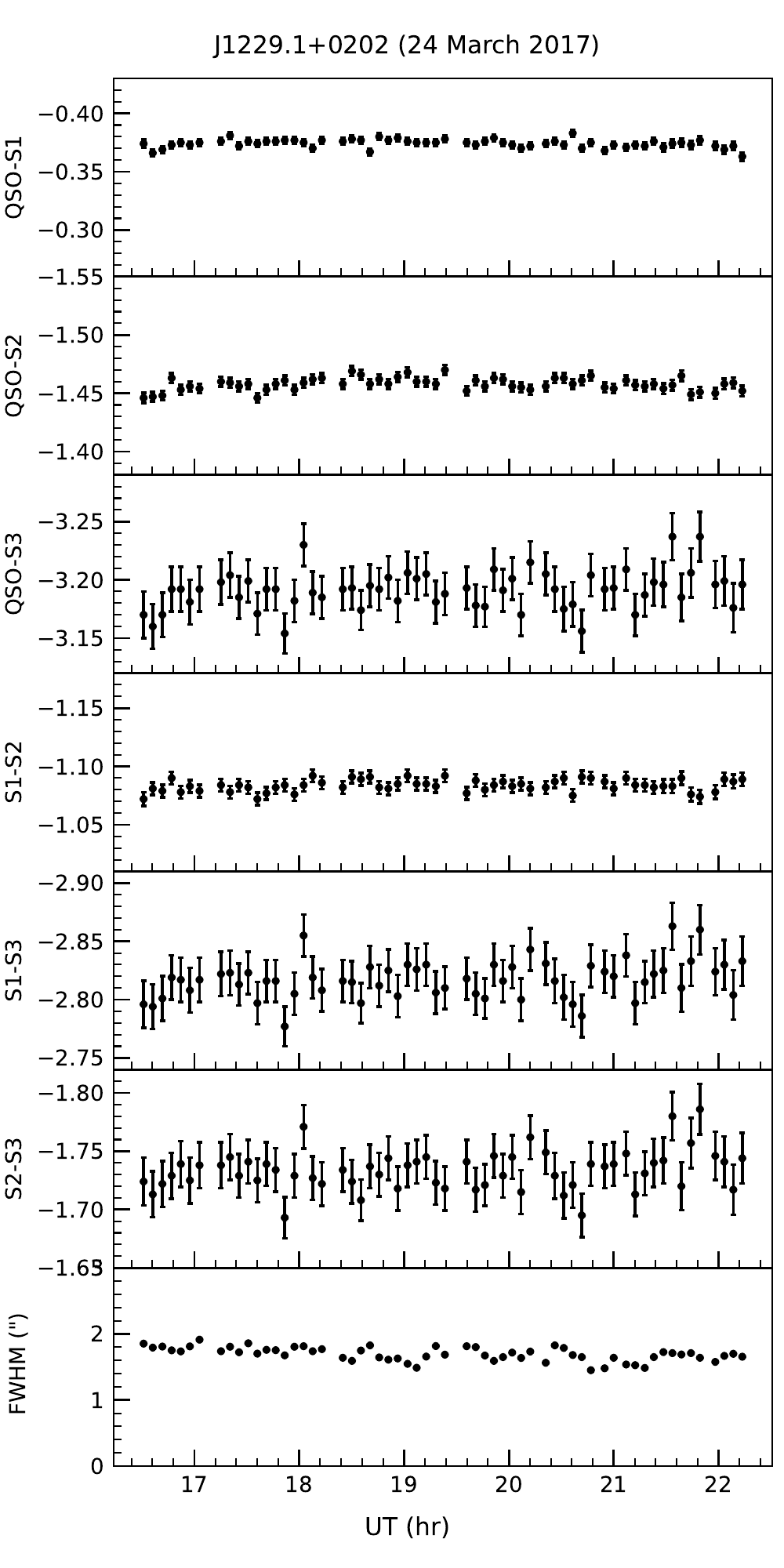}
\includegraphics[scale=0.53]{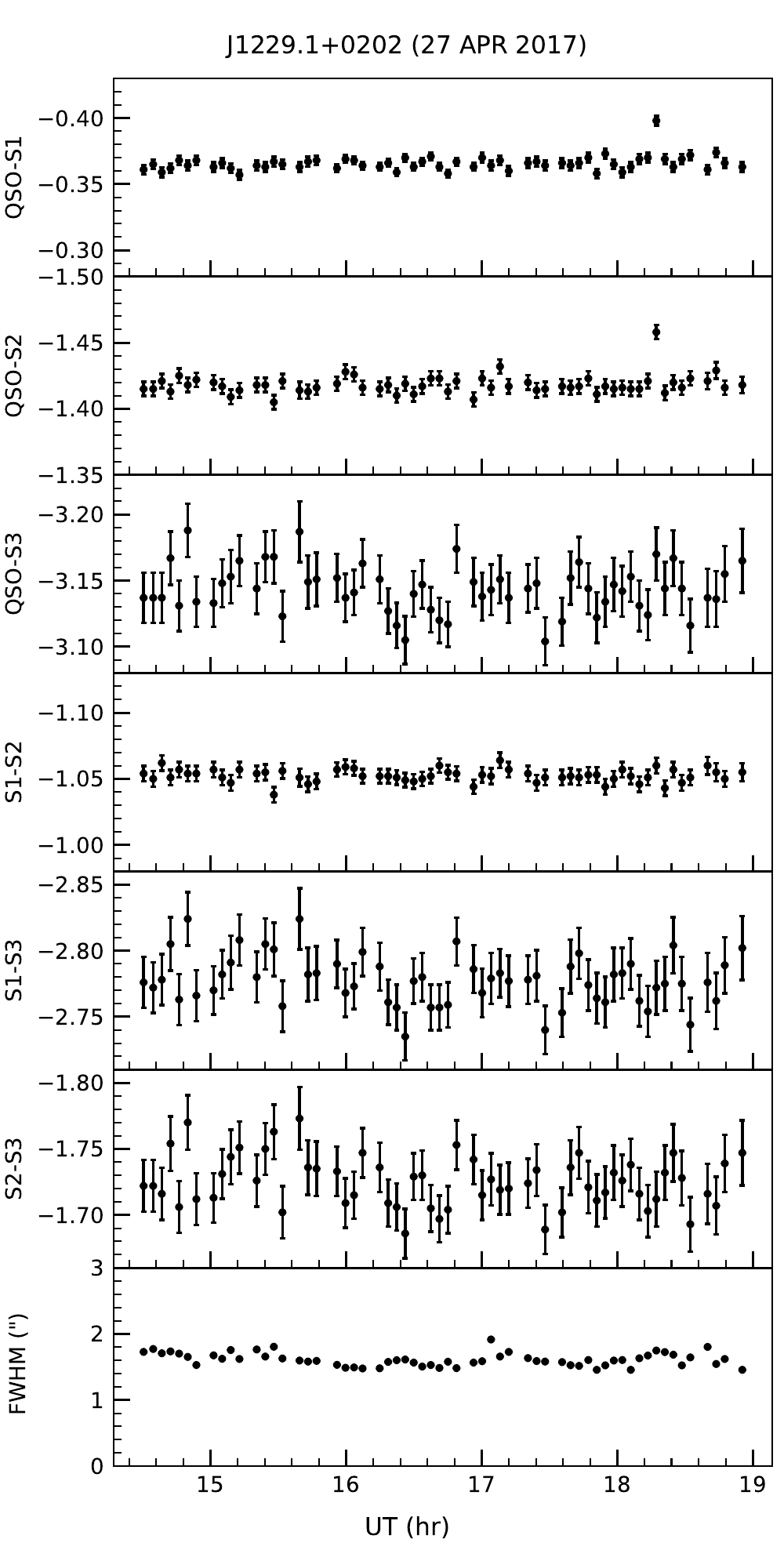}
}
\caption{Top panels: DLCs of the LSP FSRQ J1224.9+2122. Bottom panels: DLCs of the FSRQ J1229.1+0202}\label{fig-8}
\end{figure*}

\begin{figure*}
\vbox{
\hbox{
\includegraphics[scale=0.53]{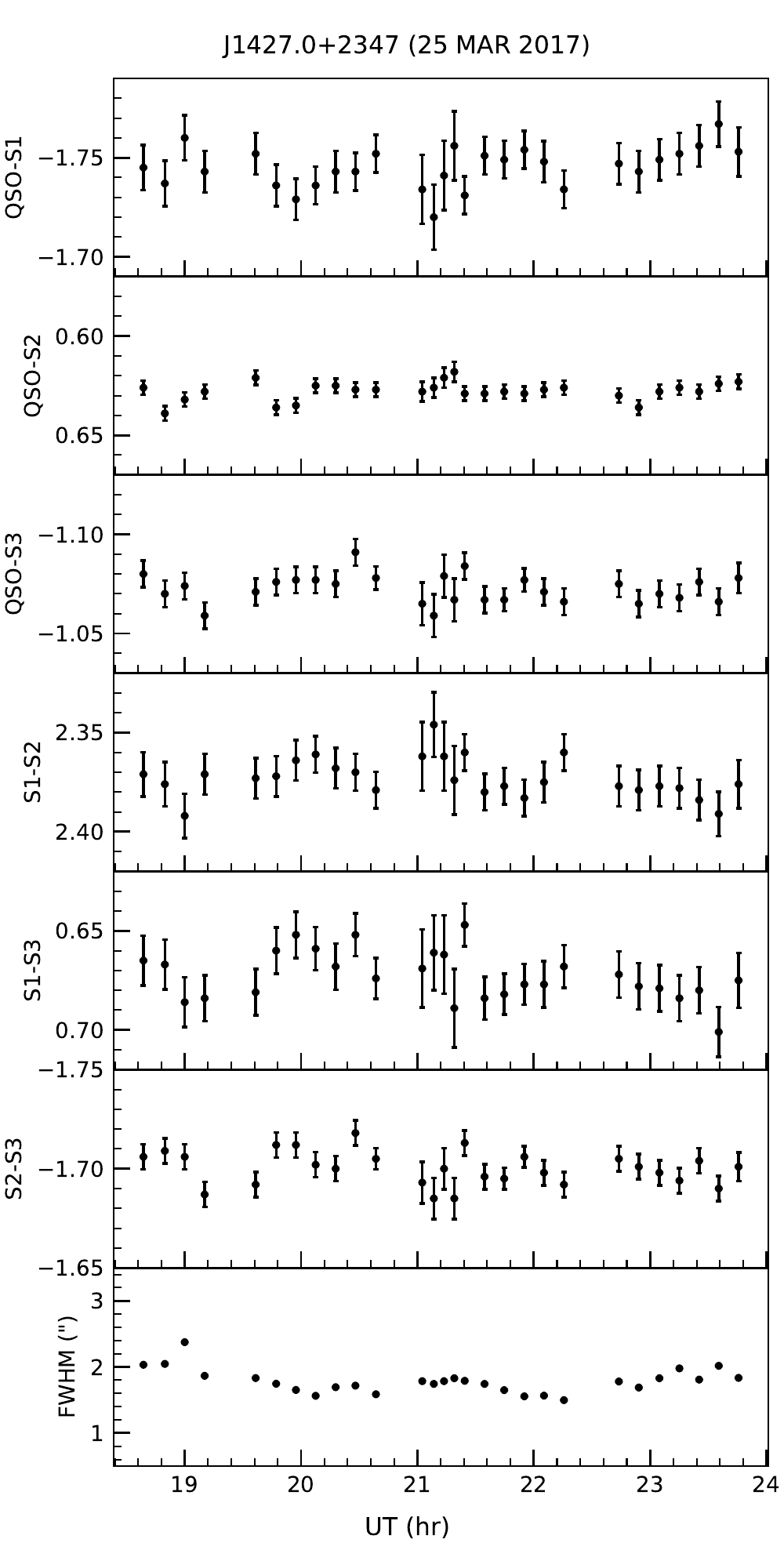}
\includegraphics[scale=0.53]{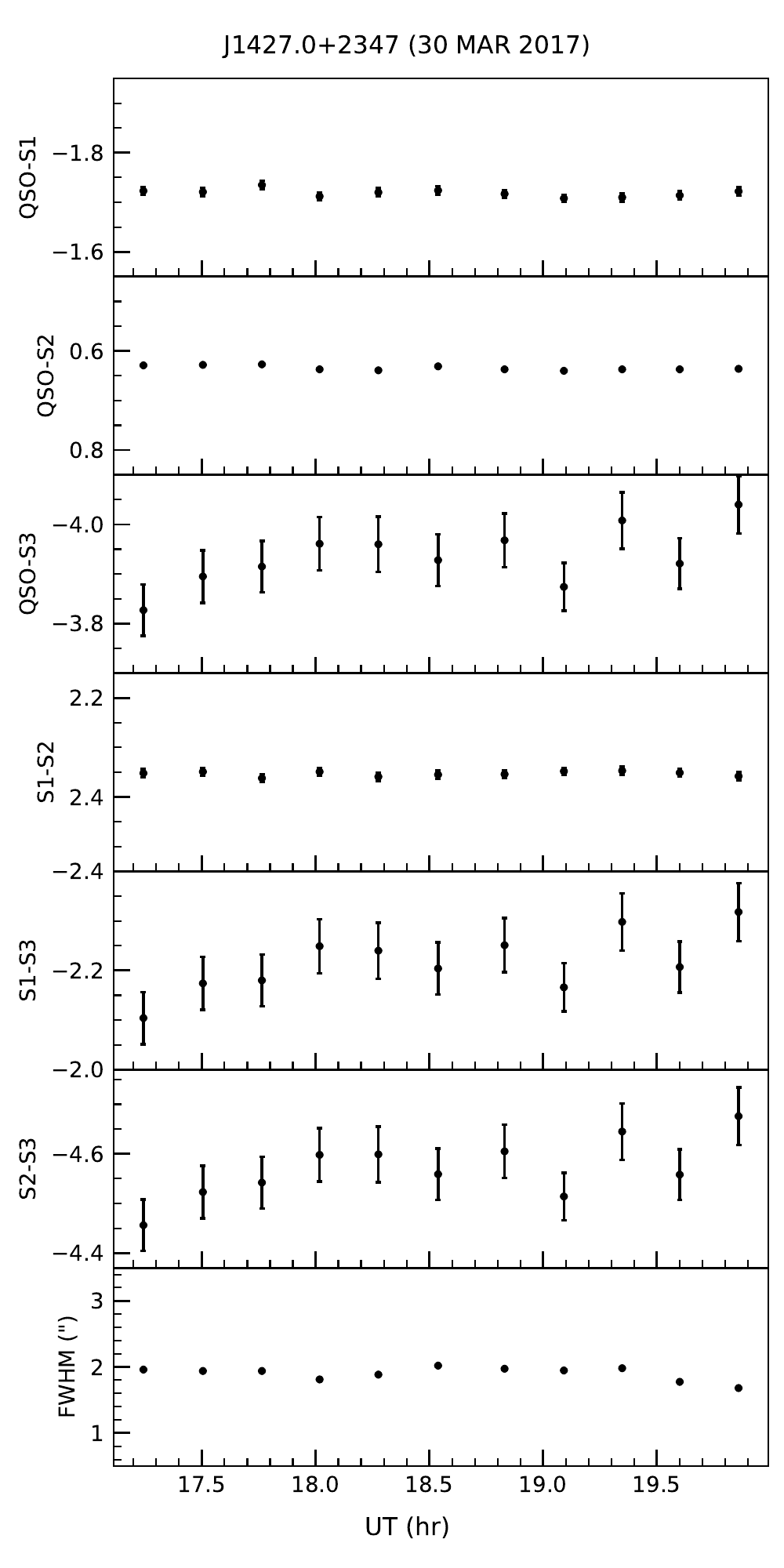}
\includegraphics[scale=0.53]{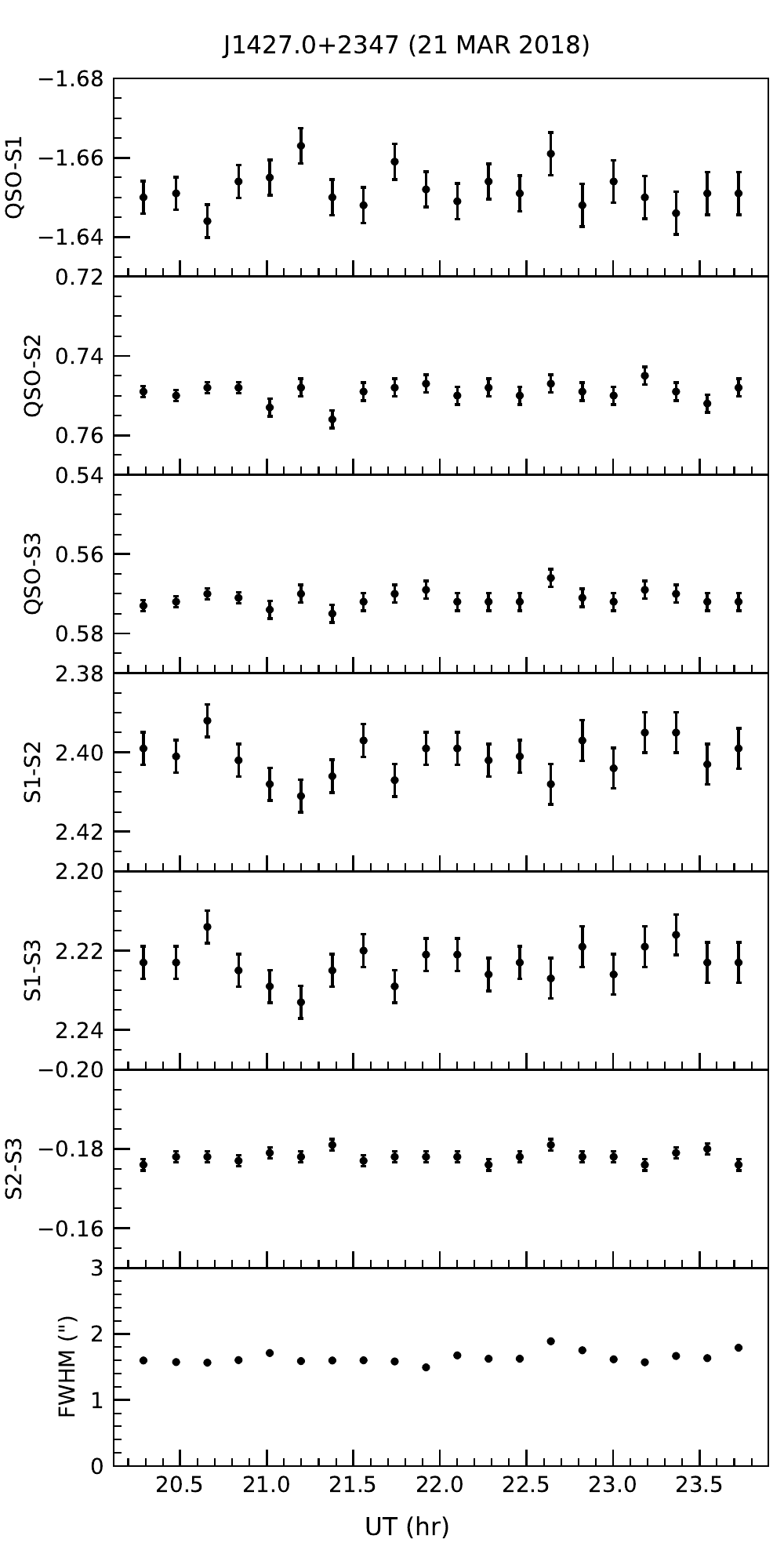}
}
\includegraphics[scale=0.53]{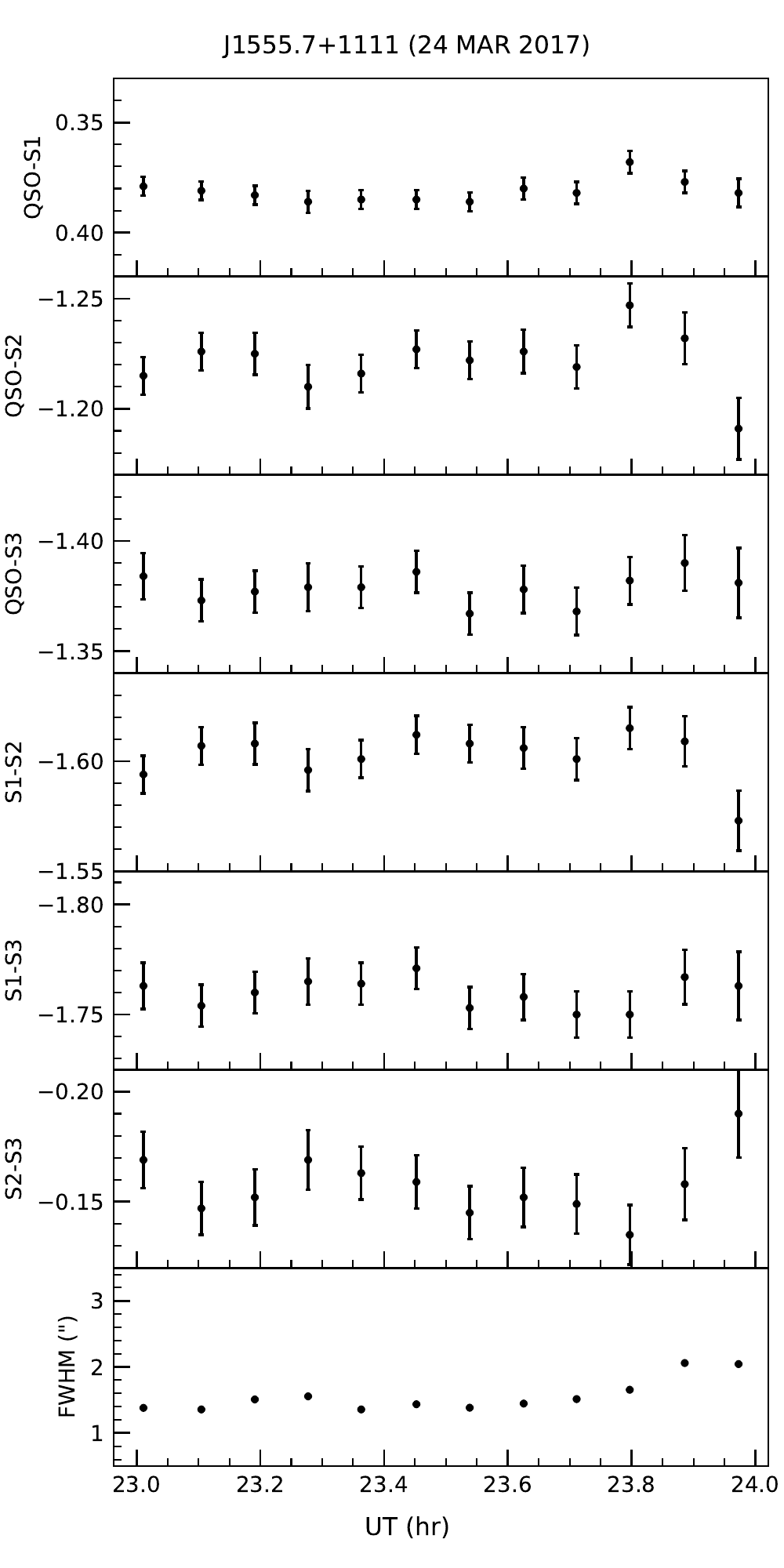}
\includegraphics[scale=0.53]{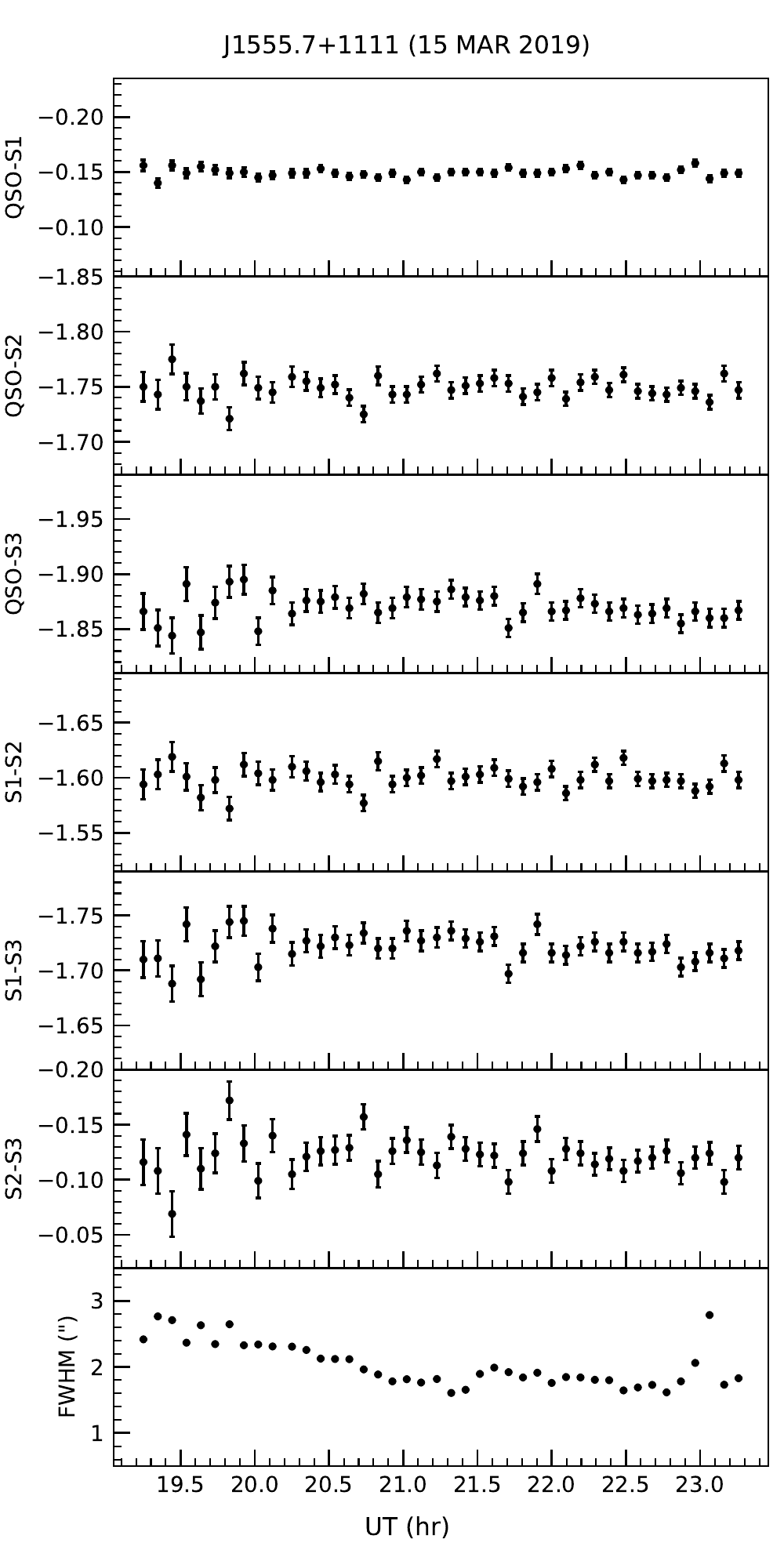}
\includegraphics[scale=0.53]{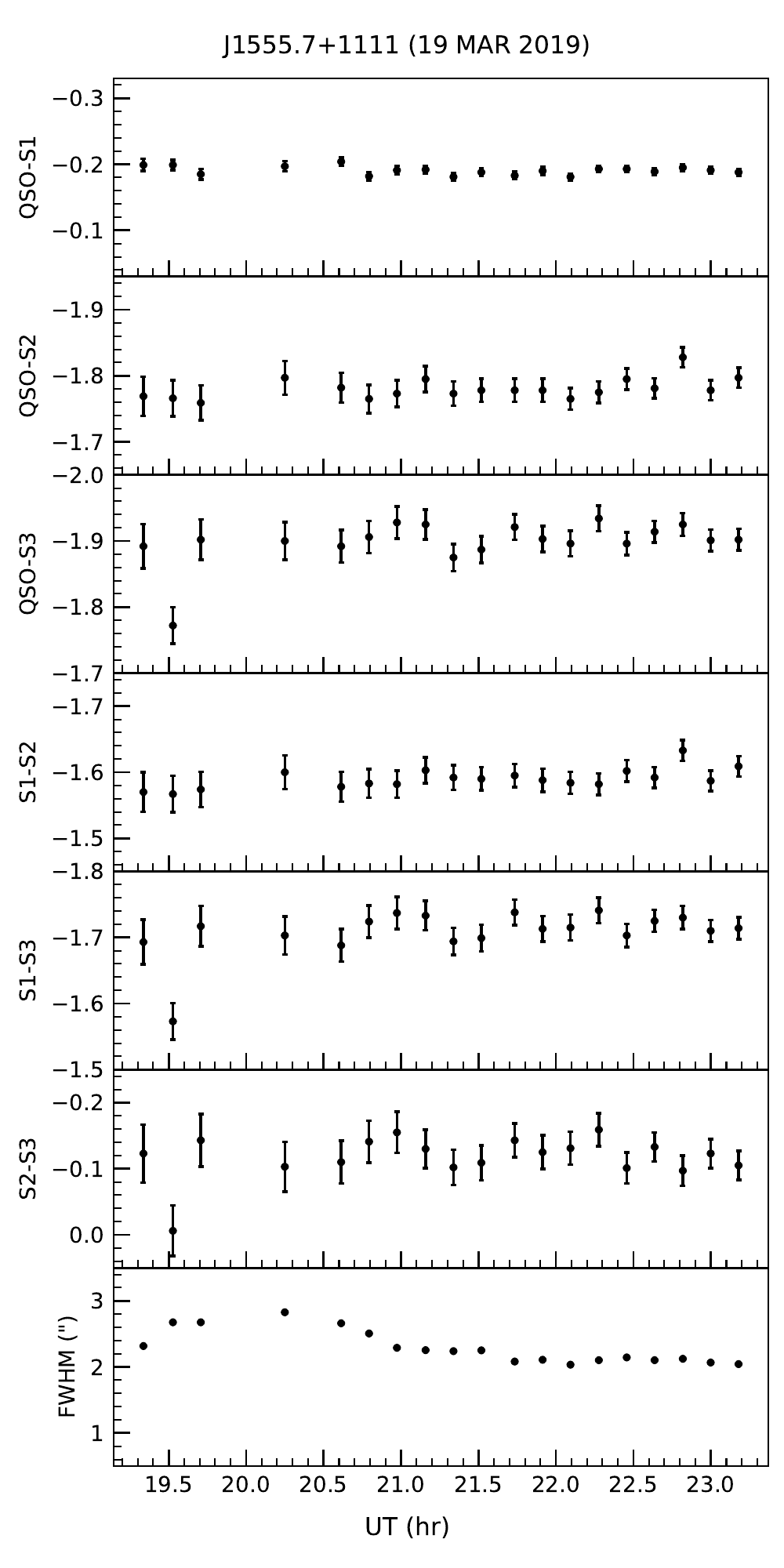}
}
\caption{DLCs of the HSP BL Lacs J1427.0+2347 (top panels) and J1555.7+1111 (bottom panels).}\label{fig-9}
\end{figure*}

\section{Sample, Observation and Data Reduction}
\subsection{Sample set}
We selected our sample of sources from the third catalog of AGN 
detected by the {\it Fermi}-LAT (3LAC; \citealt{2015ApJ...810...14A}) with the
criteria that they must be observable from the Vainu Bappu Observatory, 
Kavalur and relatively bright with R-band brightness $<$ 18 mag. Utilizing this
criteria we arrived at a sample of about 100 blazars.\ This also includes sources without redshift information. However, our 
final sample of sources were driven by the availability of telescope time.
With the above constraints we arrived at a sample of 18 blazars, that   
comprises of 5 FSRQs and 13 BL Lac objects. Further classifying them
on the basis of the SED type, our sample consists of 8 LSPs, 4 
ISPs and 6 HSPs.  They span the redshift between $z$ = 0.085 and 1.184.  
Details of the sources are given in Table 1.

\subsection{Observations}
The observations were carried out using the 1.3 m J.C. Bhattacharya 
telescope located at the Vainu Bappu Observatory (VBO), Kavalur, 
Tamil Nadu, India. Two charge coupled devices (CCDs) were
used for the observations, one a 1k $\times$ 1k ProEM CCD (peltier cooled)
and the other a 2k $\times$ 4k normal CCD (liquid nitrogen cooled).
For observations
carried out using the 2k $\times$ 4k CCD, only the central 2k $\times$ 2k region
was used. All the observations were carried out using the R-filter, with an
effective wavelength of 6400 \AA ~as the CCDs have the maximum response in this
wave band, thereby also enabling acquisition of data with better
time resolution. More details of the CCDs used in this work can be found in 
\cite{2017ApJ...844...32P}. Observations were carried out on a total
of 40 nights during the period 2016 December to 2020 March. We also aimed to acquire
dense and long duration observations for each object. However, it was
not achieved on all the nights, due to weather and sky transparency, and
we were able to acquire data for duration ranging from 1.5 to 6.9 hrs.
On one night only one  object was monitored except for one instance 
where two objects were observed on a night. The integration time to 
acquire one frame on an object was dictated by the apparent brightness 
state of the object, the phase of the moon and sky transparency. Also, the 
field of view of a source was suitably adjusted so as to have at least
three comparison stars in the CCD chip with brightness close to the quasar.
The complete log of observations is given in Table 2.

\subsection{Data Reduction}
The raw image frames were processed using the standard tasks available in IRAF\footnote{IRAF stands for Image
Reduction and Analysis Facility. It is distributed by the National Astronomy Observatories, which is operated
by the Association of Universities for Research in Astronomy, Inc. under co-operative agreement with
the National Science Foundation} which involves bias subtraction, flat fielding and cosmic ray removal. 
Bias correction was carried out by subtracting a mean bias frame from all the images (flat frames and science frames) 
acquired during the night. The mean bias frame was generated by averaging several bias frames taken during the night
using the task {\it zerocombine}.  Flat fielding was done, by generating a flat field frame, which is a median
combination of several twilight flat field frames taken during an observing night. This median combined flat
field frame (generated using the task {\it flatcombine}) was used to generate 
the final science ready image frames. Cosmic rays from the cleaned image
frames were removed using the task {\it cosmicrays} in IRAF. 

To carry out aperture photometry on the blazar and the comparison stars present on the same CCD frame, we used the
task {\it phot} in IRAF. A crucial parameter  for aperture photometry is the radius of the aperture used for
photometry, which determines the S/N of the photometric points. Firstly, among the suitable comparison
stars, we selected three stars that give the steadiest differential light curve (DLC). Once this is determined we
generated may DLCs considering a range of aperture radii, and we selected 
the aperture that minimizes the 
scatter in the DLC. This aperture was then used for generating the DLCs of the blazar relative to the 
comparison stars as well as the DLCs between comparison stars. 
The positions of the comparison stars 
are given in Table 3. 

\section{Results}
The DLCs of the blazars relative to the comparison stars as well as the DLCs of 
the comparison stars along with the variation of the FWHM of the stellar light 
distribution are given in Figs. 1 - 9. The star - star DLC of any steady pair
of comparison stars represents the observational uncertainties on any 
particular 
night, whereas the DLC of the blazar with respect to the steady comparison stars
indicate the intrinsic variability of the blazar. We consider a source to be
variable if it shows correlated variations both in amplitude and time relative
to all the comparison stars. To check for the presence of INOV statistically, we
used the power enhanced F ($F_{enh}$) test of \cite{2014AJ....148...93D}, as this statistical test
is found to be devoid of the difficulties associated with the generally
used criteria such as the C and F-statistics. Since the work of 
\cite{2014AJ....148...93D}, $F_{enc}$ has found increased use in 
characterising the INOV nature of AGN. \citep{2015MNRAS.452.4263G,2017MNRAS.466.2679K}.

\subsection{Power enhanced F ($F_{enh}$) test}
Brightness difference between the blazar and the comparison stars as well as small variations in the comparison
stars my lead to incorrect characterisation of the variability of the blazar. Such difficulties are overcome
in $F_{enh}$, test that uses multiple comparison stars, thereby reducing the possibility of false INOV detections, compared
to that using one comparison star light curve. $F_{enh}$ is defined as 
\begin{equation}
F_{enh} = \frac{Var_{blazar}^2}{Var_{star}^2}
\end{equation}
Here, $Var_{blazar}^2$ is the variance of the DLC generated between the blazar and the reference star, while
$Var_{star}^2$ is the stacked variance of the DLCs comprising the reference star and the comparison stars, which is 
given as
\begin{equation}
Var_{star}^2 = \frac{1}{(\sum_{j=1}^k N_j - k)} \sum_{j=1}^k \sum_{i=1}^{N_{j}} \sigma^2_{j,i}
\end{equation}
Here, $N_j$ is the number of data points on the $j^{th}$ star, $k$ refers to the total
number of comparison stars and $\sigma^2_{j,i}$ is the scaled square deviation and is given as
\begin{equation}
\sigma^2_{j,i} = s_j (m_{j,i} - \overline{m_j})^2
\end{equation}
where, $s_j$ is the scaling factor, $m_{j,i}$ are the  differential magnitudes of the reference
star and the $j^{th}$ comparison star and $\overline{m_j}$ is the mean differential magnitude of the 
reference star and the comparison star DLC. The  scaling factor $s_j$ is the ratio of the
average square error of the points in the DLC of the blazar - reference star ($\overline{\sigma^2_{blazar}}$) to the 
average square error of the points in the DLC of the comparison - reference star ($\overline{\sigma^2_{sj}}$). 
For the $j^{th}$ star DLC, $s_j$ is given as
\begin{equation}
s_j = \frac{\overline{\sigma^2_{blazar}}}{\overline{\sigma^2_{sj}}}
\end{equation}
For all the blazars observed in this work, we have three comparison stars, and the star having brightness
similar to the blazar is taken as the reference star. The details of the comparison stars
are given in Table 3. The blazar is considered to be variable, if 
the calculated $F_{enh}$ is greater than a critical value $F_c$ for $\alpha$ = 0.05, which corresponds to 
95\% confidence. The results of variability are given in Table 4.
\subsection{Variability amplitude}
For blazars that were found to show INOV based on the $F_{enh}$ criteria, we calculated the 
amplitude of variability ($A_{var}$)  given by  \cite{1996A&A...305...42H}. $A_{var}$ is defined
as
\begin{equation}
A_{var} = \sqrt{(A_{max} - A_{min})^2 - 2\sigma^2}
\end{equation}
Here, $A_{max}$ and $A_{min}$ are the maximum and minimum in the blazar - reference star
DLC, and $\sigma^2$ is the variance of the steadiest reference star - comparison star DLC.
The amplitude of variability calculated for the variable blazars is given in Table 4.

\subsection{Duty cycle of variability}
Blazars are not found to show INOV on all the nights of observations. Therefore, to further characterize
variability, we calculated the duty cycle (DC) of variability. DC is defined as the ratio of the time over
which a blazar is variable to the total time spent on observing the blazar. According to 
\cite{1999A&AS..135..477R}, DC is defied as
\begin{equation}
DC = 100 \frac{\sum_{i=1}^n N_i(1/\Delta t_i)}{\sum_{i=1}^n(1/\Delta t_i)}
\end{equation}
Here, $N_i$ is equal to 1, if the source is found to show INOV and zero otherwise and
$\Delta t_i$ = $\Delta t_(i,obs)(1+z)^{-1}$, is the redshift corrected time for which 
a source is observed. For FSRQs we found a DC of 10.9\%, while for BL Lacs we found a 
DC of 12.2\%. Thus, BL Lacs are found to show a higher DC of variability than 
FSRQs. Separating
the blazars into different spectral classes we found DCs of 16\%, 10.4\% and 
7.5\% for
LSP, ISP and HSP blazars. Thus, among the blazar sub-classes, LSPs are found to show high
DC of variability.

\begin{table}
\tabularfont
\caption{Log of observations.Columns are (i) the name of the source, (ii) date 
of observation, (iii) duration of observation in hours, (iv) the total number 
of data points in a night, (v) the exposure time in seconds and (vi) the mode of 
the CCD used (N=normal, EM=electron multiplying).}\label{tableExample} 
\resizebox{0.45\textwidth} {!}{ 
\begin{tabular}{lccrrc}
\topline

3FGL Name    & Date         & $\Delta$ t   & Pts &  Time  & mode \\ 
             &              & (hrs.) &      & (sec)     &      \\ \midline
 J0050.6$-$0929 & 31.12.2016  & 2.75   & 14  &  600 & N \\
                & 25.01.2019  & 2.01   &  9  &  300 & N \\
                & 26.01.2019  & 2.23   & 15  &  300 & N \\
 J0109.1+1816  &  28.12.2016  & 3.25   & 10  &  900 & N \\
               &  27.11.2018  & 3.01   & 14  &  900 & N \\
               &  31.12.2018  & 3.50    & 13  &  900 & N \\
 J0112.1+2245  &  29.12.2016  & 2.24   &  8  & 600  & N \\
               &  29.10.2017  & 2.01   & 15  & 300  & N \\
               &  20.12.2017  & 3.10   & 30  & 300  & N \\
 J0217.2+0837  &  27.12.2016  & 4.17   & 19  & 600  & N \\
               &  23.12.2017  & 4.38   & 42  & 300  & N \\
               &  24.12.2017  & 4.22   & 41  & 300  & N \\
 J0303.4$-$2407&  26.12.2016  & 6.01   & 27  & 600  & N \\
               &  01.02.2018  & 2.03   & 11  & 600  & N \\
               &  02.02.2018  & 2.27   & 15  & 600  & N \\
 J0738.1+1741  &  15.02.2020  & 2.13   &  7  &  600 & N \\
               &  16.02.2020  & 1.81   &  6  &  900 & N \\
               &  17.02.2020  & 4.29   &  8  &  900 & N \\
 J0739.4+0137  &  24.02.2017  & 5.58   &  8  & 2400 & EM     \\
               &  24.02.2019  & 5.11   & 18  & 900  & N \\
               &  15.03.2019  & 4.63   &  9  & 1200 & N \\
 J0825.9$-$2230 & 16.03.2019  & 5.01   & 16  &  900 & N \\
                & 17.03.2019  & 2.73   &  7  &  900 & N \\
                & 19.03.2019  & 3.18   &  8  &  900 & N \\
 J0846.7-0651  &  16.02.2020  & 4.58   & 9   & 1800 & N \\
               &  17.02.2020  & 3.66   & 7   & 1800 & N \\
               &  17.03.2020  & 4.88   & 9   & 1800 & N \\
 J0854.8+2006  &  25.02.2017  & 5.01   & 25  & 600  & EM     \\
               &  02.02.2018  & 2.75   & 15  & 600  & N \\
               &  20.03.2018  & 4.05   & 22  & 600  & N \\
 J0912.9$-$2104&  02.03.2017  & 2.25   & 8   &  900 & EM     \\
               &  26.01.2019  & 3.51   & 16  &  600 & EM     \\
               &  25.02.2019  & 6.87   & 13  & 1800 & N \\
 J0927.9$-$2037&  03.03.2017  & 2.04   &  7  &  900 & EM     \\
               &  29.03.2017  & 2.06   &  9  &  900 & EM     \\
               &  21.03.2018  & 5.03   & 20  &  900 & N \\
 J1015.0+4925  &  27.03.2017  & 5.22   & 20  & 900  & EM     \\
               &  21.04.2019  & 3.65   & 13  & 1200 & N \\
               &  15.02.2020  & 6.52   & 24  &  900 & N \\
 J1129.9$-$1446 & 17.03.2020  & 3.12   &  6  & 1800 & N \\
                & 18.03.2020  & 6.01   & 11  & 1800 & N \\
 J1224.9+2122   &  24.02.2019  & 1.66   &  6  & 900  &N \\
               &  17.03.2019  & 1.67   &  7  & 1200 & N \\
               &  24.04.2019  & 2.51   &  8  &  900 & N \\
 J1229.1+0202  &  02.03.2017  & 5.98   & 46  & 300  & EM \\
               &  24.03.2017  & 6.24   & 65  & 300  & EM \\
               &  27.04.2017  & 4.5    & 69  & 300  & N \\
 J1427.0+2347   & 25.03.2017  & 5.28   & 28  &  600 & EM     \\
                & 30.03.2017  & 2.88   & 11  &  900 & EM     \\
                & 21.03.2018  & 3.82   & 20  &  600 & N \\
 J1555.7+1111   & 24.03.2017   & 1.50   & 12  &  300 & EM     \\
               & 15.03.2019   & 4.2    & 42  &  300 & N \\
               & 19.03.2019   & 4.02   & 19  &  600 & N \\     
\hline
\end{tabular}}
\end{table}

\section{Notes on individual sources }
\noindent {\bf 3FGL J00506$-$0929}:  The source, an ISP BL Lac was observed on 
three nights, once is December 2016 and twice in January 2019.  On all the three 
nights the source was found to be non variable.  \\

\noindent{\bf 3FGL J0109.1+1816}: This HSP BL Lac at a redshift of $z=0.443$ 
was 
observed on a night in December 2016 and again on two nights in November and  
December 2018.\ This source has been studied for INOV for the first time. In December 2016, there is some indication of the source 
to have shown flux variations, however, statistical tests show the source to be
non-variable. 
On the other two 
nights too, the source  was found not to show INOV, however, between 27 
November 2018 and 31 December 2018, the source was found to increase in 
brightness by $\sim$0.1 mag.  \\

\noindent{\bf 3FGL J0112.1+2245:} The source is an ISP BL Lac. It was observed
on three epochs over a period of about an year. Of the three epochs,
the source was found to show INOV on 20 December 2017, with an
amplitude of variability of about 2\%.\ This source has not been studied for INOV before.\\

\noindent {\bf 3FGL J0217.2+0837:}
This source is a LSP BL Lac. It was observed for three nights between 20 December 2016 and
24 December 2017. It was found to show INOV on two nights with amplitude
of variability of about 10\% and 5\% respectively. \\

\noindent{\bf 3FGL J0303.4$-$2407:} This source was first observed on 26 December 2016 
and again on 01 and 02 February 2018. No INOV was detected in this source on 
all the three epochs.\ The observations reported here are the first time measurements for INOV.\\

\noindent {\bf 3FGL J0738.1+1741:} This source was observed for INOV on 15, 16 and 17
February 2020. Of the three nights, it was found to show INOV on 17 February 2020
with an amplitude of variability of about 11\%. \\

\noindent {\bf 3FGL J0739.4+0137:} It is a FSRQ and belongs to the spectral class ISP. It was observed for INOV
on three nights over a two year period. It was found to show low amplitude INOV
of about 4\% on one night (24 February 2019), while on the remaining two nights
INOV was not detected in this source. \\

\noindent{{\bf 3FGL J0825.9$-$2230:}} This source is an ISP BL Lac and has never been studied for INOV before. It was observed 
for INOV on three nights during March 2019. It was found not to show INOV on 
all the three nights. \\

\noindent {\bf 3FGL J0846.7$-$0651:} This source was observed for three epochs 
in the year 2020 within a span of about a month. It was found to be 
non-variable on all the three nights of observations.\ The observations reported here are the first time
measurements for INOV on this source. \\

\noindent{\bf 3FGL J0854.8+2007:} It is a well known  BL Lac and belongs
to the LSP type. It was observed on 25 February 2017 and again in February and 
March 2018. It showed INOV on one of the three nights of observations (20 
March 2018) with an amplitude of variability of about $\sim$2\%. \\

\noindent{\bf 3FGL J0912.9$-$2104:}. This source is a HSP BL Lac.\ Reports on the INOV nature of this source are not available in literature. It was observed for three epochs between March 2017 and February 2019. The source was 
not found to show INOV on all the three epochs.  \\

\noindent{\bf 3FGL J0927.9$-$2037:} This source was not found to show INOV on 
all the three nights it was observed spanning about an year. \\

\noindent{\bf 3FGL J1015.0+4925:} This source was observed for three nights on 
21 March 2017, 21 April 2019 and 15 February 2020.  It was found to be 
non-variable on all the three nights. \\

\noindent{\bf 3FGL J1129.9$-$1446:} It is a FSRQ and is the highest 
redshift source in our sample. It was observed on 17 and 18 Mach 2020, 
however, no INOV was detected. \\

\noindent{\bf 3FGL J1224.9+2122:} It is a FSRQ. 
No INOV was found from observations carried out on all the three nights. It has not been studied for INOV before.\\

\noindent{\bf 3FGL J1229.1+0202:} This source a FSRQ was observed on three 
nights for INOV over a period of two months. Of the three nights of 
observations, INOV was detected on two nights (24 March 2017 
and 27 April 2017) with amplitude of INOV of about 2\% and 4\% respectively. \\

\noindent{\bf 3FGL J1427.0+2347:} Of the three nights of observations carried 
out on this source over a period of about an year, INOV was detected on one 
night (30 March 2017) with an amplitude of variability of about 1\%. \\

\noindent{\bf 3FGL J1555.7+1111:}. No INOV was detected in this source on all 
the three nights of observations carried out  over a period of about 2 years. \\

\begin{table}
\tabularfont
\caption{Details of the comparison stars used in this work.}\label{table2} 
\resizebox{0.45\textwidth} {!}{ 
\begin{tabular}{cccr}
\topline
3FGL Name &  Stars & $\alpha_{2000}$ & $\delta_{2000}$ \\ \midline
 J0050.6$-$0929 & S1  & 00:50:51.28  & $-$09:25:49.80 \\
                & S2  & 00:50:47.11  & $-$09:30:15.70 \\
                & S3  & 00:50:59.01  & $-$09:30:46.00 \\
 J0109.1+1816   & S1  & 01:09:01.83  & 18:19:57.28   \\
                & S2  & 01:08:49.76  & 18:18:33.37   \\
                & S3  & 01:08:48.00  & 18:13:52.42   \\
 J0112.1+2245   & S1   & 01:12:00.56  & 22:45:17.60    \\
                & S2   & 01:12:10.21  & 22:44:35.20   \\
                & S3   & 01:12:20.20  & 22:43:00.48   \\
 J0217.2+0837   & S1   & 02:17:08.20  & 08:37:16.70    \\
                & S2   & 02:17:20.72  & 08:39:02.30    \\
                & S3   & 02:17:16.38  & 08:36:34.20    \\
 J0303.4$-$2407 & S1   & 03:03:21.28  & $-$24:06:17.60 \\
                & S2   & 03:03:15.52  & $-$24:05:39.10 \\
                & S3   & 03:03:25.63  & $-$24:09:23.60 \\
 J0738.1+1741   & S1   & 07:38:08.67  & 17:40:28.10   \\
                & S2   & 07:38:02.46  & 17:41:23.70   \\
                & S3   & 07:38:00.55  & 17:41:21.20   \\
 J0739.4+0137   & S1   & 07:39:16.04  & 01:37:35.61    \\
                & S2   & 07:39:11.98  & 01:37:09.86    \\
                & S3   & 07:39:13.34  & 01:35:43.85    \\
 J0825.9$-$2230  & S1  & 08:25:53.54  & $-$22:30:45.38 \\
                & S2  & 08:25:40.99  & $-$22:31:46.39 \\
                & S3  & 08:26:11.36  & $-$22:33:42.86 \\
 J0846.7$-$0651 & S1   & 08:47:55.70  & $-$07:03:09.87 \\
                & S2   & 08:47:55.59  & $-$07:03:30.35 \\
                & S3   & 08:48:02.71  & $-$07:04:30.03 \\
 J0854.8+2006   & S1   & 08:54:54.41  & 20:06:12.90    \\
                & S2   & 08:54:55.18  & 20:05:41.80    \\
                & S3   & 08:54:53.27  & 20:04:45.30    \\
 J0912.9$-$2104 & S1   & 09:13:06.17  & $-$21:02:23.65 \\
                & S2   & 09:13:01.06  & $-$21:01:21.78 \\
                & S3   & 09:12:53.26  & $-$21:05:20.78 \\
 J0927.9$-$2037 & S1  & 09:27:54.11  & $-$20:34:47.30 \\
                & S2  & 09:27:44.69  & $-$20:33:20.20 \\
                & S3  & 09:27:50.11  & $-$20:35:32.40 \\
 J1015.0+4925   & S1   & 10:15:08.02  & 49:25:41.80    \\
                & S2   & 10:14:53.81  & 49:25:32.80    \\
                & S3   & 10:15:08.89  & 49:27:15.60    \\
 J1129.9$-$1446 & S1  & 11:30:08.71  & $-$14:51:41.40 \\
                & S2  & 11:29:59.87  & $-$14:49:42.54 \\
                & S3  & 11:29:58.07  & $-$14:50:44.90 \\
 J1224.9+2122    & S1   & 12:25:01.34  & 21:23:21.33    \\
                & S2   & 12:24:55.83  & 21:25:53.34    \\
                & S3   & 12:24:43.95  & 21:19:18.47   \\
 J1229.1+0202   & S1   & 12:29:03.15  & 02:03:12.20  \\
                & S2   & 12:29:02.76  & 02:02:16.10  \\
                & S3   & 12:29:07.82  & 02:03:35.40  \\
 J1427.0+2347   & S1  & 14:26:57.02  & 23:48:02.30    \\
                & S2  & 14:26:52.97  & 23:49:11.10    \\
                & S3  & 14:27:03.36  & 23:45:45.81    \\
 J1555.7+1111    & S1  & 15:55:46.08  & 11:11:19.80    \\
                & S2  & 15:55:49.97  & 11:09:55.49    \\
                & S3  & 15:55:42.42  & 11:09:21.06    \\
\hline
\end{tabular}}
\end{table}

\begin{table*}
\caption{Results of variability. Columns are Name, reference star used, date of
observation, degrees of freedom, $F_{enh}$, critical F value,   
variability status(NV:non-variable, V:variable) and the amplitude of variability.} 
\label{tab:var_res}                   
\centering 
\resizebox{0.85\textwidth} {!}{                     
\begin{tabular}{lccccccccccc}           
\hline                		 
3FGL Name  & Reference star  &  Date &  DoF($\nu_1$,$\nu_2$ ) & $F_{\rm enh}$ & $F_c$  &  Status  &  Amplitude ($\%$)\\
\hline

J0050.6-0929  &  S1   	& 31-12-2016 &  13, 26 & 0.55 & 2.90 & NV & - \\ 
	      &	& 25-01-2019 &   8, 16 & 1.28 & 3.89 & NV & - \\ 
	      &	& 26-01-2019 &  14, 28 & 0.63 & 2.79 & NV & - \\ 
		 		
J0109.1+1816  &  S1   	& 28-12-2016 &  9, 18 & 2.14 & 3.60 & NV  & - \\ 
	      &	& 27-11-2018 &  13, 26 & 0.30 & 2.90 & NV & - \\ 
	      &	& 31-12-2018 &  12, 24 & 0.39 & 3.03 & NV & - \\ 
		 		
J0112.1+2245  &  S3 	& 29-12-2016 &   7, 14 & 1.24 & 4.28 & NV & - \\ 
	      &	& 29-10-2017 &  14, 28 & 0.31 & 2.79 & NV & - \\ 
	      &	& 20-12-2017 &  29, 58 & 2.42 & 2.05 & V  & 1.60 \\ 
		 		
J0217.2+0837  &  S2   	& 20-12-2016 &  19, 38 & 1.35 & 2.42 & NV & - \\ 
	      &	& 23-12-2017 &  41, 82 & 54.28 & 1.84 & V  & 9.78 \\ 
	      &	& 24-12-2017 &  40, 80 & 2.05 & 1.85 & V  & 5.33 \\ 
		 		
J0303.4-2407  &  S1   	& 26-12-2016 &  26, 52 & 1.10 & 2.14 & NV & - \\ 
	      &	& 01-02-2018 &  10, 20 & 2.29 & 3.37 & NV & - \\ 
	      &	& 02-02-2018 &  14, 28 & 1.28 & 2.79 & NV & - \\ 
		 		
J0738.1+1741  &  S1   	& 15-02-2020 &   5, 10 & 1.94 & 5.64 & NV & - \\ 
	      &	& 16-02-2020 &   4, 8 & 1.14 & 7.01 & NV & - \\ 
	      &	& 17-02-2020 &   7, 14 & 4.41 & 4.28 & V  & 11.48 \\ 
		 		
J0739.4+0137  &  S2 	& 24-02-2017 &   7, 14 & 0.06 & 4.28 & NV & - \\ 
	      &	& 24-02-2019 &  17, 34 & 3.39 & 2.54 & V  & 4.05 \\ 
	      &	& 15-03-2019 &   8, 16 & 1.19 & 3.89 & NV & - \\ 
		 		
J0825.9-2230  &  S2 	& 16-03-2019 &  16, 32 & 1.12 & 2.62 & NV & - \\ 
	      &	& 17-03-2019 &   6, 12 & 0.11 & 4.82 & NV & - \\ 
	      &	& 19-03-2019 &   7, 14 & 1.22 & 4.28 & NV & - \\ 
		 		
J0846.7-0651  &  S3   	& 16-02-2020 &   8, 16 & 0.38 & 3.89 & NV & - \\ 
	      &	& 17-02-2020 &   6, 12 & 0.34 & 4.82 & NV & - \\ 
	      &	& 17-03-2020 &   8, 16 & 0.57 & 3.89 & NV & - \\ 
		 		
J0854.8+2006  &  S3   	& 25-02-2017 &  24, 48 & 2.00 & 2.20 & NV & - \\ 
	      &	& 02-02-2018 &  14, 28 & 0.72 & 2.79 & NV & - \\ 
	      &	& 20-03-2018 &  21, 42 & 2.79 & 2.32 & V  & 1.62 \\ 
		 		
J0912.9-2104  &  S3     & 02-03-2017 &   7, 14 & 1.77 & 4.28 & NV & - \\ 
	      &         & 26-01-2019 &  15, 30 & 0.76 & 2.70 & NV & - \\ 
	      &         & 25-02-2019 &  12, 24 & 0.89 & 3.03 & NV & - \\ 	
		 		
J0927.9-2037  &  S2   	& 03-03-2017 &   6, 12 & 0.31 & 4.82 & NV & - \\ 
	      &	& 29-03-2017 &   8, 16 & 0.99 & 3.89 & NV & - \\ 
	      &	& 21-03-2018 &  19, 38 & 0.78 & 2.42 & NV & - \\ 
		 		
J1015.0+4925  &  S2  	& 21-03-2017 &  19, 38 & 0.56 & 2.42 & NV & - \\ 
	      &	& 21-04-2019 &  11, 22 & 2.50 & 3.18 & NV & - \\ 
	      &	& 15-02-2020 &  23, 46 & 0.49 & 2.24 & NV & - \\ 	
		 		
J1129.9-1446  &  S3   	& 17-03-2020 &   5, 10 & 2.23 & 5.64 & NV & - \\ 
	      &	& 18-03-2020 &  10, 20 & 0.28 & 3.37 & NV & - \\ 
		 		
J1224.9+2122   &  S2   	& 24-02-2019 &   5, 10 & 0.40 & 5.64 & NV & - \\ 
	      &	& 17-03-2019 &   5, 10 & 1.30 & 5.64 & NV  & - \\ 
	      &	& 24-04-2019 &   7, 14 & 2.38 & 4.28 & NV & - \\ 
		 		
J1229.1+0202  &  S1   	& 02-03-2017 &  42, 84 & 0.80 & 1.82 & NV & - \\ 
	      &	& 24-03-2017 &  59,118 & 1.78 & 1.66 & V  & 1.90 \\ 
	      &	& 27-04-2017 &  59,118 & 3.38 & 1.66 & V  & 4.04 \\ 
		 		
J1427.0+2347  &  S2   	& 25-03-2017 &  27, 54 & 1.30 & 2.11 & NV & - \\ 
	      &	& 30-03-2017 &  10, 20 & 5.11 & 3.37 & V  & 1.21 \\ 
	      &	& 21-03-2018 &  19, 38 & 1.12 & 2.42 & NV & - \\ 
		 		
J1555.7+1111  &  S1     & 24-03-2017 &  11, 22 & 1.29 & 3.18 & NV & - \\ 
	      &	& 15-03-2019 &  41, 82 & 0.94 & 1.84 & NV & - \\ 
	      &	& 19-03-2019 &  18, 36 & 0.70 & 2.48 & NV & - \\ 
\hline 
\end{tabular}}
\end{table*}						

\section{Summary}
We have presented here the results of our monitoring observations of 18 
blazars that include 5 FSRQs and 13 BL Lacs.\ For seven sources in our sample, INOV characteristics have been investigated for the first time. Dividing the sources based on the position of the synchrotron peak in their broad band SED, our
sample consists of 8 LSPs, 4 ISPs and 6 HSPs. Observations
were carried out on this sample for a total of 40 nights with the duration
of observations on a night ranging from 1.5  to 6.9 hrs. The presence or
absence of INOV during an epoch of observation was characterised by the
$F_{enh}$ statistical criteria. Using this criteria we could detect INOV in 
few sources. We further characterized the INOV properties of the different classes
of sources by calculating their DC of variability. For BL Lacs we
found  a high DC of about $\sim$12\%, while for FSRQs we found
a low DC of INOV of $\sim$11\%.  Using densely sampled long duration
monitoring observations within a night, \cite{2004JApA...25....1S} 
found very high DC in BL Lacs relative to FSRQs. This has been 
attributed to high Doppler boosting in the case of BL Lacs. 
However, \cite{2017ApJ...844...32P} reported high DC of INOV in FSRQs 
compared to BL Lacs. In this work, though BL Lacs have a high DC of INOV
than FSRQs, they are not significantly different. This could be due to
the poor timing resolution as well as the shorter duration of observations. 
 Dividing the sample based on the 
position of the synchrotron peak in their broad band SED, we
found DC of variability of about 16\%, 10\% and 7\% for LSP, ISP and
HSP blazars respectively. Our observations thus point to BL Lacs having 
marginally higher DC of INOV relative to FSRQs and LSP blazars having high 
DC of INOV relative to other spectral classes of blazars.  Such
high incidence of INOV in LSP blazars related to other spectral classes
of blazars has also been reported by \cite{2017ApJ...844...32P}.  
These new observations can be well explained in the leptonic model of 
emission from blazar jets (see \citealt{2017ApJ...844...32P}). 
In the typical low energy synchrotron
component of the broad band SED of blazars, the optical R-band used
in the observations reported in this work traces the falling part of the
SED in the case of LSP sources (majority of FSRQs are LSP sources; of 
the five FSRQs, four are LSP sources and one is an ISP source) and
the rising part of the SED in the case of BL Lac sources (majority of BL Lacs 
are HSP sources; of the thirteen BL Lacs, four belong to LSP type, 
three belong to ISP category and six are HSP sources). This means, in the case 
of LSP sources, the 
R-band traces the emission contributed mostly by the high energy
electron population, leading to faster variations in them.
Similarly, in the case of HSP blazars, in the region covered by R-band, the emission
is dominated by low energy electrons, leading to low INOV. Thus, the 
observed INOV characteristics of the blazars studied in this work is 
understandable in the leptonic scenario of emission from blazar jets. 
The observations reported in this work are of moderate quality and also
have low duration of monitoring in a night. We note that the
chances of detecting INOV increases with continuous observations of 
three to four hours \citep{1990PhDT.......263C} and 
thus good quality long duration observations will be able to detect INOV in 
more blazars. Though the observations reported here are 
broadly consistent with the jet based models, it would be of interest to 
find minute scale variations in the optical to further refine models of flux 
variations in AGN.


\section*{Acknowledgements}
The authors thank the referee for his/her comments on the manuscript. They also thank P. Anbazhagan, G. Selvakumar, V. Moorthy, R.Surendar, 
S. Venkatesh, B. Rahul, and the staff members of VBO, for their assistance 
during the observing run. K.S.P. and A.N. acknowledge the warm hospitality 
received at VBO, where this work was completed and the facilities at VBO,kavalur 
operated by the Indian Institute of Astrophysics, Bangalore. This research made 
use of data provided by the NASA/ IPAC Extragalactic Database (NED) and 
SIMBAD database(the Set of Identifications, Measurements and Bibliography for 
Astronomical Data).

\end{document}